\documentclass[hyper,12pt,letterpaper]{JHEP3}

\usepackage{latexsym,amsmath,amsfonts,amssymb}
\usepackage[dvips]{graphicx}
\usepackage{bbm}
\usepackage{subfigure}

\newcommand{\eg}{\textit{e.g.}}

\newcommand{\ie}{\textit{i.e.}}

\newcommand{\mr}{\mathrm}
\newcommand{\vev}[1]{\langle #1 \rangle}

\numberwithin{equation}{section}

\newcommand{\nn}{\nonumber}

\newcommand{\be}{\begin{equation}} \newcommand{\ee}{\end{equation}}
\newcommand{\bea}{\begin{equation} \begin{aligned}} \newcommand{\eea}{\end{aligned} \end{equation}}
\newcommand{\tabs}{\rule[-1.2ex]{0pt}{3.9ex}}

\newcommand{\calC}{\mathcal{C}}

\newcommand{\calN}{\mathcal{N}}

\newcommand{\bbC}{\mathbb{C}}

\newcommand{\bbZ}{\mathbb{Z}}

\newcommand{\cC}{\mathcal{C}}
\newcommand{\cD}{\mathcal{D}}

\newcommand{\cM}{\mathcal{M}}
\newcommand{\cN}{\mathcal{N}}
\newcommand{\cO}{\mathcal{O}}

\newcommand{\cZ}{\mathcal{Z}}

\newcommand{\bC}{\mathbb{C}}

\newcommand{\bP}{\mathbb{P}}

\newcommand{\bR}{\mathbb{R}}
\newcommand{\bZ}{\mathbb{Z}}

\newcommand{\unit}{\mathbbm{1}}

\DeclareMathOperator{\sgn}{Sgn}
\DeclareMathOperator{\Tr}{Tr}
\DeclareMathOperator{\perm}{perm}

\title{Chiral flavors and M2-branes at toric CY$_4$ singularities}

\let\CC\spadesuit
\let\AA\blacklozenge
\let\BB\clubsuit
\author{Francesco Benini$^\CC$, Cyril Closset$^\AA$, Stefano Cremonesi$^\BB$\\

$^\CC$ Department of Physics, Princeton University, \\
Princeton, NJ 08544, USA\\
$^\AA$ Physique Th\'eorique et Math\'ematique and International Solvay
Institutes \\
Universit\'e Libre de Bruxelles, C.P. 231, 1050
Bruxelles, Belgium  \\
$^\BB$ Raymond and Beverly Sackler School of Physics and Astronomy \\
Tel-Aviv University, Ramat-Aviv 69978, Israel \\
}

\preprint{PUTP-2324 \\ TAUP-2906/09}

\abstract{
We extend the stringy derivation of $\cN=2$ AdS$_4$/CFT$_3$ dualities to cases where the M-theory circle degenerates at complex codimension-two submanifolds of a toric conical CY$_4$. The type IIA backgrounds include D6-branes, and the dual $\cN=2$ quiver gauge theories contain chiral flavors. We provide a general recipe to derive the geometric moduli space of flavored versions of Abelian toric quiver gauge theories. The CY$_4$ cone is reproduced thanks to a non-trivial quantum F-term relation between diagonal monopole operators and bifundamental fields. We find new field theory duals to many geometries, including $Q^{111}$.}

\begin{document}



\section{Introduction}
\label{sec: intro}


Conformal field theories (CFT) living on a stack of $N$ M2-branes at the tip of eight-dimensional cones are attracting a lot of attention.
In the large $N$ limit, such theories are holographically dual to $AdS_4 \times H_7$ Freund-Rubin solutions of M-theory, where $H_7$ is the seven-dimensional base of the cone.
In a seminal paper \cite{Aharony:2008ug}, Aharony, Bergman, Jafferis and Maldacena (ABJM) proposed an $\cN=6$ supersymmetric three-dimensional quiver Chern-Simons (CS) theory with two gauge groups at levels $k$ and $-k$ and bifundamental matter as a dual of M-theory on $AdS_4\times S^7/\bZ_k$. When $k=1,2$, non-perturbative effects enhance the supersymmetry to $\cN=8$. The proposal was gradually extended to lower supersymmetry. If $\cN\geq 3$, $H_7$ is 3-Sasakian and hyper-K\"ahler geometry can be used to construct dual pairs \cite{Benna:2008zy, Imamura:2008nn, Hosomichi:2008jb, Aharony:2008gk, Jafferis:2008qz}.
If $\cN=2$, the cone is a Calabi-Yau (CY) four-fold and $H_7$ is a Sasaki-Einstein manifold. The study of $\cN = 2$ quiver Chern-Simons theories dual to Calabi-Yau cones was initiated in \cite{Martelli:2008si, Hanany:2008cd}, followed by a large number of works \cite{Ueda:2008hx, Imamura:2008qs, Hanany:2008fj, Franco:2008um, Hanany:2008gx, Amariti:2009rb, Franco:2009sp, Davey:2009sr, Aganagic:2009zk, Davey:2009qx, Martelli:2009ga, Davey:2009bp, Hewlett:2009bx, Taki:2009wf}.
Examples with minimal $\cN=1$ supersymmetry \cite{Ooguri:2008dk, Gaiotto:2009mv, Bobev:2009ms, Forcella:2009jj}
have also been proposed.

All these CFTs involve gauge groups with only adjoint and bifundamental matter, like conformal quiver gauge theories in 3+1 dimensions.
More recently, it has been proposed that the dynamics of M2-branes on some hyper-K\"ahler cones ($\cN=3$ SUSY) is described by flavored quiver CS theories, including matter in the fundamental and antifundamental representation of the gauge groups \cite{Gaiotto:2009tk, Hohenegger:2009as, Hikida:2009tp}. Flavors were further studied in \cite{Fujita:2009xz, Jensen:2009xh, Ammon:2009wc, JafferisToAppear}. We aim to extend this program to M2-branes probing toric CY$_4$ singularities ($\cN=2$ SUSY).

One of the problems we want to address is what happens when the four-fold has conical complex codimension-two singularities, which means that the base $H_7$ itself has codimension-two singularities: this is related to the addition of \emph{flavors} -- fields in the fundamental representation of the gauge groups. An $A_{h-1}$ complex codimension-two singularity locally looks like $\bC^2 \times \bC^2/\bZ_h$. M-theory on such a background develops $SU(h)$ gauge fields living along the singularity, and by the AdS/CFT map there must be an $SU(h)$ global symmetry in the boundary theory. Many models in the literature have such singularities, however the large non-Abelian symmetry is not manifest. It is natural to look for a description in terms of flavors in the quiver theory.

Another way to understand the issue is to select a $U(1)$ isometry of the CY$_4$ that preserves the holomorphic 4-form $\Omega_4$, quotient the geometry by $\bZ_k \subset U(1)$ and reduce along the circle to type IIA. The resulting background is a warped product $AdS_4 \times_w H_6$, with RR fluxes and varying dilaton.
If $1\ll k \ll N^{1/5}$ type IIA is weakly coupled, whereas for $k\gg N$ one expects a Lagrangian description for the 2-brane theory with weakly coupled gauge groups. If the $U(1)$ circle shrinks on a complex codimension-two surface in the CY$_4$, we get
D6-branes in the type IIA background, filling $AdS_4$ and wrapping a 3-cycle in $H_6$. In fact, $\bC^2\times \bC^2/\bZ_h$ is the complex structure of a multi-Taub-NUT which, if reduced along its $U(1)$ isometry, gives rise to $h$ D6-branes.
It is known that the D2-D6 system introduces flavors in the theory living on D2-branes, as happens in the $\cN=3$ case \cite{Gaiotto:2009tk, Hohenegger:2009as, Hikida:2009tp}.

A more systematic tool to derive the theory on M2-branes probing a CY$_4$ geometry is the K\"ahler quotient developed in \cite{Aganagic:2009zk}. Every toric conical CY$_4$
can be written as a $U(1)$ fibration over a seven-manifold, which is a toric conical CY$_3$ fibered along $\bR$. Under some regularity conditions (stressed in Section \ref{sec: U(1) fibration}), the theory living on M2-branes on the CY$_4$ can be written as the theory living on D3-branes on the CY$_3$, dimensionally reduced and refined by Chern-Simons couplings, which encode the details of the fibration. This approach is powerful because it does not need metric details of the  four-fold, but only algebraic geometric data. When the $U(1)$ fiber shrinks on codimension-two submanifolds of the CY$_4$, we get D6-branes wrapping divisors of the CY$_3$, and the theory on M2-branes has the same quiver and superpotential as the theory on D3-branes on the CY$_3$ in the presence of D7-branes wrapping the same divisors. For $h$ D7-branes wrapping an irreducible divisor, the effect is that of introducing $h$ pairs of quarks $(p,q)$ coupled via the superpotential term
\be \nn
\delta W = p \, (\text{divisor equation}) \, q \;,
\ee
where the divisor equation is written in terms of the bifundamental matter fields in the theory.

Led by these considerations, we can study what happens if we start with an $\cN=2$ quiver Chern-Simons theory, dual to a toric CY$_4$ geometry, and we flavor it.
We mean that we select a subset $\{X_\alpha\}$ of bifundamental fields in the quiver, and for each of them we introduce $h_\alpha$ pairs of chiral multiplets $(p_\alpha, q_\alpha)$ in the (anti)fundamental representation of the gauge groups, coupled by the superpotential term
\be \nn
W = W_0 + \sum\nolimits_\alpha p_\alpha X_\alpha q_\alpha \;,
\ee
$W_0$ being the ``unflavored'' superpotential. Because of the parity anomaly, this has to be accompanied by a shift of Chern-Simons levels. A concept of ``chirality'' is induced by $\cN=2$ supersymmetry, and inherited from four dimensions.

To study the chiral ring and moduli space of this theory, a crucial r\^ole is played by BPS diagonal monopole operators $T^{(n)}$ \cite{'tHooft:1977hy, Polyakov:1976fu, Moore:1989yh, Borokhov:2002ib, Borokhov:2002cg, Borokhov:2003yu, Itzhaki:2002rc, Benna:2009xd}. Due to quantum corrections,
they acquire global and gauge charges in the presence of flavors.
Generically there is only one possible non-trivial OPE compatible with all the symmetries, that in the Abelian case reads
\be \nn
T \tilde T = \prod\nolimits_\alpha (X_\alpha)^{h_\alpha} \;.
\ee
We conjecture that this quantum F-term relation holds, since our results strongly support this claim from the AdS/CFT point of view. The moduli space has Higgs and Coulomb branches. We show that the geometric branch, in which $p_\alpha = q_\alpha=0$, is described by the matter fields $X_a$ plus the two monopole operators $T$ and $\tilde T$, subject to the classical F-term relations from $W$ plus this quantum F-term relation, modded out by the full gauge group $U(1)^G$.
The geometric moduli space is still a toric CY$_4$, that we precisely identify. Similar ideas appeared in \cite{JafferisTalk}.

The paper is organized as follows. In Section \ref{sec: M-theory} we start with a top-down perspective, and analyze the K\"ahler quotient reduction of M-theory in the presence of KK monopoles. In Section \ref{sec: flavoring} we turn to a bottom-up approach and flavor quiver Chern-Simons theories; their moduli space is studied in Section \ref{sec: moduli space}. Section \ref{sec: geometry} is devoted to deformations by real and complex masses. In Section \ref{sec: examples} we work out many examples. We conclude in Section \ref{sec: conclusions}, followed by two Appendices.

\textbf{Note added:} While writing up our results, we became aware of the related work of Daniel Jafferis \cite{JafferisPaper}, with whom we coordinated the release of the paper.


\section{M-theory reduction and D6-branes \\ A top-down perspective}
\label{sec: M-theory}


Let us consider M2-branes probing a toric conical Calabi-Yau four-fold $Y_4$ in M-theory.%
\footnote{We refer the reader to \cite{Franco:2005sm, Kennaway:2007tq, Yamazaki:2008bt, Closset:2009sv} for a simple introduction to basic facts about toric geometry and its relevance for quiver gauge theories.}
We are interested in the type IIA string theory background that one obtains by reducing along a $U(1)$ isometry, in particular in the case that the four-fold contains KK monopoles and the $U(1)$ shrinks along them.
The isometry group of a toric four-fold contains $U(1)^4$. A specific $U(1)$ or more generally $\bR$ subgroup is the superconformal R-symmetry, while the remaining commuting $U(1)_F^3$ leaves the holomorphic 4-form invariant.
Reduction along a circle in $U(1)_F^3$ manifestly preserves eight supercharges in type IIA.

The toric data of the four-fold are specified by a Lagrangian $U(1)^4$ fibration over a strictly convex rational polyhedral cone. Each facet of the cone represents a toric divisor. In fact the normal vector to a facet, normalized to have integer components, represents the $U(1)$ cycle that shrinks on the facet. The collection $\{\vec v_s\}$ in $\bZ^4$ of the normal vectors to all facets is called the toric fan.
The Calabi-Yau condition is equivalent to the end-points of all vectors in the toric fan being coplanar; one can then use an $SL(4,\bZ)$ transformation to rewrite them as $\vec v_s = (1, \vec w_s)$, with $\{\vec w_s\}$ vectors in $\bZ^3$. The information encoded in the toric fan can be summarized by the 3d toric diagram: a 3d convex polyhedron whose strictly external points are $\vec w_s$. We will
call \emph{strictly} external, among the external points, a point which does not lie along a line connecting two external points -- this means that strictly external points are not inside an edge nor a face of the toric diagram. Each strictly external point represents a conical toric divisor.
The elements $(0,1,0,0)$, $(0,0,1,0)$, $(0,0,0,1)$ in the $\bZ^4$ ambient space of the toric fan generate the flavor $U(1)_F^3$ symmetry group that commutes with the R-symmetry.

Over the intersection of two adjacent facets, two $U(1)$ cycles in the fiber shrink. Suppose that the shrinking cycles are $(1,x,y,z)$ and $(1,x,y,z+1)$ (two points vertically aligned) in $\bZ^4$: at the intersection of the two facets the $U(1)_M$ cycle $(0,0,0,1)$, linear difference of the previous ones, shrinks as well. This happens along a complex codimension-two conical submanifold of the four-fold, and one can locally view the M-theory background as a KK monopole for that $U(1)_M$ action. Reducing along $U(1)_M$, one gets a D6-brane on some type IIA background.

As shown in \cite{Aganagic:2009zk}, the type IIA background can be written as the fibration of a CY$_3$ cone $Y_3$ over a real line $r_0 \in \bR$, where the K\"ahler moduli of $Y_3$ vary linearly (according to the CY$_4$ being conical) along the line. The full type IIA background has also RR fluxes, as required by $\cN=2$ supersymmetry and corresponding to the fibration of $U(1)_M$, as well as non-trivial dilaton and warping. Degeneration loci of the $U(1)_M$ fiber result in various ``objects'' in the type IIA background.
More precisely, the three-fold is the result of the K\"ahler quotient $\text{CY}_4 // U(1)_M = \text{CY}_3$.
$Y_3$ is toric, and defined by a 2d toric diagram which is the projection of the 3d toric diagram to a plane orthogonal to the primitive vector%
\footnote{A primitive vector is a vector in $\bZ^d$ with coprime components.}
$\vec v_M$ that represents the cycle $U(1)_M$ used for the reduction (more details below).
Equivalently, we can always perform an $SL(3,\bZ)$ transformation of the 3d toric diagram and map $\vec v_M$ to $(0,0,0,1)$; then the 2d toric diagram of $Y_3$ is the ``vertical'' projection of the 3d diagram to the plane $z=0$.

\begin{figure}[t]
\begin{center}
\includegraphics[height=7cm]{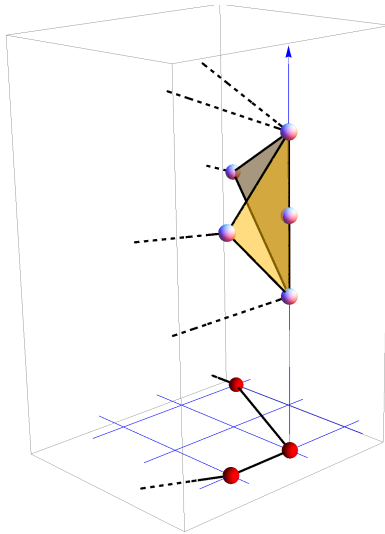}
\caption{Vertical projection from the 3d toric diagram of $Y_4$ to the 2d toric diagram of $Y_3$. The three aligned points give rise to two D6-branes. \label{fig: toric projection}}
\end{center}
\end{figure}

In our example, the fact that two adjacent external points%
\footnote{Points are adjacent if the line connecting them is strictly external, not contained in a face.}
in the 3d toric diagram project to the same point (which is then necessarily strictly external) in the 2d toric diagram, implies the presence of a D6-brane wrapping a toric divisor of the CY$_3$ in type IIA (spanning the spacetime $\bR^{2,1}$ and localized at $r_0=0$). The toric divisor is specified by the projected point.
More generally, if the 3d toric diagram has a collection of $h+1$ aligned adjacent external points $(1,x,y,z_j)$ with $z_j = z$, $z+1$, \dots, $z+h$, and we reduce along the $U(1)_M$ cycle $(0,0,0,1)$, all $h+1$ points project down to the same strictly external point in the 2d toric diagram and give rise to $h$ coincident D6-branes wrapping a toric divisor of the CY$_3$ (see Figure \ref{fig: toric projection}).

Along the D6-branes lives a $U(h)$ gauge theory, which by the AdS/CFT map corresponds to a $U(h)$ global symmetry on the boundary. We could then expect the boundary theory to admit a description in which such a symmetry is manifest. The same conclusion can be reached in M-theory: $h+1$ adjacent external points in the 3d toric diagram indicate $h$ KK monopoles in $Y_4$ (a multi-Taub-NUT geometry), whose complex structure is locally $\bC^2 \times \bC^2/\bZ_h$; the $A_{h-1}$ singularity carries $SU(h)$ gauge fields, besides the $U(1)$ gauge field from the KK reduction of the M-theory potential $C_3$.
More precisely, in the near core limit the latter $U(1)$ KK mode is non-normalizable -- correspondingly in the dual theory the diagonal $U(1)$ in $U(h)$ is actually gauged.

\subsection{CY$_4$ as a $U(1)$ fibration}
\label{sec: U(1) fibration}

Let us explain how to rewrite a toric CY $d$-fold as a (possibly singular) $U(1)$ fibration over a manifold, which in turn is the fibration of a toric CY $(d-1)$-fold along a real line, with K\"ahler moduli that vary linearly along the line. In the first part, we follow the exposition in \cite{Aganagic:2009zk}.

Consider a toric CY $d$-fold, realized as the moduli space of a gauged linear sigma-model (GLSM). There are $N+d$ chiral superfields $\phi_s$ with $s = 1, \dots, N+d$, and $N$ $U(1)$ gauge groups with integer charges $Q^a_s$ (of maximal rank) with $a = 1, \dots, N$. The CY condition is $\sum_s Q^a_s = 0$ for all $a$.
The number $N+d$ of fields can be taken to be equal to the number of dots in the $(d-1)$-dimensional toric diagram. Then the charge matrix $Q^a_s$ encodes the $N$ linear relations $\sum_s Q^a_s \, \vec v_s = 0$ between the vectors $\{\vec v_s\}$ in the toric fan.
The CY $Y_d$ is simply the K\"ahler quotient $\bC^{N+d}//U(1)^N$, which corresponds to imposing the moment map (D-term) equations
\be
\sum\nolimits_s Q^a_s \, |\phi_s|^2 = r^a
\ee
and quotienting by the gauge group
\be
\phi_s \;\to\; e^{i \sum_a \lambda_a Q^a_s} \, \phi_s \;.
\ee
The moment map (or FI) parameters $r^a$ are the resolution parameters of $Y_d$. We will be mainly interested in the conical case $r^a=0$. Moreover, for each $a$ the charges $Q^a_s$ can be taken coprime without loss of generality.

To exhibit the fibered structure, we add the complex variable $r_0 + i\theta_0$ and choose a set of charges $Q^0_s$ satisfying the CY condition $\sum_s Q^0_s = 0$. Then we impose one more equation and divide by one more gauge symmetry:
\be
\label{r0 theta0}
\sum\nolimits_s Q^0_s \, |\phi_s|^2 = r_0 \;; \qquad\qquad
\theta_0 \;\to\; \theta_0 + \lambda \;,\qquad \phi_s \;\to\; e^{i \lambda \,Q^0_s} \, \phi_s \;.
\ee
It is easy to check that the manifold is the same as before: using (\ref{r0 theta0}) $r_0$ can be eliminated while $\theta_0$ can be gauged away (without leaving any residual gauge transformation).

On the other hand, we can fix $r_0$ and think of $\theta_0$ as a $U(1)$ fibration. The base manifold $Y_{d-1}$ is then the K\"ahler quotient $\bC^{N+d} // U(1)^{N+1}$:
\be
\label{Xd-1}
\sum\nolimits_s Q^0_s \, |\phi_s|^2 = r_0 \;, \qquad\qquad \sum\nolimits_s Q^a_s \, |\phi_s|^2 = r^a \qquad \forall\, a=1,\dots,N \;,
\ee
modded out by
\be
\phi_s \;\to\; e^{i \lambda \, Q^0_s \;+\; i\sum_a \lambda_a Q^a_s} \, \phi_s \;.
\ee
$Y_{d-1}$ is toric and Calabi-Yau. Moreover, $Y_{d-1}$ is fibered over the real line $r_0$, with a particular combination of the resolution parameters (set by $Q^0_s$) varying linearly with $r_0$. The tip of $Y_d$ is at $r_0 = 0$.

\paragraph{Projecting the toric diagram.} Given the set of charges $Q^a_s$, the toric fan of $Y_d$ is given by $N+d$ primitive vectors $\{\vec v_s\}$ in $\bZ^d$ which solve the $N$ linear conditions $\sum_s Q^a_s \, \vec v_s = 0$ for all $a=1,\dots,N$. We can collect the vectors as columns of a matrix $(G_K)^i_s$, with $i = 1,\dots,d$, of maximal rank. Then
\be
Q^a_s \, (G_K^T)^s_i = 0
\ee
and the rows of $G_K$ span the kernel of $Q^a_s$ as a map from $\bR^{N+d}$ to $\bR^{N}$. We can use a transformation of $SL(d,\bZ)$ to map the vectors to $\vec v_s = (1, \vec w_s)$. The same equation can be used to obtain the charges of a GLSM, given the matrix $G_K$ of all vectors in the toric fan.

The toric diagram of $Y_{d-1}$ can be obtained in the same way. We add the extra condition $\sum_s Q_s^0 \, \vec v_s = 0$. The vectors $\vec v_s$ do not satisfy it, because the rows of $(G_K)^i_s$ are linearly independent. In order to satisfy the extra relation, we must project the vectors on a hyperplane in such a way that the linear combination%
\footnote{We mean that $\vec v_M$ is the primitive vector in $\bZ^d$ which is parallel to $\sum_s Q^0_s \, \vec v_s$.}
\be
\vec v_M \equiv \text{primitive } \sum\nolimits_s Q^0_s \, \vec v_s
\ee
vanishes, that is a hyperplane orthogonal to $\vec v_M$. Notice that the CY condition on $Q^0_s$ plus the particular chosen frame $\vec v_s = (1, \vec w_s)$ assures that $\vec v_M = (0, \vec w_M)$. To make the projection clearer, we can perform an $SL(d-1,\bZ)$ transformation that maps $\vec v_M$ to $(0, \dots, 0,1)$, and changes the toric diagram of $Y_d$ accordingly. Then the toric diagram of $Y_{d-1}$ is obtained from the one of $Y_d$ with the ``vertical'' projection that forgets the last component (Figure \ref{fig: toric projection}).

\paragraph{Fixed points.} The reduction $Y_{d-1} = Y_d // U(1)_M$ can always be done. However, whenever the $U(1)_M$ fiber degenerates, we should expect some extra object or singularity in the type IIA background, on top of any possible geometric toric singularity (even non-isolated) of $Y_{d-1}$.

A first class of singularities arises from loci where the fiber $U(1)_M$ shrinks:
\begin{itemize}
\item each strictly external dot $\vec v_s$ in the $(d-1)$-dimensional toric diagram of $Y_d$ is a conical toric divisor (complex codimension one) where the circle $\vec v_s$ shrinks;
\item each external edge $v_{sr}$ connecting two adjacent dots $\vec v_s$ and $\vec v_r$ is a conical codimension 2 surface where the span in $U(1)^d$ of the two circles shrinks;
\item each external polyhedron $v_{s_1 \dots s_n}$ of dimension $n-1$ constructed between the strictly external dots $\vec v_{s_1}, \dots, \vec v_{s_n}$ is a conical codimension $n$ surface where the span in $U(1)^d$ of the $n$ circles shrinks.
\end{itemize}
In order to have a non-singular K\"ahler quotient for the projection, we should make sure that the circle $\vec v_M$ is not contained in any of the spans above (the first case is automatically excluded). Practically, we require $\vec w_M$ not to be parallel to any external sub-object in the convex polyhedron of the $(d-1)$-dimensional toric diagram. We stress that we are not worried about singularities in the quotient $Y_{d-1}$, but rather about degenerations of the fiber.

There is a second class of possible singularities, where the $U(1)_M$ fiber degenerates to $U(1)/\bZ_p$ for some $p$. This happens if some of the charges in $Q^0_s$ have modulus larger than 1. In this case, there could be a conical surface where the fiber $U(1)_M$ degenerates: we have to make sure that the only point where this happens is the tip of $Y_d$.

\paragraph{The case of CY$_4$.} Specializing to the case of interest -- $Y_4$ and $Y_3 = Y_4 // U(1)_M$ --, whenever none of the singularities above arises in the quotienting, we are sure that the reduction of M-theory on $Y_4$ along $U(1)_M$ gives a pure IIA background (to which the arguments in \cite{Aganagic:2009zk} can be applied), without extra objects on top of it.

In particular, we should make sure that: 1) there are no external edges in the 3d toric diagram parallel to $\vec w_M$; 2) there are no external faces parallel to $\vec w_M$; 3) once $\vec w_M$ is expressed as an integer sum of the $\vec w_s$ in the 3d toric diagram, if some coefficients have modulus larger than 1, the fiber does not degenerate outside the tip of $Y_4$.

On the contrary, whenever the fiber degenerates, we should expect some extra objects in type IIA that have to be taken into account. In this paper we study what happens if the fiber shrinks on a complex codimension-two submanifold of the four-fold (giving rise to D6-branes). The other cases deserve a separate study.

\subsection{IIA background as a CY$_3$ fibration with D6-branes}
\label{sec: IIA background}

The symplectic reduction of $Y_4$ to a CY$_3$ is useful because it allows to exploit all the powerful techniques available for D3-branes probing toric singularities, to get information about the field theory. Given a toric CY$_3$ singularity in type IIB and $N$ D3-branes probing it, the dual SCFT in 3+1 dimensions can be generically found with the algorithm in \cite{Hanany:2005ss} (see also \cite{Feng:2005gw}).%
\footnote{On top of that, a huge number of examples has been explicitly worked out, see \cite{Kennaway:2007tq, Yamazaki:2008bt} and references therein.}
As explained in \cite{Aganagic:2009zk}, we can consider the same CY$_3$ singularity in type IIA and probe it with D2-branes: the dual theory is the same quiver (with same superpotential) as before, but in 2+1 dimensions. We can further include RR fluxes and fiber the CY$_3$ over a real line (in an $\cN=2$ supersymmetric way). This corresponds to switching on $\cN=2$ Chern-Simons couplings in the Yang-Mills (YM) quiver gauge theory.
Conversely, whenever the symplectic reduction CY$_4 // U(1)_M$ is regular (the fiber $U(1)_M$ nowhere degenerates), the procedure allows to obtain a CS quiver theory which reproduces the CY$_4$ in its moduli space. The relation between the $CY_4$ geometric moduli space of a 3d quiver
CS theory and the $CY_3=CY_4//U(1)_M$ mesonic moduli space of a 4d
gauge theory with the same quiver and superpotential was first pointed
out in \cite{Jafferis:2008qz,Martelli:2008si,Hanany:2008cd}.

We aim to extend the correspondence to cases in which the K\"ahler quotient has complex dimension-two degeneration loci. To begin with, let us understand what the toric divisors of $Y_3$ correspond to.
Each strictly external point $\vec p$ in the 2d toric diagram corresponds to a toric divisor, to which is associated a collection of $Q$ bifundamental fields $\{X_\eta\}_{\eta=1, \dots, Q}$ in the quiver theory, that have the same charges under all global (but not gauge) symmetries. The number $Q$ is given by \cite{Hanany:2001py}
\be
\label{Q multiplicity}
Q = \left| \det \begin{pmatrix} \Delta x & \Delta y \\ \Delta x' & \Delta y' \end{pmatrix} \right| \;,
\ee
where $(\Delta x, \Delta y)$ is the vector connecting the strictly external point $\vec p$ to the next strictly external point along the perimeter, while $(\Delta x', \Delta y')$ is the vector connecting $\vec p$ to the previous strictly external point.%
\footnote{\label{foot: multiplicity} $Q$ is more conveniently defined as the modulus of the cross product of two consecutive legs in the $(p,q)$-web that is dual to the 2d toric diagram.}
A time-filling D3-brane wrapped on the 3-cycle which is the radial section of the toric divisor (such embedding is supersymmetric) corresponds to a dibaryonic operator $X_\eta^N$ \cite{Berenstein:2002ke,Franco:2005sm,Butti:2006au}. Since the 3-cycle has the topology of a Lens space with fundamental group $\bZ_Q$ \cite{Franco:2005sm}, the D3-branes admit a $\bZ_Q$ flat connection resulting in $Q$ degenerate vacua. They correspond to the $Q$ different dibaryonic operators $\{X_\eta^N\}$. An easy way to identify the set of fields is through perfect matchings (that we review in Appendix \ref{sec: pm}) in the brane tiling construction \cite{Kasteleyn, Hanany:2005ve, Franco:2005rj, Feng:2005gw}.

Instead of wrapping a D3-brane on a radial section, one can wrap $h$ spacetime-filling D7-branes on the whole toric divisor (this problem has been considered, \eg, in \cite{Ouyang:2003df, Franco:2006es, Benini:2006hh, Benini:2007gx, Benini:2007kg}). They introduce a $U(h)$ global symmetry in the field theory, and $h$ flavors of chiral fields $p_\eta$, $q_\eta$ coupled to one of the bifundamental fields $X_\eta$ through the superpotential term $W = p_\eta X_\eta q_\eta$.
A $\bZ_Q$ connection, flat everywhere but at the tip, can be specified on the D7-branes to distinguish which bifundamental is flavored.%

The same discussion holds in type IIA: D6-branes wrapping toric divisors of the CY$_3$ provide chiral flavors to the quiver gauge theory on D2-branes at the tip. Each stack of $h$ D6-branes introduces a $U(h)$ flavor group (this is not \emph{always} the case: we will be more precise in Section \ref{sec: geometry}) and flavor chiral multiplets $p_{\hat ki}$, $q_{j\hat k}$ coupled to a bifundamental $X_{ij}$ through a superpotential term%
\footnote{We absorb superpotential couplings inside chiral superfields.}
\begin{equation}
\label{DeltaW}
W = \Tr p_{\hat{k}i} X_{ij} q_{j\hat{k}} \;.
\end{equation}
Here $\hat k$ stands for a flavor group, $i,j$ for gauge groups and fields are in the fundamental (anti-fundamental) of the first (second) index; all indices are contracted. We will jump between the notations $X_a$ and $X_{ij}$ for bifundamental fields. The field $X_{ij}$ is determined as explained above.

The D6-branes are localized along $\bR$: reducing the cone $Y_4$ their position is $r_0=0$. More generally, the position along $r_0$ corresponds to a real mass for the quarks in field theory, and to a partial resolution (or K\"ahler) parameter in M-theory (see Section \ref{sec: resolutions}). Since D6-branes, possibly with worldvolume flux, are sources for RR fields, the 2- and 4-form fluxes on 2- and 4-cycles vanishing at the CY$_3$ singularity jump at $r_0$:
\be
\label{CS shift geometry}
\delta \int_{\cC_2} F_2 = \#(\cC_2, D6) \;, \qquad\qquad \delta \int_{\cC_4} F_4 = \#(\cC_4, D6, \cC_4^{(F_\mr{wv})}) \;,
\ee
where the jump depends on the intersection on $Y_3$ between the cycles, the divisor and the cycle representing the worldvolume flux. This means that moving the D6's to the left or to the right of the D2-branes, the CS levels must jump as well. We will study this in detail.

Summarizing, whenever the $U(1)_M$ action has codimension-two fixed loci which descend in type IIA to D6-branes wrapping divisors of the CY$_3$, the field theory derived using the CY$_3$ singularity is actually flavored.

\

We conclude this section with some comments.
Two important differences between chiral flavors in $AdS_5/CFT_4$ and in $AdS_4/CFT_3$ must be borne in mind. Firstly, in 4d gauge theories chiral flavors are constrained by gauge anomaly cancelation, whereas in 3d such a constraint does not exist. The dual statement is that D7-branes wrapping divisors are constrained by RR $C_0$ tadpole cancelation, whilst D6-branes are not because the RR $F_2$ flux can escape to infinity along the transverse non-compact real line.
The number of fundamental minus anti-fundamental fields for a gauge group in 3d need not vanish: if it is odd, the parity anomaly requires the presence of half-integral CS levels \cite{Niemi:1983rq, Redlich:1983kn, Redlich:1983dv}.
Secondly, in general the addition of flavors to an $AdS_5/CFT_4$ pair breaks conformal invariance and the RG flows leads the theory to a fixed point which is outside the validity of supergravity \cite{Benini:2006hh} (the dual statement is that D7-branes force the dilaton to run towards $-\infty$ at the tip). Flavoring $AdS_4/CFT_3$ pairs, the theory still flows to an interacting fixed point which however in many examples \cite{Gaiotto:2009tk, Hohenegger:2009as, Hikida:2009tp, Cherkis:2002ir, Bianchi:2009ja} (and in the ones of this paper too) is still described by type IIA/M-theory.

In the following, we will focus on the Abelian case: we will consider a single M2/D2-brane and the corresponding quiver theory will have $U(1)$ gauge groups. One expects the low energy field theory on a stack of $N$ M2/D2-branes to be described by the same quiver with $U(N)$ gauge groups, and the geometric moduli space to be the symmetric product of $N$ copies of $Y_4$. We leave the non-Abelian extension for the future.


\section{Flavoring Chern-Simons-matter theories \\ A bottom-up perspective}
\label{sec: flavoring}


In the rest of this paper we turn to a bottom-up perspective. We start with a generic toric CY$_4$ geometry and a regular (as described in Section \ref{sec: U(1) fibration}) IIA reduction along $U(1)_M$, such that the Chern-Simons-matter theory dual to M2-branes probing $Y_4$ can be read off \cite{Aganagic:2009zk}. Then we study the effect of chirally flavoring such a theory in a very general way, and in particular we study how the flavoring deforms the moduli space of the quiver theory.
Alternatively, we can start with a toric CY$_3$ geometry and its dual quiver theory (which in 3+1 dimensions is the theory dual to D3-branes probing $Y_3$), add to it generic $\cN=2$ Chern-Simons couplings (which corresponds to fibering $Y_3$ over $\bR$ and adding RR fluxes) and flavors (D6-branes), and study what is the resulting CY$_4$ geometry seen by M2-branes.

To begin with, let us specify the flavoring procedure. The starting point is an $\cN=2$ quiver Chern-Simons theory in 2+1 dimensions. The matter fields are chiral multiplets $X_a$ in the adjoint or bifundamental representation,
and we restrict ourselves to the Abelian case. Then we introduce $B$ families of flavor chiral multiplets $(p_\alpha, q_\alpha)$, each coupled to a matter field $X_\alpha$ via the superpotential
\be
\label{flavoring W}
W = W_0 + \sum\nolimits_\alpha p_\alpha X_\alpha q_\alpha \;.
\ee
Here $p_\alpha$ ($q_\alpha$) transform in the anti-fundamental (fundamental) of the gauge group under which $X_\alpha$ is in the fundamental (anti-fundamental).
Each pair $(p_\alpha, q_\alpha)$ really represents $h_\alpha$ fields, and introduces a $U(h_\alpha)$ flavor symmetry.

We could be more general and couple a flavor pair to a bifundamental operator $\cO_\alpha = \prod_{\beta=1}^n X_\beta$ constructed from a string of matter fields $X_\beta$. This is equivalent to coupling each of the $X_\beta$ to its own flavor pair $(p_\beta, q_\beta)$, and then introducing complex masses

\be
W = W_0 + \sum_{\beta=1}^n p_\beta X_\beta q_\beta + \sum_{\beta = 1}^{n-1} m_\beta \, p_{\beta+1} q_\beta  \;.
\ee
Integrating out the massive fields, we flavor the operator $\cO_\alpha$ (see Section \ref{sec: complex masses}).

Every time we introduce two new flavor fields $(p,q)$ coupled to $X_\alpha$, the parity anomaly \cite{Niemi:1983rq, Redlich:1983kn, Redlich:1983dv} requires to shift two CS levels as
\be
\label{CS shift}
\delta k_i = \pm \frac12\, g_i[X_\alpha] \;,
\ee
$g_i$ being the gauge charges. The sign is a choice of theory. If we add $h_\alpha$ flavors $(p_\alpha, q_\alpha)$, we choose sign $h_\alpha$ times, so that the shift
\be
\label{CS shift h}
\delta k_i = \Big( \frac{h_\alpha}2 - \gamma_\alpha \Big) \, g_i[X_\alpha]
\ee
is parametrized by an integer $\gamma_\alpha$ with $0 \leq \gamma_\alpha \leq h_\alpha$.

The reason for this is that gauge invariance requires
\be
k_i + \frac12 \sum\nolimits_\psi \big( g_i[\psi] \big)^2 \,\in \bZ \;,
\ee
where the sum runs over all fermions charged under the $i$-th gauge group. When the second term is half-integral, the fermion determinant is multiplied by $(-1)$ under certain gauge transformations, and the lack of gauge invariance of the CS terms cures it. In our setup the gauge charges of flavors are $g_i = \pm 1$, so consistency requires that each addition of two flavor fields is accompanied by a half-integral opposite shift of two CS levels (unless $X_\alpha$ is in the adjoint).

We can proceed in the opposite way and integrate the flavors out. In 2+1 dimensions it is possible to give \emph{real} mass $\tilde m$ to a chiral multiplet $Z$ through the Lagrangian term \cite{deBoer:1997kr, Aharony:1997bx}
\be
\int d^4\theta \, Z^\dag e^{\tilde m \theta \bar\theta} Z \;.
\ee
To give real masses to the flavors, we promote the $U(h)$ flavor symmetry to a background gauge symmetry; then a VEV $\vev{\sigma_F}$ for the real adjoint background scalar field $\sigma_F$ in the $U(h)$ vector multiplet provides real masses to all flavors charged under $U(h)$. After diagonalization of $\vev{\sigma_F}$ by a flavor rotation, each flavor of charge $q_\psi$ acquires a real mass $M_\psi = q_\psi \vev{\sigma_F}$.
When integrating out massive fermions $\{\psi\}$, the CS levels $k_i$ are shifted by one-loop diagrams as
\be
k_i \to k_i + \frac12 \sum\nolimits_\psi \big( g_i[\psi] \big)^2 \sgn(M_\psi) \;.
\ee
Integrating out just two flavor fields $(p,q)$, $M_p = \vev{\sigma_F}$ and $M_q = - \vev{\sigma_F}$; we can then write $\delta k_i = \frac12 g_i[X_\alpha] \, \sgn(\vev{\sigma_F})$. The choice of $\sgn(\vev{\sigma_F})$ corresponds to the choice of sign in (\ref{CS shift}): a choice of positive (negative) sign in (\ref{CS shift}) is undone by $\vev{\sigma_F}<0$ ($\vev{\sigma_F}>0$).

In the next subsection we compute the effect of flavors on monopole operators, while in Section \ref{sec: moduli space} we study the moduli space of the flavored theories.

\subsection{Monopole operators and flavors}
\label{sec: monopoles}

A fundamental r\^ ole in the study of the quantum moduli space of the flavored theories is played by monopole operators \cite{'tHooft:1977hy, Polyakov:1976fu, Moore:1989yh, Borokhov:2002ib, Borokhov:2002cg, Borokhov:2003yu, Itzhaki:2002rc, Benna:2009xd}: in the Euclidean theory, their insertion creates quantized flux in a $U(1)$ subgroup of the gauge group through a sphere surrounding the insertion point. In radial quantization, they correspond to states with flux on $S^2$. For a $U(1)$ gauge group:
\be\label{state with n units of fluxes}
\frac{1}{2\pi}\int_{S^2} F = n \;.
\ee
In a Chern-Simons theory, monopole operators pick up an electric charge since
\be
\label{CS charge}
S \;\supset\; \frac{k}{4\pi}\int A\wedge dA = k\, n \int A_0\, dt \;,
\ee
where $k$ is the Chern-Simons level. In a Chern-Simons-matter theory, fermionic matter fields can correct these charges from 1-loop diagrams.
A special r\^ole is played by ``diagonal'' monopole operators, that we will denote $T^{(n)}$ with $n \in \bZ$: they have the same flux $n$ along all $U(1)$ gauge groups in the quiver. They pick up electric charges $(nk_1, \dots, nk_G)$ under $U(1)^G$, where $G$ is the number of gauge factors and $k_i$ are the CS levels. They were studied in detail in \cite{Benna:2009xd} and shown to be BPS (after having been dressed by scalar modes) in the ABJM theory \cite{Aharony:2008ug}; we expect them to be BPS in generic $\cN=2$ theories describing M2-branes on CY$_4$, since they correspond to modes of eleven-dimensional supergravity in short multiplets.

The monopole operators $T^{(n)}$ can acquire a charge under any $U(1)$ symmetry of the theory, both global and gauged, from quantum corrections \cite{Borokhov:2002ib, Borokhov:2002cg, Borokhov:2003yu, Benna:2009xd}. In the case of global symmetries the charge comes entirely from fermionic modes, while in the case of gauge symmetries the quantum contribution sum up with (\ref{CS charge}). The quantum correction (in the Abelian case) to the charge $Q$ from fermionic modes is
\be
\label{monopole charge correction}
\delta Q[T^{(n)}] = -\frac{|n|}{2} \, \sum\nolimits_\psi Q[\psi] \;,
\ee
where we sum over all fermions $\psi$ in the theory. Notice that only fermions in chiral representations contribute. The result is proportional to the mixed $Q$-gravitational anomaly that the same theory would have in 3+1 dimensions.

Formula (\ref{monopole charge correction}) implies that in Chern-Simons quiver theories satisfying the toric condition, diagonal monopole charges do not receive any quantum correction. Quiver theories have matter chiral multiplets in the adjoint and bifundamental representation only; the toric condition is that all gauge ranks are the same (here 1), each matter field appears in the superpotential in exactly two monomials, and the number $G$ of gauge groups plus the number $P$ of monomials in the superpotential equals the number $E$ of matter fields. Let $Q$ be a non-R global or gauge symmetry: each monomial $W_\mu$ in the superpotential must have vanishing charge. Summing over all monomials: $2\sum_\psi Q[\psi] = 2 \sum_a Q[X_a] = \sum_\mu \sum_{a\in \mu} Q[X_a] = 0$,
where $X_a$ are all matter fields, and gaugini must be chargeless. In the case of the R-symmetry, each monomial $W_{\mu}$ must have R-charge 2, so that: $2 \sum_\psi R[\psi] = 2G + 2\sum_a R[\psi_a] = 2G-2E + 2\sum_a R[X_a] = 2G - 2E + 2P = 0$, where we used the fact that gaugini have R-charge 1.

Therefore, let us start with a quiver theory in which the monopole fields $T^{(n)}$ have only gauge charges $(n k_1, \dots, n k_G)$. Then we flavor the theory as in (\ref{flavoring W}): we couple a set of flavor pairs $(p_\alpha, q_\alpha)$, each in number $h_\alpha$, to some bifundamental operators $X_\alpha$ in the quiver, constructed as products of bifundamental fields,%
\footnote{We are mainly interested in the case that $X_\alpha$ are pure bifundamental fields, but the arguments that follow apply as well to composite bifundamental fields, \emph{i.e.} connected open paths in the quiver.}
via
\be
W = W_0 + \sum\nolimits_\alpha p_\alpha X_\alpha q_\alpha \;.
\ee
We are interested in the charges induced on the monopole operators by flavors.
Let us start with non-R symmetries. First, there are the new flavor symmetries $U(h_\alpha)$ of which $p_\alpha$ and $q_\alpha$ are in conjugate representations, so that the diagonal monopole operators cannot get a charge under $U(h_\alpha)$.%
\footnote{To apply (\ref{monopole charge correction}), take any generator of $U(h_\alpha)$ and consider the $U(1)$ subgroup it generates.}
Next, for any $U(1)$ flavor symmetry of $W_0$ under which $X_\alpha$ has charge $Q_\alpha$, $(qp)_\alpha$ must have charge $-Q_\alpha$. Then, according to (\ref{monopole charge correction}), the diagonal monopoles pick up a charge
\be
\label{flav charges}
Q[T^{(n)}] = \frac{|n|}{2} \sum\nolimits_\alpha h_\alpha \, Q[X_\alpha]
\ee
in the flavored quiver. In the case of \emph{gauge} charges, the contribution from fermions has to be summed with the contribution from Chern-Simons couplings:
\be
\label{gauge charge of diag monopole}
g_i[T^{(n)}] = nk_i + \frac{|n|}{2} \sum\nolimits_\alpha h_\alpha \, g_i[X_\alpha] \;,
\ee
where $g_i$ are the gauge charges under $U(1)^G$.
Eventually, consider the R-symmetry: $R[p_\alpha] + R[q_\alpha] = 2 - R[X_\alpha]$ at the IR fixed point, so that the monopoles get an R-charge
\be
\label{R charge}
R[T^{(n)}] = -\frac{|n|}{2} \sum\nolimits_\alpha h_\alpha \, \big( R[\psi_{p_\alpha}] + R[\psi_{q_\alpha}] \big) = \frac{|n|}{2} \sum\nolimits_\alpha h_\alpha \, R[X_\alpha] \;.
\ee
These charges allow us to conjecture the following holomorphic quantum relation:
\be
\label{quantum rel}
T^{(n)} \, T^{(-n)} = \Big( \prod\nolimits_\alpha X_\alpha^{h_\alpha} \Big)^{|n|} \;,
\ee
which is consistent with all manifest symmetries in the action.
This is understood as an operator statement: the equation must be multiplied on both sides by the necessary fields to form gauge-invariant operators.
In Section \ref{sec: moduli space} we show that in the usual unflavored case (where quantum corrections seem not to play a r\^ole) the relation $T^{(n)} T^{(-n)} = 1$ can be inferred from the form of the moduli space. Moreover, (\ref{quantum rel}) is analogous to the quantum relation which appeared in the $\calN=3$ setup of \cite{Gaiotto:2009tk} (see also \cite{Borokhov:2002cg}), and we will show that it reproduces the CY$_4$ moduli spaces as expected from the M-theory reduction, as we also check in several examples in Section \ref{sec: examples}.
In the following we will use the notation
\be
T^{(1)} \equiv T \;,\qquad T^{(-1)} \equiv \tilde T \;,
\ee
for the simplest diagonal monopole operators.


\section{Moduli space of flavored quivers}
\label{sec: moduli space}


\subsection{Unflavored quivers and monopoles}
\label{sec: no flav moduli space}

The moduli space of any (unflavored) $\cN=2$ Chern-Simons quiver theory was worked out in \cite{Martelli:2008si}. Let us review the analysis here, and show how the monopole operators $T$, $\tilde T$ can be included at the classical level. We focus on the Abelian case, and impose the further condition $\sum k_i = 0$.

The F-term equations $\partial W/\partial X_a = 0$, where $X_a$ are all the chiral superfields in the quiver and $a = 1,\dots, M$, define an algebraic variety
\be
\cZ = \{ X_a \;|\; dW = 0 \} \subset \bC^M \;.
\ee
This is exactly the same as in the corresponding quiver theory in 3+1 dimensions. The D-term equations are
\be
\label{D-terms}
\cD_i = \frac{k_i \sigma_i}{2\pi} \qquad \forall\, i \qquad \text{ and } \qquad |X_a|^2 \, \Big( \sum\nolimits_i g_i[X_a] \, \sigma_i \Big)^2 = 0 \qquad \forall\, a \;,
\ee
where $\cD_i$ with $i=1,\dots, G$ are the D-terms expressed in terms of the scalars in chiral multiplets
\be
\label{def D-terms}
\cD_i \equiv \sum\nolimits_a g_i[X_a] \, |X_a|^2 \;,
\ee
while $\sigma_i$ are the real scalars in the $\cN=2$ vector multiplets. We focus on the particular branch $\sigma_1 = \dots = \sigma_G \equiv \sigma$, which is a CY four-fold \cite{Martelli:2008si}. The second set of equations in (\ref{D-terms}) is then automatically solved. We can rewrite the first set as
\be
\label{D-terms 2}
\sum\nolimits_i c_i \, \cD_i = 0 \qquad \forall\, \{ c_i \} \text{ such that } \sum\nolimits_i c_i k_i = 0 \;.
\ee
If $k_i$ are not all vanishing, these are $G-2$ independent equations (the equation with $c_i=1$ is trivial, as follows from (\ref{def D-terms})). The scalar $\sigma$ can be eliminated by
\be
\label{D-term CS}
\frac\sigma{2\pi} = \sum\nolimits_i \frac{k_i}{|k|^2} \, \cD_i \;,
\ee
where $|k|^2 = \sum_i k_i^2$.

Naively, one would divide by $U(1)^{G-1}$ gauge transformations (no scalar transforms under the diagonal $U(1)_\mr{diag}$), but this would give an odd dimensional space. In fact, the diagonal photon $A_\mr{diag} \equiv \sum_i A_i$ is only coupled to the other photons via a $bf$ term, whilst it is not coupled to any matter current. Hence it can be dualized into a scalar $\tau$, which is invariant under $U(1)_\mr{diag}$ but transforms under the remaining group,
\be
A_i \;\to\; A_i + d\theta_i \;, \qquad\qquad \tau \;\to\; \tau + \frac1 G \sum\nolimits_i k_i \, \theta_i \;.
\ee
We can use a gauge transformation to gauge $\tau$ away: this leaves us with $U(1)^{G-2}$ gauge transformations (the ones with $\sum k_i\theta_i=0$), which precisely correspond to the $G-2$ D-term equations in (\ref{D-terms 2}), plus a residual discrete group from the gauge fixing of $\tau$. On the branch we consider, in Euclidean signature the energy of the vacuum vanishes when $A_1 = \dots = A_G$ (which is a diagonal flux), hence the quantization $\frac{1}{2\pi} \int dA_\mr{diag} \in G\, \bZ$ \cite{Martelli:2008si}, which implies that $\tau$ has period $2\pi/G$. The gauge fixing of $\tau$ leaves a residual $\bZ_q$ symmetry, where $q=\gcd\{k_i\}$, to divide by. As a result, the moduli space is a $\bZ_q$ quotient of the K\"ahler quotient:
\be
\cM = (\cZ // U(1)^{G-2}) / \bZ_q \;.
\ee

\paragraph{Including monopole operators.} We can describe the same moduli space in a slightly different way. Instead of gauge fixing $\tau$, we keep it in the description of the moduli space. Given the periodicity of $\tau$, we can construct the two complex fields
\be
T = \rho \, e^{iG \, \tau} \;, \qquad\qquad \tilde T = \tilde \rho \, e^{-iG \, \tau} \;,
\ee
where their dimensionless moduli $\rho$ and $\tilde\rho$ are not specified yet. The gauge transformations are
\be
T \;\to\; e^{ i\sum k_i \theta_i} \, T \;,\qquad\qquad \tilde T \;\to\; e^{-i\sum k_i \theta_i} \, \tilde T \;,
\ee
so that their gauge charges are $\pm(k_1, \dots, k_G)$ respectively. Keeping $T$, $\tilde T$ in the description, we will have to divide by the full gauge group $U(1)^G$ (still nothing is charged under $U(1)_\mr{diag}$). We can rewrite the D-term equations (\ref{D-terms 2}) and (\ref{D-term CS}) as
\be
0 = \tilde \cD_i = k_i \, g^2 \, |T|^2 - k_i \, g^2 \, |\tilde T|^2 + \cD_i \qquad \forall\, i
\ee
with the extra complex constraint
\be
\label{class monopole rel}
T \tilde T = 1 \;,
\ee
where $\tilde \cD_i$ are ``improved D-terms''. Here $g^2$ is some mass scale, discussed below. The improved D-term equations can be thought of as arising in the presence of extra chiral fields $T$, $\tilde T$ with charges $\pm (k_1, \dots, k_G)$.

The equivalence works as follows:
\bea
0 &= \sum_i c_i \tilde \cD_i = \sum _i c_i \cD_i \qquad\qquad \forall \, \{c_i\} \text{ s.t. } \sum_i c_i k_i = 0 \\
0 &= \sum_i \frac{k_i}{|k|^2} \, \tilde \cD_i = g^2 |T|^2 - g^2 |\tilde T|^2 + \sum_i \frac{k_i}{|k|^2} \, \cD_i \;.
\eea
The first set is exactly (\ref{D-terms 2}). The second equation is equivalent to (\ref{D-term CS}) if we express $\rho$ and $\tilde\rho$ in terms of $\sigma$ through the equations (\ref{class monopole rel}) and
\be
|T|^2 - |\tilde T|^2 + \frac \sigma{2\pi g^2} = 0.
\ee
These two equations have one and only one solution in terms of $\sigma$.

As a result, the \emph{same} moduli space can be obtained by adding $T$, $\tilde T$ to the set $\{X_a\}$ of chiral fields, adding (\ref{class monopole rel}) to the set of classical F-term relations derived from the superpotential, and dividing by the full gauge group $U(1)^G$. Rephrasing, we start with a larger algebraic variety
\be
\tilde \cZ = \{ X_a, T, \tilde T \;|\; dW = 0 ,\, T \tilde T=1 \} \subset \bC^{M+2} \;,
\ee
and construct the geometric moduli space as the K\"ahler quotient
\be
\cM = \tilde\cZ // U(1)^G \;.
\ee

It is natural to associate $T$ and $\tilde T$ with the monopole operators.
In fact, following \cite{Aharony:1997bx}, it is natural to combine the vector multiplet scalar $\sigma$ and the scalar dual to the photon in a chiral multiplet. The mass scale $g^2$ does not affect the moduli space, and we can use the coupling of the diagonal photon $A_\mr{diag}$ in a YM-CS UV completion of the theory.
The relation (\ref{class monopole rel}) is a particular case of the quantum relation  (\ref{quantum rel}): we see that in the unflavored case it appears at the classical level, in the parametrization of the moduli space. In fact $T$ and $\tilde T$ are necessary to parametrize the moduli space with operators invariant under the full $U(1)^G$ gauge group.

\paragraph{The toric case.} In case the CS-matter quiver theory is a \emph{brane tiling} \cite{Kasteleyn, Hanany:2005ve, Franco:2005rj, Feng:2005gw}, this branch of the moduli space is a toric CY$_4$. A brane tiling (more details in Appendix \ref{sec: pm}) is a bipartite graph on the torus which encodes both the matter content (the quiver) and the superpotential of the gauge theory, and imposes particular constraints on them. Brane tilings describe the quiver gauge theories dual to D3-branes probing toric CY$_3$ singularities and so, by the construction of \cite{Aganagic:2009zk} reviewed in Section \ref{sec: M-theory}, they become relevant for M2-branes probing toric CY$_4$ singularities as well. In this case, the brane tiling can be refined to include the Chern-Simons levels $k_i$ (with the constraint $\sum_i k_i =0$). One assigns an integer $n_{ij}$ to each bifundamental field $X_{ij}$ -- the CS levels are then defined to be $k_i = \sum_j (n_{ij} - n_{ji})$ \cite{Ueda:2008hx,Imamura:2008qs}.

The geometric moduli space of a CS-matter brane tiling theory can be easily computed with the Kasteleyn matrix algorithm, that we review in Appendix \ref{sec: pm}. The algorithm furnishes the following output:
\begin{itemize}
\item A set of fields $t_\rho$, called perfect matchings, in terms of which the bifundamental fields can be parametrized:
\be
\label{pm redefinitions}
X_a = \prod_{\rho \,\in\, R(a)} t_\rho \;,
\ee
where $R(a)$ are subsets of the perfect matchings. This parametrization is such that the F-term relations $dW=0$ are automatically solved.
\item The 3d toric diagram of the CY$_4$. Each perfect matching $t_\rho$ is mapped to a point of the toric diagram, even though several perfect matchings can be mapped to the same point. Therefore perfect matchings can be used as fields of an auxiliary GLSM, whose moduli space reproduces the toric manifold.

If $q=\gcd\{k_i\}>1$, some internal points are not represented by any perfect matching, and the result of the GLSM has to be quotiented by $\bZ_q$. Alternatively we can include all points of the toric diagram in the GLSM, at the price of adding new fields and gauge symmetries.
\end{itemize}

\subsection{Flavored quivers}
\label{sec: flav moduli space}

Let us study the geometric moduli space of a quiver theory, flavored along the lines of Section \ref{sec: flavoring}. Let $\{X_\alpha\}$ be the set of bifundamental fields which are flavored, with superpotential $W = W_0 + \sum_\alpha p_\alpha X_\alpha q_\alpha$, and $h_\alpha$ the number of flavors in each family.

The F-term equations $dW=0$ are clearly modified. In particular there could be Higgs branches where $p_\alpha$, $q_\alpha$ get a VEV. This can happen when $X_\alpha = 0$, which, in the dual gravitational theory, corresponds either in IIA to the D2-brane ending on the D6-branes and turning on instanton field-strength configurations on their worldvolume, or in M-theory to the M2-brane ending on the local $\bC^2 \times \bC^2/\bZ_{h_\alpha}$ singularity. However we will not study Higgs branches. Therefore on the branch where
\be
p_\alpha = 0 \;,\qquad q_\alpha = 0 \qquad \forall\, \alpha
\ee
the F-term equations $dW=0$ are the same as in the unflavored case. To those, we add the conjectured quantum relation (\ref{quantum rel}):
\be
T \tilde T = \prod\nolimits_\alpha X_\alpha^{h_\alpha} \;.
\ee
We get an algebraic variety
\be
\tilde \cZ = \{ X_a, T, \tilde T \;|\; dW = 0 ,\, T \tilde T= \prod_\alpha X_\alpha^{h_\alpha} \} \subset \bC^{M+2} \;,
\ee
where $M$ is the total number of bifundamental chiral fields. $\tilde \cZ$ has to be divided by the complexified gauge group $U(1)^G$, so that the moduli space of the flavored quiver is
\be
\label{flavored moduli space}
\cM_\mr{flav} = \tilde \cZ // U(1)^G \;.
\ee
The gauge charges of $T$ and $\tilde T$ are in (\ref{gauge charge of diag monopole}), and recall that, generically, in the flavoring process the Chern-Simons levels have to be shifted as explained in Section \ref{sec: flavoring}.

Notice that, even though not discussed in this paper, the same construction goes through if we couple a flavor group not to a bifundamental field $X_\alpha$ but to a bifundamental operator $\cO_\alpha = \prod_\beta X_\beta$ built out of a connected open path in the quiver.

\paragraph{The toric case.} In case the CS-matter quiver theory is a brane tiling, and thus its geometric moduli space is a toric CY$_4$, the flavoring of Section \ref{sec: flavoring} produces a new theory whose geometric moduli space is still a toric CY$_4$, and we can explicitly provide its toric diagram.

Toricity is easy to understand: if we interpret the tiling as a quiver theory in 3+1 dimensions, its mesonic moduli space is $\cM_{3+1} = \{ X_a | dW=0\}//U(1)^G$, which is a toric threefold and thus has (at least) $U(1)^3$ symmetry. The space $\cM_\mr{flav}$ in (\ref{flavored moduli space}) is then a fourfold, has an extra $U(1)$ symmetry acting on $T, \tilde T$ and is then toric.

The strategy is to consider a \emph{different theory} -- that we call the A(uxiliary)-theory, as opposed to the flavored theory under consideration%
\footnote{We call the A-theory ``auxiliary'' because it is not our primary object of study, but rather a tool to compute the toric diagram of $\cM_\mr{flav}$. In fact, one might suspect the two theories to be dual.}
-- of which we can easily construct the toric diagram, and then show that its geometric moduli space is the same as $\cM_\mr{flav}$ in (\ref{flavored moduli space}). The A-theory is a usual CS-matter brane tiling theory, and its geometric moduli space can be computed with the Kasteleyn matrix algorithm. It is constructed as follows.

\begin{figure}[t]
\begin{center}
\hspace{\stretch{1}}
\includegraphics[height=2.8cm]{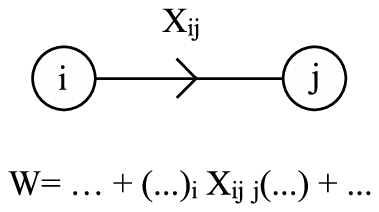}
\hspace{\stretch{1}} \raisebox{1.5cm}{$\longrightarrow$} \hspace{\stretch{1}}
\includegraphics[height=2.8cm]{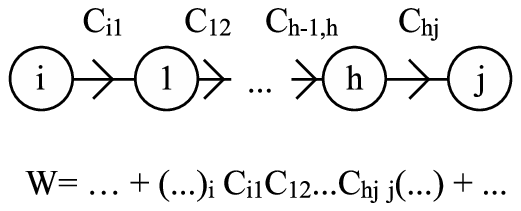}
\hspace{\stretch{1}}
\caption{Deformation of the unflavored theory to construct the A-theory. \label{fig: multibond}}
\end{center}
\end{figure}

We start with the brane tiling of the unflavored theory, refined by numbers $n_{ij}$ that encode the CS levels as $k_i = \sum_j (n_{ij} - n_{ji})$ (see Appendix \ref{sec: pm}). Every time in the flavored theory we add $h_\alpha$ flavors $(p_\alpha, q_\alpha)$ coupled to a bifundamental $X_\alpha \equiv X_{ij}$, in the A-theory we introduce $h_\alpha$ new gauge groups $U(1)_1^{(l)}$ with $l=1, \dots, h_\alpha$ and substitute $X_{ij}$ by $h_\alpha + 1$ bifundamental fields $C_{i1}, C_{12}, \dots , C_{h_\alpha j}$ coupled to the new groups in a chain as in Figure \ref{fig: multibond}. The new superpotential of the A-theory is equal to the old one, but with the substitution $X_{ij} \to C_{i1} C_{12} \dots C_{h_\alpha j}$. In the tiling this corresponds to substituting the edge $X_{ij}$ by $h_\alpha +1$ nearby edges $C_{i1}, C_{12}, \dots , C_{h_\alpha j}$, connecting the same two superpotential nodes as $X_{ij}$, and enclosing $h_\alpha$ new faces $U(1)^{(l)}_1$ between them.%
\footnote{Such a feature of the tiling has been dubbed ``multi-bond'' and studied in \cite{Hanany:2008fj, Hanany:2008gx, Davey:2009sr}.}

Then we assign integers to the $C$ fields: going from $C_{i1}$ to $C_{h_\alpha j}$, they must be a sequence of increasing consecutive integers including $n_{ij}$ (the old integer of $X_{ij}$). This means that we can choose an integer $\gamma_\alpha$, with $0 \leq \gamma_\alpha \leq h_\alpha$, and then the numbers $n$ are:
\be
\label{integer assignment}
(C_{i1}, C_{12}, \dots , C_{h_\alpha j}) \;\to\; (n_{ij} - \gamma_\alpha,\; n_{ij} - \gamma_\alpha + 1, \dots, n_{ij} - \gamma_\alpha + h_\alpha) \;.
\ee
The parameter $\gamma_\alpha$, that represents the choice of theory, must be taken equal to the one in (\ref{CS shift h}).
The CS levels of the new gauge groups $U(1)^{(l)}_1$ are all 1; the CS levels of $U(1)^{(i)}$ and $U(1)^{(j)}$ are shifted as $k_i \to k_i - \gamma_\alpha$ and $k_j \to k_j + \gamma_\alpha - h_\alpha$.

We claim that the moduli space $\cM_\mr{flav}$ in (\ref{flavored moduli space}) is a CY$_4$, and its 3d toric diagram is the toric diagram obtained from the A-theory, for instance by the Kasteleyn matrix algorithm. The proof is given in Appendix \ref{sec: proof}.

\

The deformation of the unflavored moduli space at the level of toric diagram is readily understood. The perfect matchings $t_\rho$ that define the unflavored 3d toric diagram, output of the Kasteleyn matrix algorithm as in Appendix \ref{sec: pm}, have ``horizontal'' coordinates $(x,y)$ and ``height'' $z$. For each perfect matching $t_\rho$, we add a number of consecutive points above and below $t_\rho$, with the same horizontal coordinates $(x,y)$ as $t_\rho$. The points are added to the perfect matchings which appear in the parametrization (\ref{pm redefinitions}) of flavored fields $X_\alpha$. To be precise, the number of consecutive points above and below $t_\rho$ is:
\be
\nn
t_\rho \quad\to\quad \sum_{\alpha \,\in\, R^{-1}(\rho)} (h_\alpha - \gamma_\alpha) \quad \text{above} \quad  \text{and } \sum_{\alpha \,\in\, R^{-1}(\rho)} \gamma_\alpha \quad \text{below} \;,
\ee
where $R^{-1}(\rho)$ is the set of fields which contains $t_\rho$ in their parametrization (\ref{pm redefinitions}). A rich zoology of examples is provided in Section \ref{sec: examples}.

\

The reason for the addition of points goes as follows. In constructing the tiling of the A-theory, we substitute the edges $X_\alpha$ with $h_\alpha + 1$ new edges connecting the same two superpotential nodes, and assign them the integers in (\ref{integer assignment}). Therefore, for each perfect matching that was constructed using $X_\alpha$, we get $h_\alpha$ new perfect matchings with the same horizontal coordinates and consecutive heights determined by their integers. It is easy to check that the net result on the toric diagram is the one claimed above.

Finally, since each field $X_a$ appears in at least one strictly external perfect matching, the deformed 3d toric diagram of the flavored theory has external ``columns of vertically aligned points'', which correspond to local KK monopoles in the CY$_4$ that is local $\bC^2 \times \bC^2/\bZ_{h_\alpha}$ singularities. Thus the bottom-up approach gives results in perfect agreement with the top-down analysis of Section \ref{sec: M-theory}.


\section{Back to geometry: real and complex masses}
\label{sec: geometry}


Each non-compact toric divisor of a toric CY$_3$ is a strictly external point of its 2d toric diagram. In the field theory it corresponds to a set of fields $\{X_\eta\}_{\eta = 1,\dots,Q}$ (with the same global charges), where $Q$ is determined by (\ref{Q multiplicity}): the equation $X_\eta = 0$, for any of the $Q$ fields, defines the divisor as a submanifold of the mesonic moduli space. Placing a stack of $h$ D6-branes on the divisor introduces $h$ flavors coupled to one of the fields $\{X_\eta\}$ via the superpotential $\delta W = p X_\eta q$. This follows from the fact that the modes from 2-6 strings described by $(p,q)$ become massless when some D2-branes are on top of the D6-branes. Moreover the D6-branes carry $U(h)$ gauge fields, which by the AdS/CFT map give rise to $U(h)$ global symmetry in the boundary theory.

There are $Q$ fields such that the equation $X_\eta = 0$ describes the same irreducible divisor. The reason is that the radial section of the divisor can have non-trivial fundamental group (in the toric case $\pi_1(S^3/\bZ_Q) = \bZ_Q$); therefore a flat connection can be specified as boundary condition on the D6 worldvolume, distinguishing which of the $Q$ fields it is coupled to. The connection is then flat everywhere but at the tip, where its flux can affect the shift of CS levels via (\ref{CS shift geometry}). Indeed, flavoring different fields in the set $\{X_\eta\}$ implies shifting different CS levels (\ref{CS shift}). Clearly we can pile up D6-branes with different flat connection.

The converse is not true: a generic field $X_a$ corresponds -- via the equation $X_a = 0$ -- to a collection of pairwise intersecting toric divisors, rather than to a single irreducible divisor. More precisely, each field is part of a set $\{X_\eta \}_{1,\dots,Q}$ which corresponds to a collection of consecutive strictly external points along the perimeter of the toric diagram. The number $Q$ of fields in the set is still given by the formula in footnote \ref{foot: multiplicity}, but taking the cross product between two non-consecutive legs (in the $(p,q)$-web) that enclose the sequence of points \cite{Hanany:2001py}. Flavoring one of the fields $X_\eta$ via $\delta W = p X_\eta q$ is accomplished by placing a stack of D6-branes on the collection of intersecting divisors, described by $X_\eta = 0$. The map is easily worked out with perfect matchings and the Kasteleyn matrix algorithm; we give an example in Appendix \ref{sec: pm divisors}.

All these statements translate to M-theory. A stack of $h$ D6-branes on the fibered CY$_3$ uplift to a CY$_4$ with $h$ KK monopoles, which locally have complex structure $\bC^2 \times \bC^2/\bZ_h$ and the geometry of a multi-Taub-NUT. The equation $X_\alpha = 0$ describes the location of the core of the multi-Taub-NUT. Such a singularity in M-theory carries $SU(h)$ gauge fields, while the extra $U(1)$ comes from the KK reduction of the bulk potential $C_3$. In fact the geometry of $h$ coincident KK monopoles is
\be
ds_{KK}^2 = U \, d\vec x \cdot d\vec x + \frac1U \, (d\theta + A_\omega)^2 \qquad\qquad \text{with} \qquad
U = \frac1{|\vec x|} + \frac1{\lambda^2} \;,
\ee
where $\vec x \in \bR^3$, $U$ is a harmonic function
on $\bR^3$, $A_\omega = \vec \omega \cdot d\vec x$ is a $U(1)$ connection on $\bR^3$ such that $dU = *_3 \, dA_\omega$, $\theta$ has period $4\pi/h$ and $\lambda$ is the asymptotic radius of the circle.
For $h =1$ the metric is smooth, otherwise it has an $A_{h-1}$ singularity. The 2-form
\be
B = d\Lambda = d \bigg[ \, \frac{|\vec x|}{|\vec x| + \lambda^2} \, \big( d\theta + A_\omega \big) \, \bigg]
\ee
is closed, anti-self-dual, regular and integrable. Thus a local KK reduction $C_3 = A \wedge B$ gives an extra $U(1)$ gauge field propagating around the core of the multi-Taub-NUT.

The flat boundary condition for the connection on the D6-branes uplifts to a flat boundary condition for $C_3$ (and possibly the gauge fields at the singularity). However, since in type IIA the connection is not flat at the tip and its flux can affect the CS levels which ultimately determine the fibration of the CY$_3$ along $\bR$, in M-theory different boundary conditions can uplift to different geometries. An example will be given in subsection \ref{subsec: expl of modified CC2/Z2}.

\subsection{Real masses and partial resolutions}
\label{sec: resolutions}

We can introduce real masses for chiral fields with the term
\be
\int d^4\theta\, Z^\dag e^{\tilde m \theta \bar\theta} Z \;.
\ee
As in Section \ref{sec: flavoring}, we can think of the real mass as a VEV for a background scalar $\sigma_F$, in the $\cN=2$ vector multiplet of $U(h)$. In this way we give opposite mass to the flavors $p$ and $q$.
The VEV of $\sigma_F$ corresponds to the position of the D6-branes along the real line $\bR$ transverse to the CY$_3$. When the D6-branes at $r_0$ are displaced from the D2-branes at the tip, the flavors can be integrated out at low energy. We showed in (\ref{CS shift geometry}) that opposite signs for $\sigma_F$ affect the CS levels, consistently with the field theory discussion in Section \ref{sec: flavoring}.

Real masses, like Fayet-Iliopoulos parameters, do not affect the superpotential \cite{Aharony:1997bx}.
Uplifting to M-theory, real masses do not affect the complex structure of the CY$_4$ but rather its K\"ahler parameters: they correspond to blowing up a 2-cycle. In simple examples, integrating out a flavor pair corresponds to removing a single strictly external point from the 3d toric diagram: the local $\bC^2 \times \bC^2/ \bZ_h$ singularity manifests itself as a column of $h+1$ external points, and integrating out a quark pair with negative (positive) $\vev{\sigma_F}$ corresponds to a partial resolution of the upmost (lowest) point in the column. Only in this limit of infinite mass/resolution parameter, the effective complex structure changes, as the removal of the point in the toric diagram shows.
In more complicated situations, the partial resolution corresponding to giving infinite real mass to a flavor pair could correspond to removing more than one point: the precise map is via perfect matchings, as analyzed in Section \ref{sec: flav moduli space}.

\subsection{Complex masses}
\label{sec: complex masses}

Complex masses for the flavors can correspond to geometric deformations of the D6-brane embeddings, but not always.

Suppose we want to flavor a bifundamental operator $\cO_\alpha = \prod_\beta X_\beta$, made of an open chain of bifundamental fields. We can proceed in the following way: we flavor each field $X_\beta$ separately, and then introduce complex masses for each chiral pair:
\be
W = W_0 + \sum_{\beta=1}^n p_\beta X_\beta q_\beta + \sum_{\beta = 1}^{n-1} m_\beta \, p_{\beta+1} q_\beta  \;.
\ee
After integrating out the massive flavors, we get
\be
W = W_0 + \frac{(-1)^{n-1}}{\prod_\beta m_\beta} \, p_1 \big( \prod\nolimits_\beta X_\beta \big) q_n \equiv W_0 + p_\alpha \cO_\alpha q_\alpha \;,
\ee
with suitable redefinition of fields. Since fermions in vector-like representations do not contribute to the monopole charges, the quantum F-term relation is unmodified:
\be
T \tilde T = \prod\nolimits_\alpha (X_\alpha)^{h_\alpha} = (\cO_\alpha)^{h_\alpha} \;.
\ee
Therefore the two theories where we flavor $\cO_\alpha$ or each $X_\beta$ separately have the same geometric moduli space, and can only differ in their Higgs branches.

The complex masses $m_\beta$ do \emph{not} correspond to deformations of the D6-brane embeddings. In fact we can probe the embedding with D2-branes: the quarks become massless on $\bigcup_\beta \{X_\beta = 0\}$, which does not depend on $m_\beta$. The actual geometric meaning of such masses, which have to do with the intersections between D6's, is not clear to us.

This leads to the following natural generalization. Consider starting with a conical CY$_3$, not necessarily toric, and its dual quiver theory defined by D3-branes probing it. We can always include RR fluxes and fiber it along $\bR$, that is add $\cN=2$ Chern-Simons terms in field theory (the geometry then uplifts to a CY$_4$ in M-theory). Then consider a collection of divisors of the CY$_3$, defined by a set of ``bifundamental equations'' written in terms of bifundamental fields in the quiver theory:
\be
\bigcup\nolimits_\alpha \{ \text{equation}_\alpha = 0 \} \;.
\ee
Each equation is a bifundamental operator and, if it is an adjoint, a mass term $\mu \unit$ can be included. We place $h_\alpha$ D6-branes on the divisor $\{\text{equation}_\alpha = 0\}$. For each equation, this corresponds to introducing a pair of $h_\alpha$ flavor fields, with the correct gauge charges to couple to the bifundamental operator. They contribute to the charges of monopole operators precisely such that the only non-trivial possible quantum relation is
\be
T \tilde T = \prod\nolimits_\alpha (\text{equation}_\alpha)^{h_\alpha} \;.
\ee
It then follows that the moduli space is the CY$_4$
\be
\cM_\mr{flav} = \{ X_a, T, \tilde T \;|\; dW = 0,\; T \tilde T = \prod\nolimits_\alpha (\text{equation}_\alpha)^{h_\alpha} \} // U(1)^{G} \;.
\ee
It would be nice to check or prove this statement.

\section{Examples -- various flavored quiver gauge theories}
\label{sec: examples}

In this section we discuss various examples of three-dimensional toric quiver gauge theories with flavors. Some of the flavored quivers have Chern-Simons terms, others do not. However, even when there are no CS terms, the models have a large $N_f$ expansion ($N_f$ being generically the number of flavors) and in the large $N$ and large $N_f$ limit they are expected to be dual to type IIA string theory on a weakly curved background with D6-branes. When the CS levels do not vanish and there are flavors, two independent expansion parameters $k$ and $N_f$ may be taken large and allow a reduction to type IIA string theory.

All the YM-CS quivers we consider are expected to flow to an interacting fixed point. Using the conjectured OPE of monopole operators explained in Section \ref{sec: monopoles}, we discuss the quantum chiral ring at this fixed point.
Given any toric flavored Chern-Simons quiver, we can use
the Kasteleyn matrix algorithm in the A-theory to find the toric diagram of the geometric moduli space. We will see in various examples how this works in detail. Practically, we solve the moduli space equations of the flavored theory by introducing new perfect matching variables as suggested by the A-theory. The associated GLSM corresponds to the toric CY$_4$ of the geometric moduli space.

Recall that the gauge invariant functions of the GLSM are the affine coordinates of the toric variety, and that they satisfy an algebra which defines the geometry as an algebraic variety. It follows from our construction that the quantum chiral ring of the quiver corresponds to the ring of affine coordinates on the toric variety. This is an important point, since this equivalence is a necessary condition for the existence of an AdS/CFT correspondence.

For each example we can consider the charges $Q^0 \equiv Q_M$ of the GLSM fields under $U(1)_M$. In our convention the charges are such that $\sum_s Q^M_s \vec v_s= (0,0,0,1)$, see  Section \ref{sec: M-theory}.%
\footnote{This only defines $Q_M$ modulo the baryonic symmetries (the other $U(1)$s in the GLSM). However the $U(1)_M$ charges of the affine coordinates are unambiguous.}
Then, one can work out in each case what is the locus of fixed points of the $U(1)_M$ action, and to which divisors it corresponds to in the type IIA reduction, making the link with the top-down approach of Section \ref{sec: M-theory}.

Let us fix the notation. The perfect matching variables $t_i$ of the unflavored quiver are denoted $a_z,   b_z, c_z,\cdots$, with $z$ the vertical coordinate of the corresponding point in the toric diagram. The toric diagram of the flavored theory is obtained by adding columns of points above and below some of the original points, as explained in Section \ref{sec: flav moduli space}. By an $SL(4,\bbZ)$ transformation, we can always set the base of three of the columns of points to $z=0$.
We will always choose such a convenient frame.
Although we consider quivers with Abelian gauge groups only, we nevertheless write the non-Abelian superpotentials, in order to make the link with well-known quivers more explicit.


\subsection{Flavoring the $\bbC^3$ quiver}

Our first example is the flavoring of $\calN=8$ SYM, the low energy field theory on a D2-brane on flat $\bC^3 \times \bR$. The quiver is simply that of $\calN=4$ SYM in 3+1 dimensions. In $\calN=2$ notation, we have a single vector superfield and three adjoint chiral superfields $\Phi_1$, $\Phi_2$, $\Phi_3$, with superpotential $W= \Phi_1[\Phi_2,\Phi_3]$.

\begin{figure}[t]
\begin{center}
\subfigure[\small Quiver with one flavor group.]{
\includegraphics[width=2.6cm]{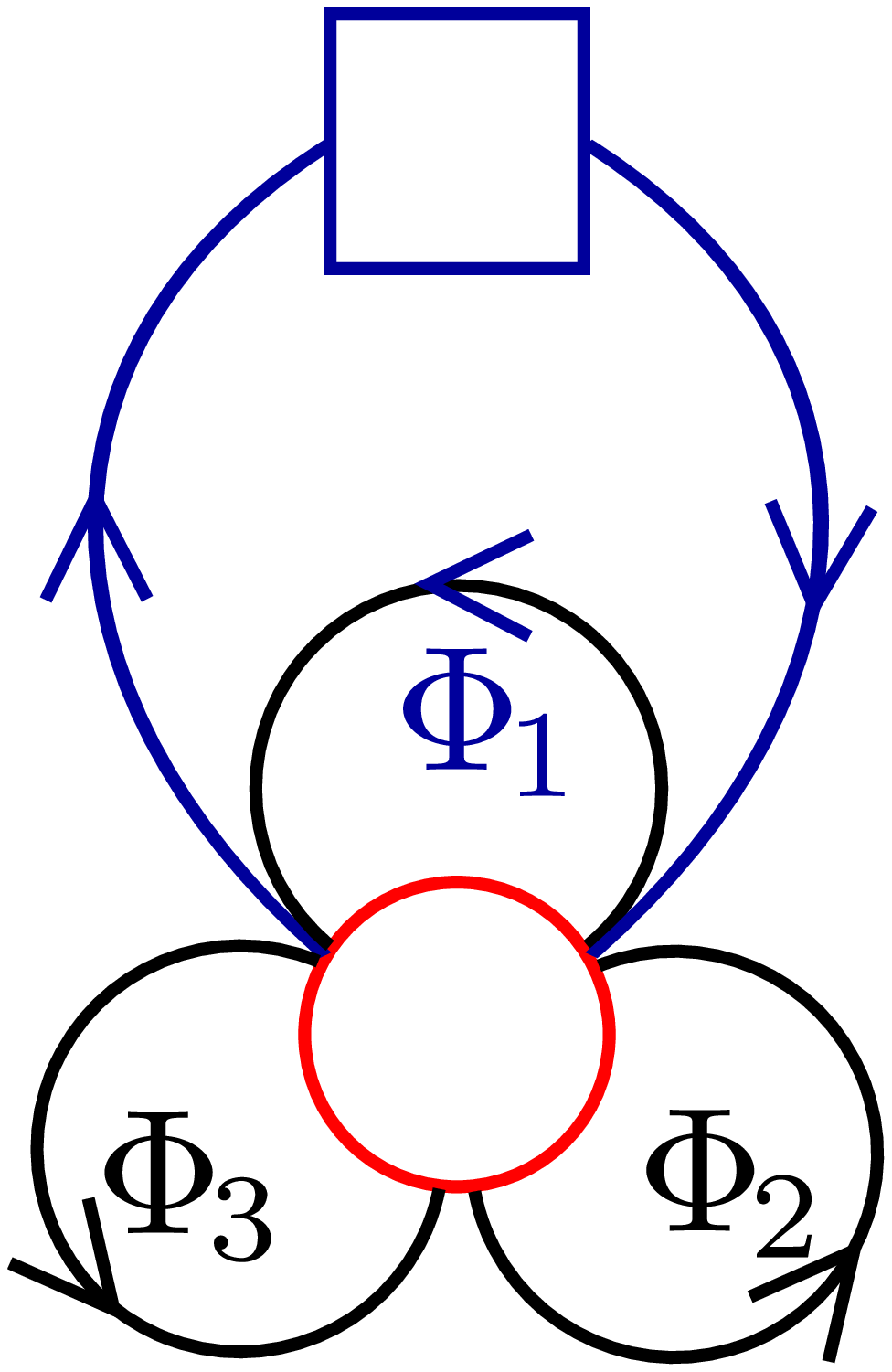}
\label{diagConi}
} \qquad
\subfigure[\small Quiver with two flavor groups.]{
\includegraphics[width=3.8cm]{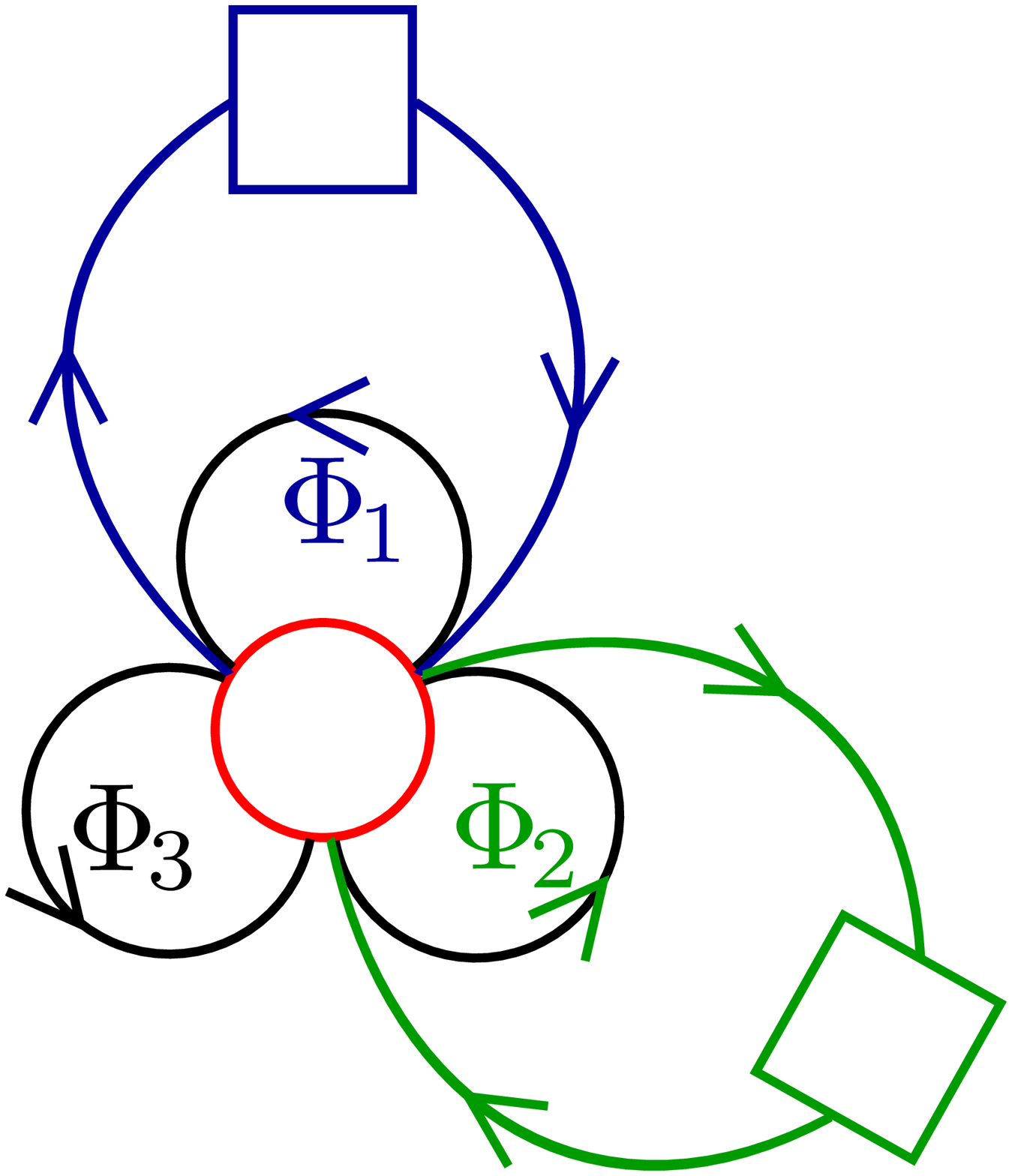}
\label{diagdP1}
} \qquad
\subfigure[\small Quiver with three flavor groups.]{
\includegraphics[width=5.0cm]{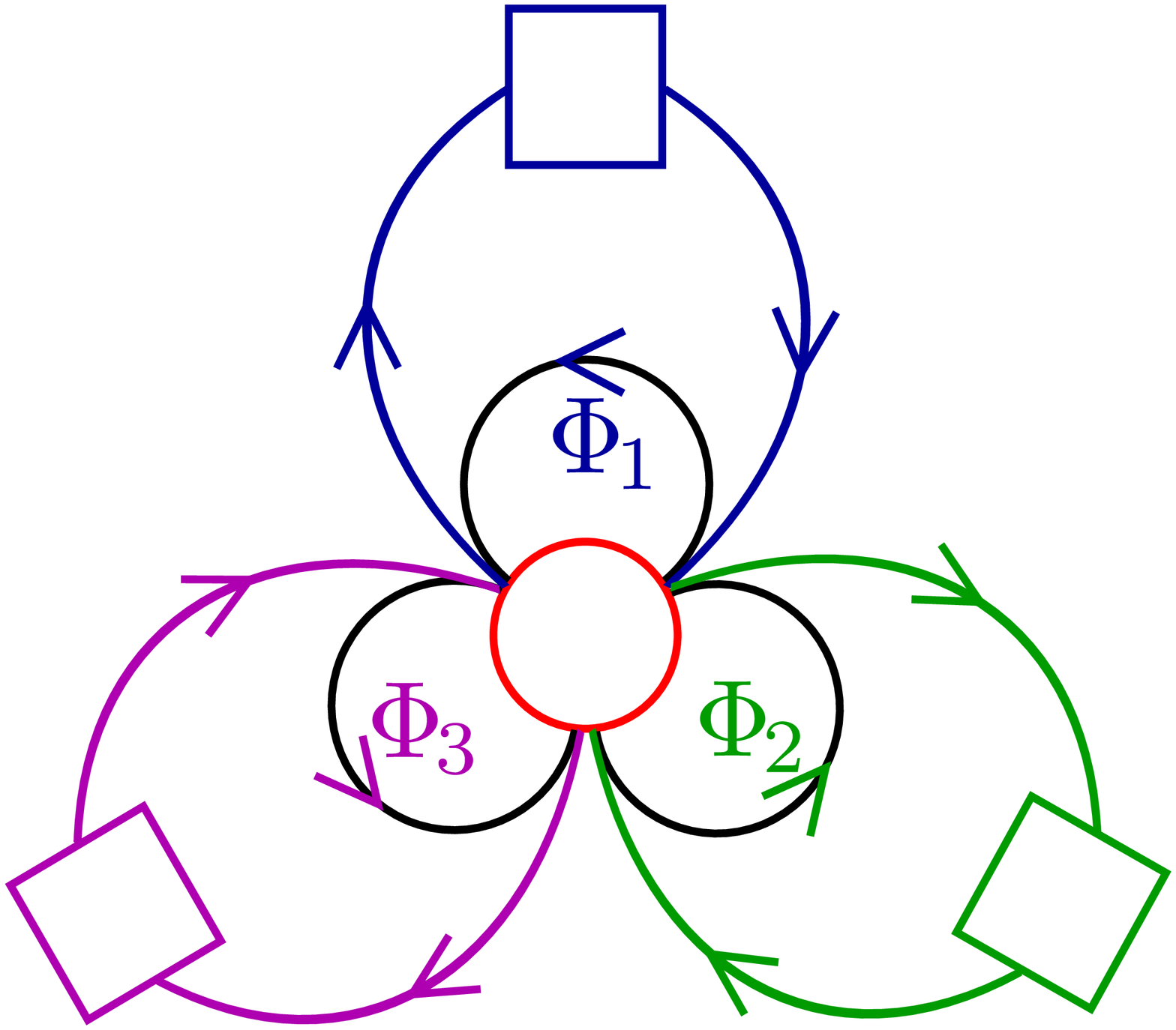}
\label{diagSPP}
}
\caption{\small Quivers for flavored SQED. Circles are gauge groups, squares are flavor groups. Colored arrows indicate bifundamental fields coupled to flavors via a superpotential term.}\label{Flavored C3 quivers}

\vspace{1ex}

\subfigure[\small $\bbC^4$]{
\includegraphics[height=3.8cm]{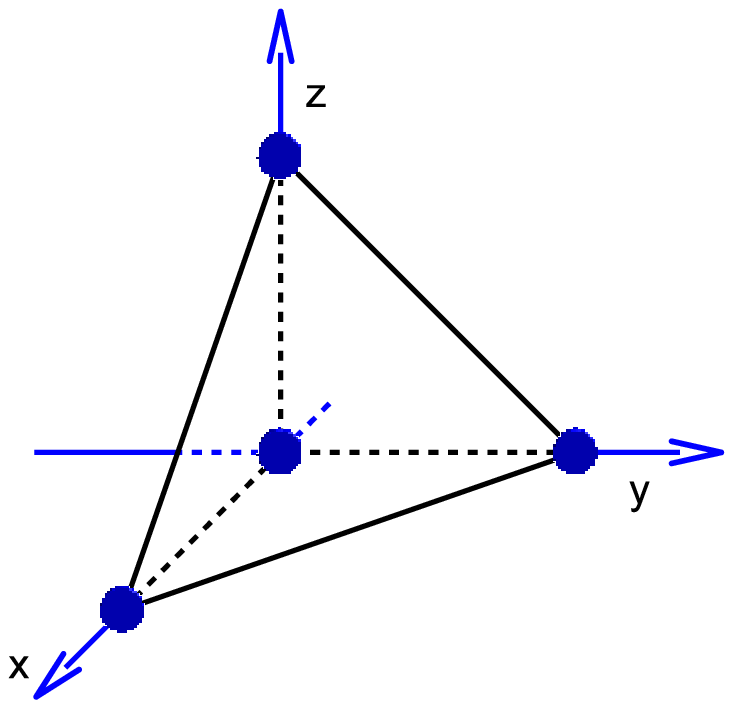}
\label{fig: ToricDiagC4}
} \quad
\subfigure[\small $\calC \times \bbC$]{
\includegraphics[height=3.8cm]{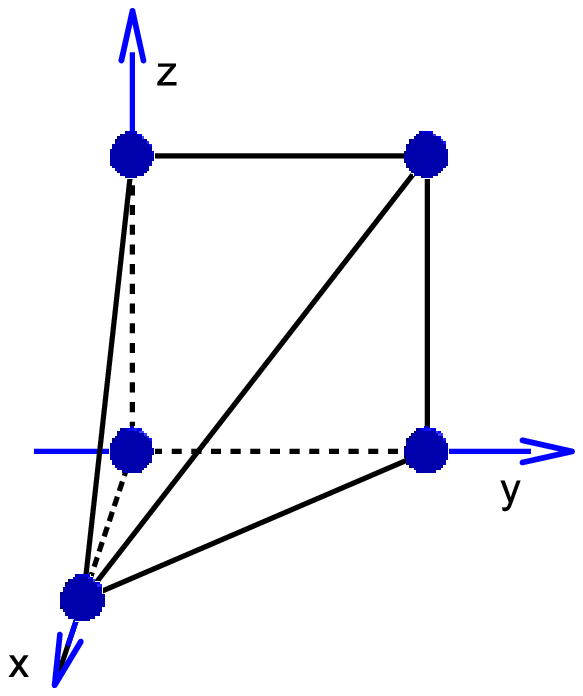}
\label{fig: ToricDiagCConiC3}
} \quad
\subfigure[\small $D_3$]{
\includegraphics[height=3.8cm]{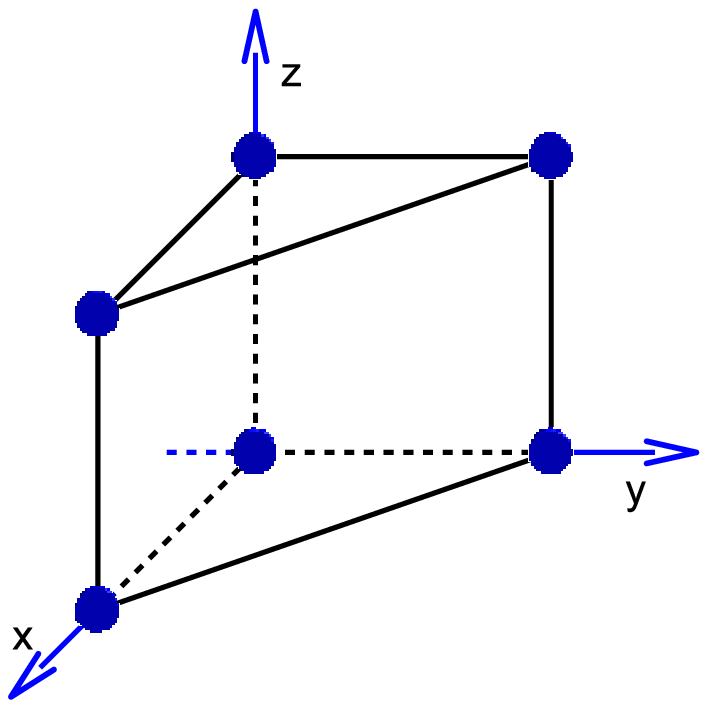}
\label{fig: ToricDiagD3C3}
}
\caption{\small Toric diagrams corresponding to some flavors for the $\bbC^3$ quiver.}
\end{center}
\end{figure}

We can add one, two or three flavor groups by coupling flavors to the appropriate chiral superfields, as shown in Figure \ref{Flavored C3 quivers}.
We denote by $p_i$ and $q_i$ the fundamental and antifundamental fields in the $i$-th flavor group coupled to the field $\Phi_i$.
The flavoring of a $\Phi_i$ corresponds to introducing D6-branes at $z_i=0$, $x_9=0$, and D2/D6-brane intersections induce the superpotential
\be
\label{W of flavored Neq8}
W = \Phi_1 [\Phi_2, \Phi_3] +  \sum_{i=1}^{h_1} p_{1,i} \Phi_1 q_{1,i} + \sum_{j=1}^{h_2} p_{2,j} \Phi_2 q_{2,j} + \sum_{l=1}^{h_3} p_{3,l} \Phi_3 q_{3,l}\;.
\ee
In the general case, the flavor group is $G_F= U(h_1)\times U(h_2)\times U(h_3)/U(1)$.
The charges of the fields under the various gauge and global symmetries are summarized in the following table:
\be
\label{charges of fields in flavored Neq8}
\begin{tabular}{l|ccccccc|cc}
 & $\Phi_i$ & $p_1$ & $q_1$ & $p_2$ & $q_2$ & $p_3$ & $q_3$ & $\tilde{T}$ & $T$ \\
\hline
 $U(1)$ & $0$ & $-1$ & $1$ & $-1$ & $1$ & $-1$ & $1$ &$0$ & $0$ \\
 $U(h_1)$ & $(1)$ & $(h_1)$ & $(\overline{h_1})$& $(1)$ & $(1)$  & $(1)$ & $(1)$  & $(1)$ & $(1)$\\
 $U(h_2)$ & $(1)$ &  $(1)$ & $(1)$&  $(h_2)$ & $(\overline{h_2})$& $(1)$ & $(1)$& $(1)$ & $(1)$\\
 $U(h_3)$ & $(1)$ & $(1)$ &$(1)$&$(1)$ &$(1)$ & $(h_3)$ & $(\overline{h_3})$& $(1)$ & $(1)$\\
\end{tabular}
\ee
In this simple case, flavor groups are non-chirally coupled and so the monopole operators $T$, $\tilde{T}$ do not acquire any gauge charge. Nevertheless, they do acquire some R-charge,
\be
R(T)= R(\tilde T) = \frac{1}{2} \, \big( h_1R(\Phi_1)+h_2R(\Phi_2)+h_2R(\Phi_2) \big) \;.
\ee
The quantum holomorphic relation (\ref{quantum rel}) is
\be
T\tilde{T} = \Phi_1^{h_1}\Phi_2^{h_2}\Phi_3^{h_3} \;.
\ee
It describes an affine variety whose affine coordinates are the  five gauge invariant operators $T$, $\tilde{T}$ and $\Phi_i$ (in the case of a $U(N)$ gauge group one should consider the eigenvalues). Let us discuss a few particular cases related to known models in the literature \cite{Hanany:2008fj, Davey:2009sr, Jensen:2009xh}.

\begin{itemize}
\item In the case $h_2=h_3=0$, The chiral ring relation is
\be
T\tilde{T}= \Phi_1^{h_1},
\ee
and the geometric branch of the moduli space is $\bbC^2\times \bbC^2/\bbZ_{h_1}$. For $h_1=1$ we have $\bbC^4$, see Figure \ref{fig: ToricDiagC4}. This model is related to the dual ABJM model of \cite{Hanany:2008fj, Davey:2009sr}. We discuss it in a bit more details in Section \ref{subsec: dual ABJM}.

\item For $h_1=h_2=1$, $h_3=0$, we have $\bbC \times \calC$ ($\calC$ the conifold), see Fig. \ref{fig: ToricDiagCConiC3}. The A-theory for this model is the so-called Phase III of $\bbC \times \calC$ discussed in \cite{Davey:2009sr}.

\item For $h_1=2$, $h_2=1$, $h_3=0$, we have $\bbC$ times the suspended pinch point (SPP). This was also noticed in \cite{Jensen:2009xh}. In general, for $h_1=a$, $h_2=b$, the geometry is $\bbC \times C(L^{aba})$.

\item For $h_1=h_2=h_3=1$, the geometry is $D_3$, see Fig. \ref{fig: ToricDiagD3C3} . The A-theory for this model is the Phase III of $D_3$ discussed in \cite{Davey:2009sr}.

\end{itemize}
When some $h_i >1$, these geometries have non-isolated singularities. Remark that we have considered the most general toric flavoring of the $\bbC^3$ quiver. The GLSM for the strictly external points is
\be
\begin{tabular}{l|c c c c c c}
            & $t_1$ & $t_2$ & $t_3$ & $t_4$ & $t_5$ & $t_6$ \\ \hline
$U(1)_{B_1}$ & $h_2$  & $-h_1$  & $0$   &  $-h_2$ &  $h_1$ &  $0$   \\
$U(1)_{B_2}$ & $0$  & $h_3$  & $-h_2$   &  $0$ &  $-h_3$ &  $h_2$   \\
$U(1)_{B_3}$ & $-h_3$  & $0$  & $h_1$   &  $h_3$ &  $0$ &  $-h_1$
\end{tabular}
\ee
This GLSM does not encode various orbifold identifications which might in general arise: for a full description of the geometry one should consider the full GLSM, encoding all the relations in the toric diagram, with $h_1+h_2+h_3+3$ homogeneous coordinates.

\subsubsection{Flavoring $\Phi_1$: the dual ABJM geometry}\label{subsec: dual ABJM}

\begin{figure}[t]
\begin{center}
\includegraphics[width=3.8cm]{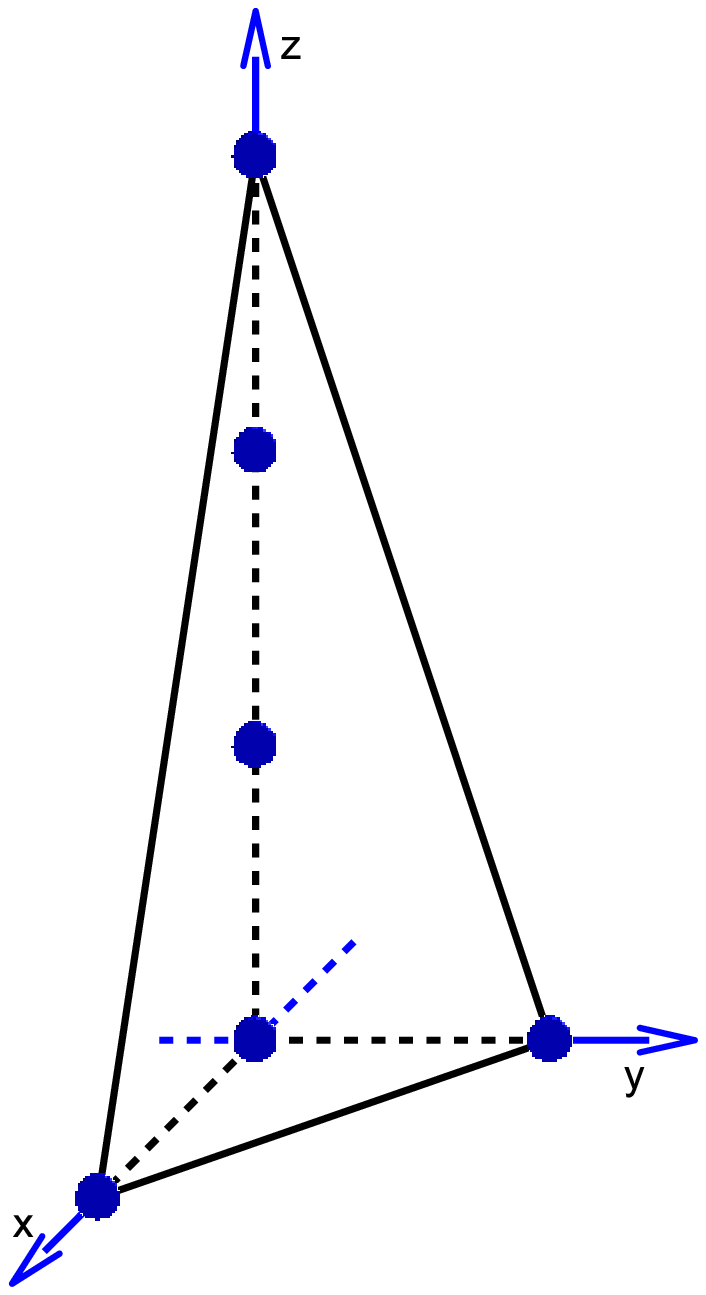}
\caption{\small Toric diagrams of $\bbC^2\times \bbC^2/\bbZ_h$, for $h=3$.}\label{fig:ToricDiagC4Zh}
\end{center}
\end{figure}

Let us discuss a bit more in detail the case $\bbC^2\times \bbC^2/\bbZ_h$.  This geometry has the toric diagram shown in Figure \ref{fig:ToricDiagC4Zh},
\be
a_0=(0,0,0), \quad \cdots,\quad a_h=(0,0,h),  \quad b_0=(0,1,0), \quad c_0=(1,0,0)
\ee
There are $h+3$ homogeneous coordinates, and GLSM
\be
\begin{tabular}{l|c c c c c c ccc}
            & $a_0$ & $b_0$ & $c_0$ & $a_1$ & $a_2$ & $a_3$ & $\cdots$  & $a_{h-1}$ & $a_h$ \\ \hline
$U(1)_{B_1}$ & $1$ &   $0$&    $0$ &  $-2$ &   $1$  & $0$    & $\cdots$  & $0$ &        $0$  \\
$U(1)_{B_2}$ & $0$ &   $0$&    $0$ &  $1$ &   $-2$  & $1$    & $\cdots$  & $0$ &        $0$  \\
$U(1)_{B_3}$ & $0$ &   $0$&    $0$ &  $0$ &   $1$  & $-2$    & $\cdots$  & $0$ &        $0$  \\
$\vdots$     &     &      &        &     &         &          &  $\ddots$  &            &\\
$U(1)_{B_{h-1}}$& $0$ &   $0$&    $0$ &  $0$ &   $0$  & $0$    & $\cdots$  & $-2$ &        $1$ \\ \hline
$U(1)_M$&     $1$ &   $0$&    $0$ &  $-1$ &   $0$  & $0$    & $\cdots$  & $0$ &        $0$
\end{tabular}
\ee
The five affine coordinates are
\bea
x_1 &= \Phi_1=  a_0a_1\dots a_{h-1}a_{h} \;, \;\;\; & x_2 &= T = a_0^h a_1^{h-1} \dots a_{h-2}^2 a_{h-1} \;, \\
x_3 &= \tilde{T}= a_1a_2^2\dots a_{h-1}^{h-1} a_{h}^h \;, & x_4 &= \Phi_2= b_0 \;, & x_5 &= \Phi_3 = c_0 \;,
\eea
and of course they satisfy
\be\label{rel C2over Zh}
x_2x_3=x_1^h.
\ee
 Also, the $Q_M$ charges of $(x_1,\cdots, x_5)$ are $(0,1,-1,0,0)$, so that $U(1)_M$ has fixed points at $x_2=x_3=0$. Gauging $U(1)_M$, we get the type IIA geometry, which is $\bbC^3$ spanned by $(z_1,z_2,z_3)= (x_1,x_4,x_5)$ since the gauge invariant coordinate $x_2x_3$ can be eliminated by (\ref{rel C2over Zh}). The locus of fixed points of $U(1)_M$ in the CY$_4$ descends to the divisor $x_1=0$ in $\bbC^3$, where we must have a stack of $h$ D6-branes. This was the argument of section \ref{sec: M-theory}, which motivates the field theory we presented.

Note that the same geometry is obtained as the moduli space of the so-called dual ABJM model of \cite{Hanany:2008fj}, at CS level $h$. This model was also studied in \cite{Davey:2009sr, Choi:2008za, RodriguezGomez:2009ae}, and some puzzles were found. At $h=1$, the dual ABJM model corresponds to the A-theory for our flavored theory with a single flavor. For $h$ flavors, our A-theory is a tiling with an $(h+1)$-ple bond. It would be interesting to compare in more details our proposal to the one of \cite{Hanany:2008fj}.

For some specific values of the superpotential couplings, the supersymmetry of our flavored quiver gets enhanced to $\calN=4$, since the geometry $\bbC^2\times \bbC^2/\bbZ_h$ is hyper-K\"ahler. Indeed, our setup is a $\calN=2$ version of the setup considered in \cite{Cherkis:2002ir}.

\subsection{Flavoring the conifold quiver}

\begin{figure}[t]
\begin{center}
\subfigure[\small The quiver.]{
\includegraphics[height=5.3cm]{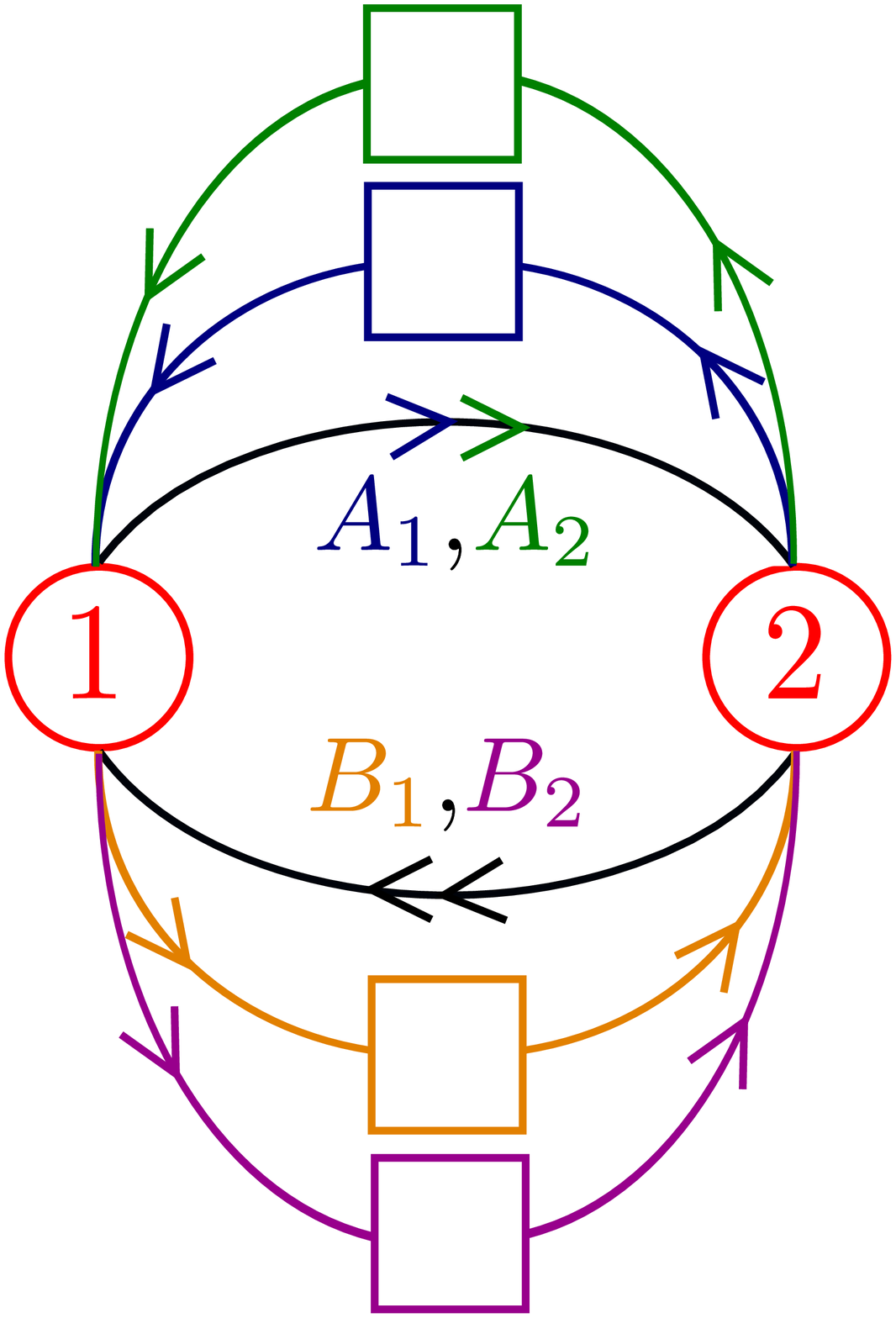}
\label{quiver_general_flav_ABJM}
} \qquad \qquad
\subfigure[\small Toric diagram, for $k=2$.]{
\qquad\includegraphics[height=5.3cm]{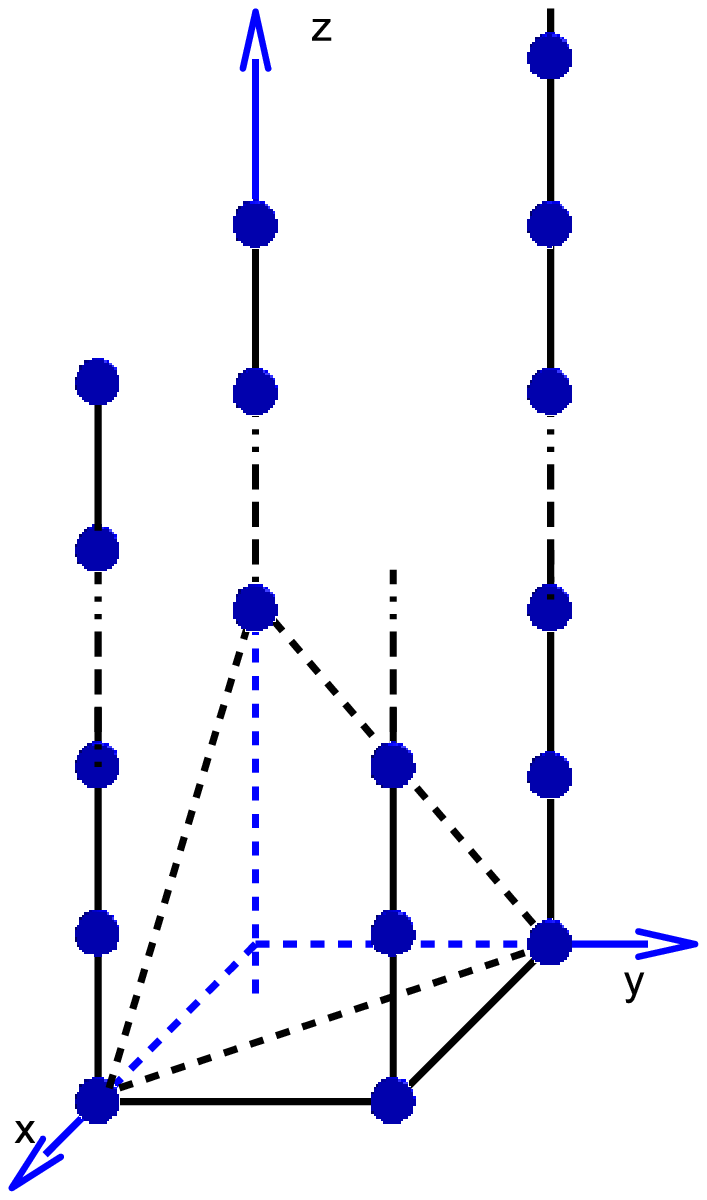}\qquad
\label{fig:Toric_generalFlav}
}
\caption{\small Quiver for a generic flavoring of ABJM, and the corresponding toric diagram, with four columns of heights $h_a$, $\cdots$, $h_d$.}
\end{center}
\end{figure}

Consider the quiver of the ABJM theory, equal to the Klebanov-Witten (KW) quiver for D-branes on the conifold $\calC$. It has two nodes, four bifundamental fields, $A_1$, $A_2$, $B_1$, $B_2$, and superpotential $W=A_1B_1A_2B_2-A_1B_2A_2B_1$.
There are four points in the toric diagram of $\bbC^4/\bbZ_k$, corresponding to the four perfect matchings in the brane tiling of the conifold theory and to the bifundamental fields: because the F-term relations are trivial in the Abelian theory, we can write (with abuse of notation)
\bea
\label{pm variables of KW theory}
a_k &= A_1 =(0,0,k)\;, \qquad\qquad & b_0 &= B_1 = (0,1,0) \;, \\
c_0 &= A_2 =(1,1,0) \;, & d_0 &= B_2 = (1,0,0) \;.
\eea
We then consider the toric diagram obtained by adding four columns of points of heights $h_a$, $h_b$, $h_c$, $h_d$ above the four base points (any other choice of adding the points above or below, is $SL(4,\bZ)$ equivalent to this up to a change in $k$):
\be
\label{generic toric diag for flavored ABJM}
a_{k+i} = (0,0,k+i) \;, \qquad  b_j= (0,1,j) \;, \qquad
c_l = (1,1,l) \;, \qquad d_m = (1,0,m) \;,
\ee
where $i=0,\dots,h_a$, $j=0,\dots,h_b$, $l=0,\dots,h_c$, $m=0,\dots,h_d$. See Figure \ref{fig:Toric_generalFlav}.

This toric geometry corresponds to a generic flavoring of the ABJM theory at level $k$, with flavor group $G_F = U(h_a)\times U(h_b)\times U(h_c)\times U(h_d)/U(1)$.
The quiver is shown in Figure \ref{quiver_general_flav_ABJM}, and the superpotential is
\be\label{W_flavored_ABJM}
\begin{split}
W &= A_1B_1A_2B_2-A_1B_2A_2B_1 +\\
&+\sum_{i=1}^{h_a} p_{1,i} A_1 q_{1,i} + \sum_{j=1}^{h_b} p_{2,j} B_1 q_{2,j} + \sum_{l=1}^{h_c} p_{3,l} A_2 q_{3,l} + \sum_{r=1}^{h_d} p_{4,r} B_2 q_{4,r}  \;.
\end{split}
\ee
Before studying several interesting cases, let us discuss the general solution for the geometric moduli space in this family of models. We have the quantum relation  (\ref{quantum rel}),
\begin{equation}
\label{quantum_flav_ABJM}
T \tilde{T} = A_1^{h_a} B_1^{h_b} A_2^{h_c} B_2^{h_d} \;,
\end{equation}
and the CS levels are $(k+f,-k-f)$, with $f=\frac{1}{2}(h_a-h_b+h_c-h_d)$.
The gauge charges of bifundamental fields and monopole operators are (schematically)
\be\label{charges_flav_ABJM}
\begin{tabular}{c|c c c c}
                & $A_i$ & $B_j$ &  $T$  & $\tilde{T}$ \\ \hline
$U(1)_{k+f}$    &  $1$  & $-1$  &  $k+2f$ &  $-k$    \\
$U(1)_{-(k+f)}$ & $-1$  &  $1$  &  $-k-2f$ & $k$
\end{tabular}
\ee
The relation (\ref{quantum_flav_ABJM}) can be solved by the perfect matching variables, as
\begin{equation}\label{bifundamentals_flav_ABJM}
A_1=\prod_{i=0}^{h_a} a_{k+i}\;,\qquad B_1=\prod_{j=0}^{h_b} b_j\;,\qquad A_2=\prod_{l=0}^{h_c} c_l\;,\qquad B_2=\prod_{m=0}^{h_d} d_m
\end{equation}
and
\begin{equation}\label{T_and_Ttilde_flav_ABJM}
\begin{split}
T &= \Big(\prod_{i=0}^{h_a} a_{k+i}^{h_a-i} \Big) \Big(\prod_{j=0}^{h_b} b_j^{h_b-j} \Big) \Big(\prod_{l=0}^{h_c} c_l^{h_c-l} \Big) \Big(\prod_{m=0}^{h_d} d_m^{h_d-m} \Big)  \\
\tilde{T} &= \Big(\prod_{i=0}^{h_a} a_{k+i}^i \Big) \Big(\prod_{j=0}^{h_b} b_j^j \Big) \Big(\prod_{l=0}^{h_c} c_l^l \Big) \Big(\prod_{m=0}^{h_d} d_m^m \Big) \;,
\end{split}
\end{equation}
Notice that each perfect matching variable (\ref{pm variables of KW theory}) of the ABJM theory is replaced by the product of all GLSM fields associated to the relevant column of points in the toric diagram. Monopole operators are instead products of fields along the four columns, with increasing or decreasing powers as we move vertically. This is to be compared to \eqref{equation T in term of R and C}.
It is easy to show that the $U(1)$ ambiguities of this parametrization reproduce the GLSM associated to the toric diagram (\ref{generic toric diag for flavored ABJM}).


\subsubsection{Flavoring the field $A_1$: the $\calC\times \bbC$ geometry}

\begin{figure}[t]
\begin{center}
\subfigure[\small The quiver.]{
\includegraphics[height=3.8cm]{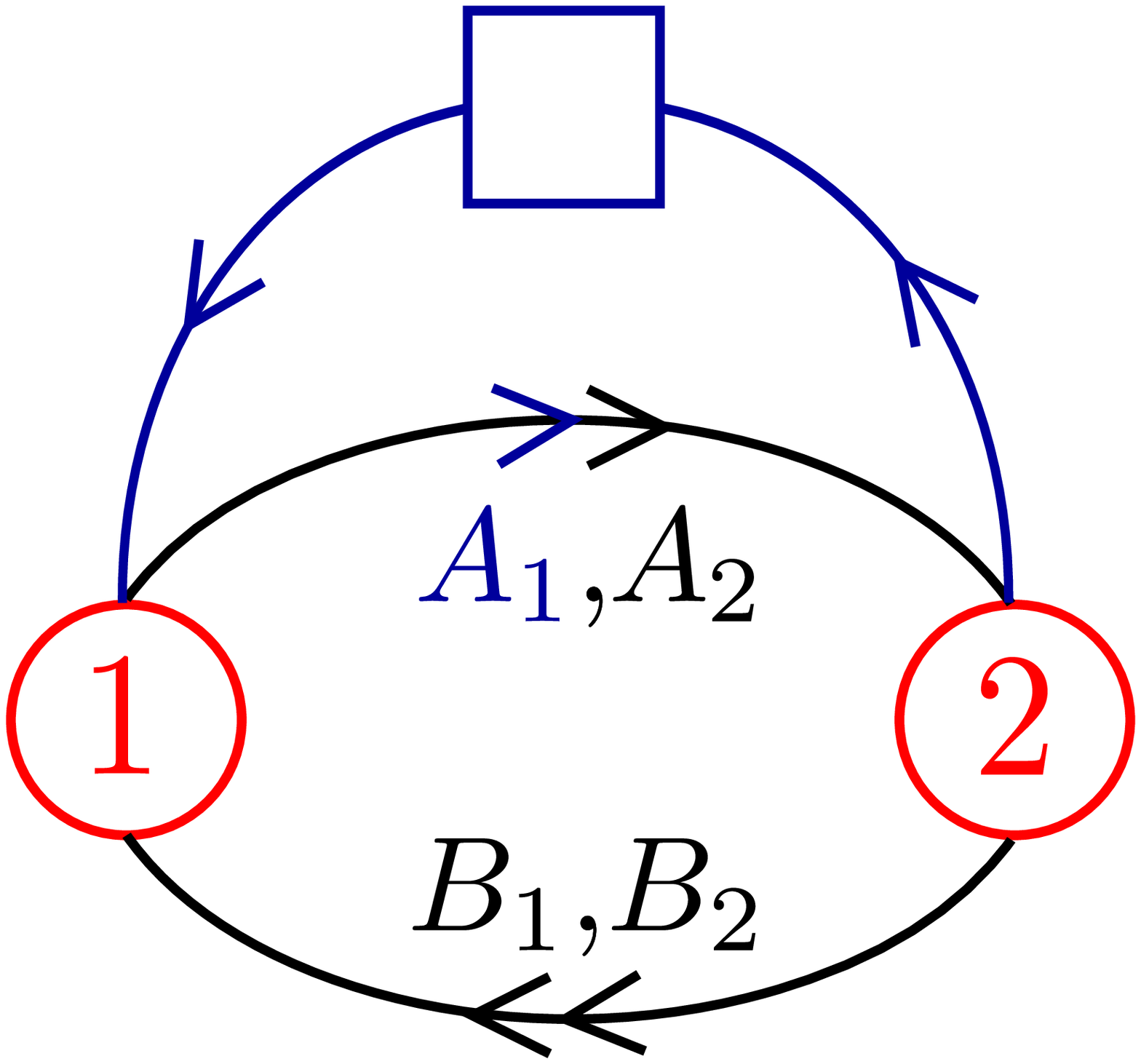}
\label{fig:KW1flav}
} \qquad \qquad
\subfigure[\small $\calC \times \bbC$]{
\includegraphics[width=4.0cm]{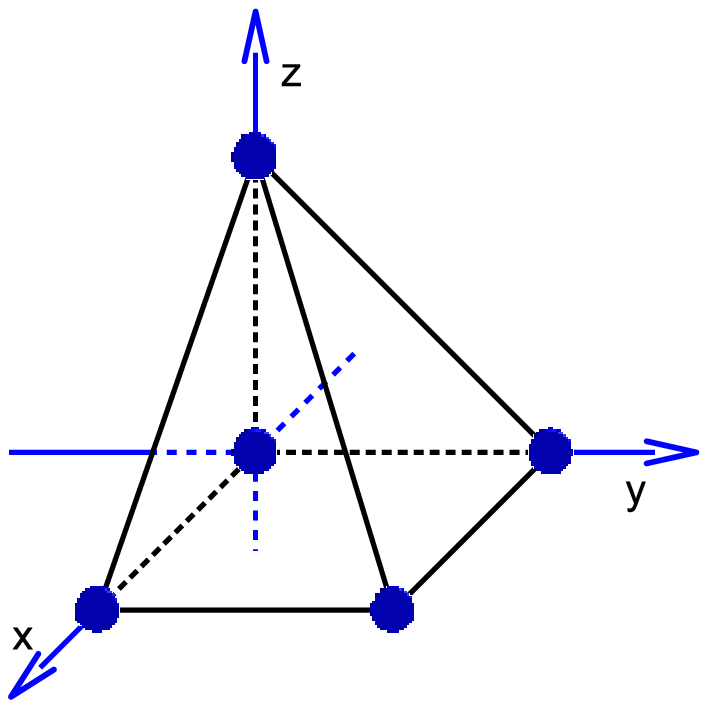}
\label{fig: ToricDiagCConifoldKW}
}
\caption{\small ABJM quiver with one chiral flavor, and its dual geometry.}
\end{center}
\end{figure}

Let us add a $U(1)$ flavor group to the 3d KW theory ($k=0$), coupled to the bifundamental field $A_1$ as in Figure \ref{fig:KW1flav}.
The superpotential is $W= A_1B_1A_2B_2 - A_1B_2A_2B_1  + pA_1q$, and the CS levels are $(\frac{1}{2},-\frac{1}{2})$. The charges of the fields under the gauge and flavor groups are
\be
\begin{tabular}{l|cccc|cc}
 & $A_i$ &$B_i$  & $p$& $q$ & $T$ & $\tilde{T}$   \\
 \hline
 $U(1)_{\frac{1}{2}}$  &$1$ &$-1$  &$0$ &$-1$ & 1 & $0$   \\
 $U(1)_{-\frac{1}{2}}$  &$-1$ &$1$  &$1$ &$0$& $-1$ & $0$  \\
 $U(1)_F$  &$0$ &$0$  &$-1$ &$1$& 0 & $0$
\end{tabular}
\ee
There are seven gauge invariant operators, namely $A_iB_j$, $TB_i$ and $\tilde T$. Using the quantum relation $T\tilde{T} = A_1$, we can however express $A_1B_i$ as $\tilde T TB_i$, so that we actually have only 5 generators of the chiral ring,
\be
  x_1=TB_1 \;, \quad x_2= A_2B_2 \;, \quad x_3=TB_2 \;, \quad x_4= A_2B_1 \;, \quad  x_5=\tilde{T} \;,
\ee
subject to the relation
\be
x_1x_2-x_3x_4=0\, .
\ee
Hence, the moduli space is $\calC\times \bbC$. Indeed, the quantum relation can be solved by $T=a_0$, $\tilde{T}=a_1$ and $A_1= a_0a_1$. The GLSM is
\be
\begin{tabular}{l|c c c c c }
            & $a_0$ & $b_0$ & $c_0$ & $d_0$ & $a_1$ \\ \hline
$U(1)_B$ & $1$  & $-1$  & $1$   &  $-1$ &  $0$   \\ \hline
$U(1)_M$ & $1$  &  $0$ &   $0$   & $0$   & $-1$
\end{tabular}
\ee
where we also specified the $U(1)_M$ charges. The toric diagram is shown in Figure \ref{fig: ToricDiagCConifoldKW}.
The locus of fixed points of the $U(1)_M$ action descends to the toric divisor $\{a_0=0\}$ in the conifold, where the D6-brane sits.

\subsubsection{Flavoring the field $A_1$: the $C(Y^{2,1}(\bC\bP^2))$ geometry}

\begin{figure}[t]
\begin{center}
\subfigure[\small $C(Y^{2,1}(\bC\bP^2))$]{
\includegraphics[width=4.1cm]{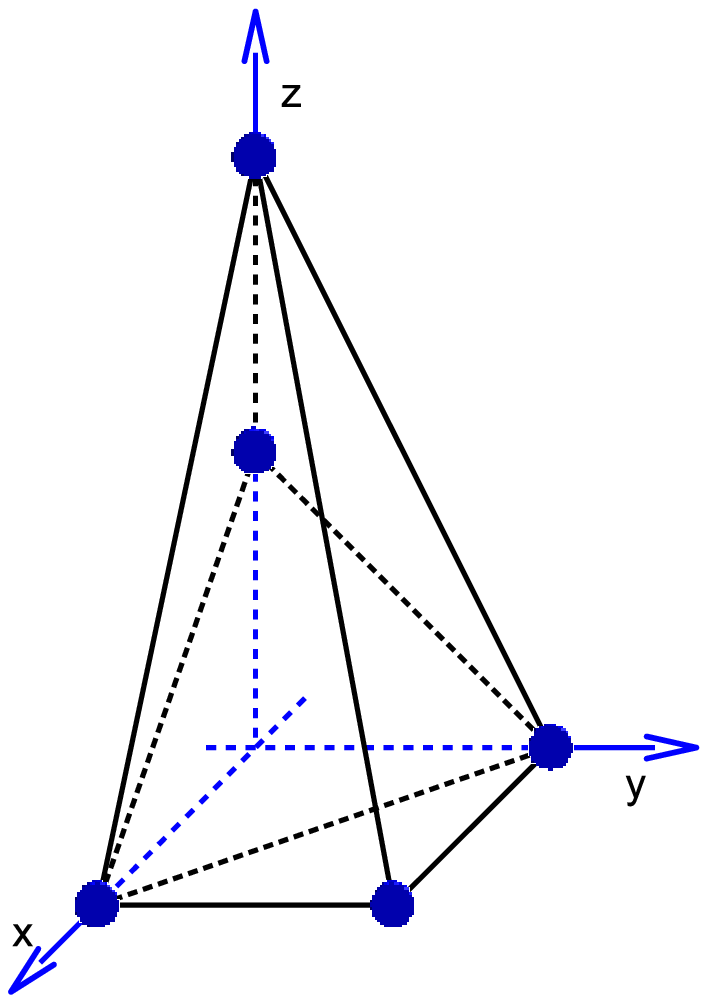}
\label{fig: ToricDiagY21CP2}
}\quad
\subfigure[\small $D_3$.]{
\includegraphics[width=4.1cm]{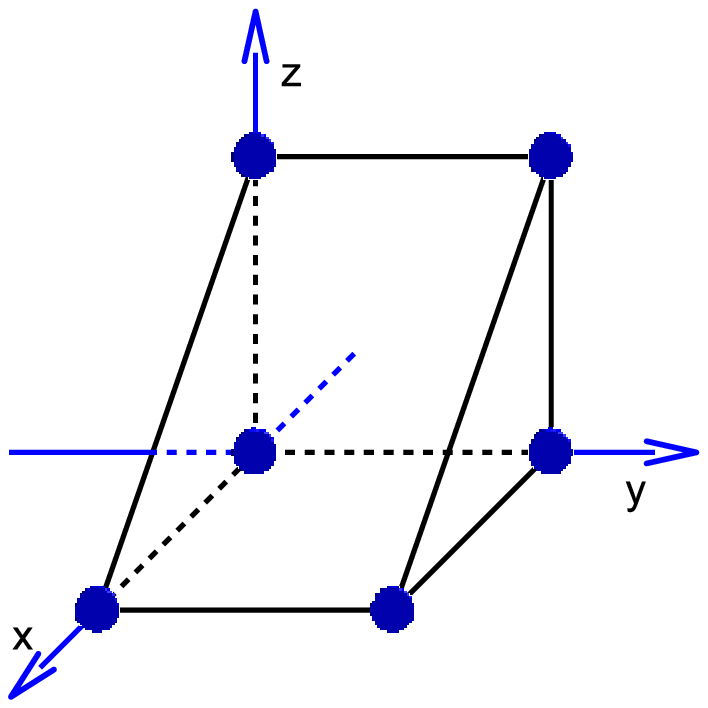}
\label{fig:ToricDiagD3}
}\quad
\subfigure[\small The cubic conifold.]{
\includegraphics[width=4.1cm]{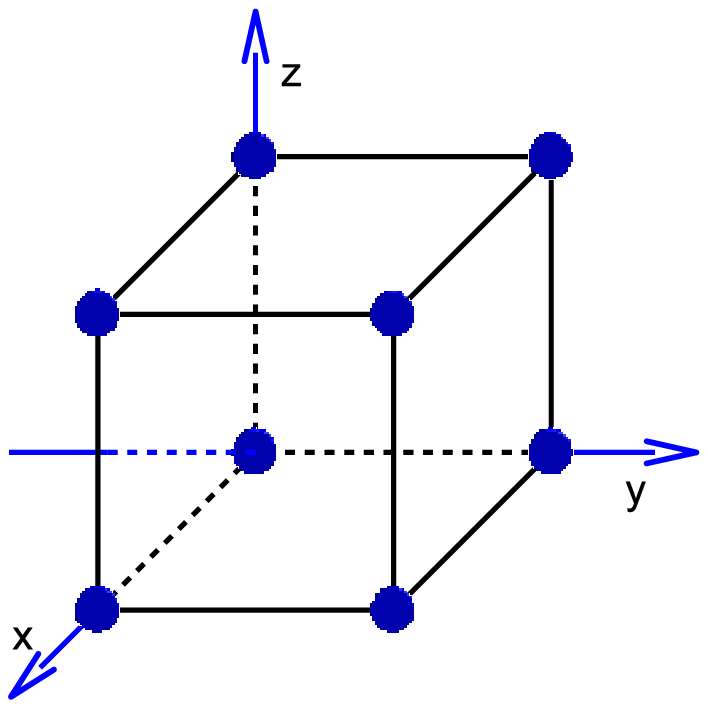}
\label{fig:ToricDiagCube}
}
\caption{\small Toric diagrams corresponding to some flavors for the ABJM quiver.}
\end{center}
\end{figure}

Let us then couple a $U(1)$ flavor group to $A_1$ in the ABJM theory at level $k=1$. Now the CS levels are $(\frac{3}{2}, -\frac{3}{2})$ and the fields have gauge charges
\be
\begin{tabular}{l|cccc|cc}
 & $A_i$ &$B_i$ & $p$&$q$ & $T$ & $\tilde{T}$  \\
 \hline
 $U(1)_{\frac{3}{2}}$  &$1$ &$-1$ & $0$& $-1$& 2 & $-1$    \\
 $U(1)_{-\frac{3}{2}}$  &$-1$ &$1$ & $1$& $0$& $-2$ & $1$
\end{tabular}
\ee
The quantum relation is solved by $T=a_1$, $\tilde{T}=a_2$,  $A_1=a_1a_2$. The GLSM is
\be
\begin{tabular}{c|c c c c c }
             & $a_1$ & $b_0$ & $c_0$ & $d_0$ & $a_2$  \\ \hline
$U(1)_{B}$ &  $2$  &  $-1$  &  $1$  &  $-1$  & $-1$   \\ \hline
$U(1)_M$ & $1$  &  $0$ &   $0$   & $0$   & $-1$
\end{tabular}
\ee
The corresponding toric diagram is shown in Fig. \ref{fig: ToricDiagY21CP2}, and it corresponds to the cone over $Y^{2,1}(\bC\bP^2)$ \cite{Martelli:2008rt}. This geometry and a related theory (actually the A-theory for our flavored theory) was discussed in \cite{Benishti:2009ky}.
There are nine gauge invariant operators for this quiver, matching the nine affine coordinates of the $C(Y^{2,1}(\bC\bP^2))$ singularity:
\be\nonumber
\begin{array}{ccc}
\qquad x_1=TB_1B_1 = a_1b_0^2,&
\qquad x_2=TB_2B_2 = a_1d_0^2,&
\qquad x_3=\tilde{T}A_1 = a_1a_2^2, \\
\qquad x_4=TB_1B_2= a_1b_0d_0,&
\qquad x_5=A_1B_1= a_1b_0a_2,&
\qquad x_6=A_1B_2=a_1d_0a_2 ,\\
\qquad x_7=A_2B_1= b_0c_0,&
\qquad x_8=A_2B_2= c_0d_0,&
\qquad x_9=\tilde{T}A_2= c_0a_2
\end{array}
\ee
The chiral ring relations are:
\be\label{affine coord relations for Y21}
\begin{array}{cccc}
x_1x_8=x_4x_7\;,\qquad& x_2x_9=x_6x_8\;, & \qquad x_3x_7=x_5x_9\;, & \qquad x_4x_9=x_5x_8\;, \\
x_1x_9=x_5x_7\;,\qquad& x_2x_7=x_4x_8\;, & \qquad x_3x_8=x_6x_9\;, & \qquad x_4x_9=x_6x_7\;, \\
x_1x_2=x_4^2\;,\qquad& x_1x_3=x_5^2\;,&\qquad x_2 x_3=x_6^2\;,\\
x_1x_6=x_4x_5\;,\qquad& x_2x_5=x_4x_6\;,&\qquad x_3x_4=x_5x_6 \;.
\end{array}
\ee

\subsubsection{Flavoring the fields $A_1$ and $A_2$: the $C(Q^{1,1,1})$ geometry}\label{subsec: Q111 from 2 flavors}

\begin{figure}[t]
\begin{center}
\subfigure[\small ABJM quiver with two chiral flavor groups.]{
	\qquad\qquad\includegraphics[height=4.5cm]{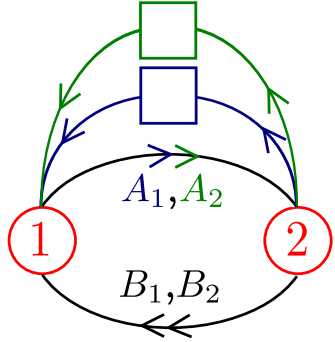}\qquad\qquad
\label{fig:KW2chiralflav}
} \qquad \qquad \quad
\subfigure[\small $C(Q^{1,1,1})$.]{
\includegraphics[width=4.0cm]{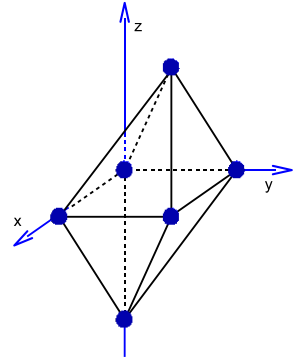}
\label{fig: ToricDiagQ111}
}
\caption{\small ABJM quiver with two chiral flavor groups, and a dual geometry.}
\end{center}
\end{figure}

Consider the conifold quiver with two $U(1)$ flavor groups coupled to $A_1$ and $A_2$ respectively, as in Fig. \ref{fig:KW2chiralflav}. The superpotential is
\be
W= A_1B_1A_2B_2 - A_1B_2A_2B_1  +p_1A_1q_1 +p_2A_2q_2 \;,
\ee
and we choose vanishing CS levels. In the toric diagram, this corresponds to adding one point below $a_0$ and one point above $c_0$, see Fig. \ref{fig: ToricDiagQ111}. The gauge charges of the fields and monopole operators are
\be
\begin{tabular}{l|cccc|cc}
 & $A_i$ &$B_i$  & $p_i$& $q_i$ & $T$ & $\tilde T$  \\
 \hline
 $U(1)_0$  &$1$ &$-1$  &$0$ &$-1$ & $1$ & $1$  \\
 $U(1)_0$  &$-1$ &$1$  &$1$ &$0$ & $-1$ & $-1$
\end{tabular}
\ee
The monopole operators satisfy the relation
\be\label{quantm rel TT eq A1A2}
T\tilde{T} = A_1 A_2\;.
\ee
We can solve it by introducing two new perfect matching variables $a_{-1}$ and $c_1$:
\bea
\label{pm variables of Q111 theory}
A_1 =a_{-1}a_0 , &\qquad B_1 =b_0, &\qquad T =a_{-1}c_0, \\
A_2 =c_0c_1, &\qquad  B_2= d_0, &\qquad \tilde{T} =a_0c_1.
\eea
The associated GLSM is a minimal presentation of the one for the real cone over $Q^{1,1,1}$:
\be
\begin{tabular}{c|c c c c c c}
             & $a_0$ & $b_0$ & $c_0$ & $d_0$ & $a_{-1}$ & $c_1$ \\ \hline
$U(1)_{B_1}$ &  $1$  &  $-1$  &  $1$  &  $-1$  & $0$  & $0$ \\
$U(1)_{B_2}$ &  $1$  &  $0$  &  $1$ &  $0$ &  $-1$  &  $-1$ \\ \hline
$U(1)_M$     &  $0$  &  $0$  &  $1$ &  $0$ &  $0$  &  $-1$
\end{tabular}
\ee
The gauge invariant operators generating the chiral ring are:
\be\nn
\begin{array}{ccc}
\qquad x_1= A_1B_1= a_{-1}a_0b_0 ,&
\qquad x_2= A_2B_2= c_0c_1d_0 ,&
\qquad  x_3= A_2B_1= b_0c_0c_1 ,\\
\qquad x_4= A_1B_2= a_{-1}a_0d_0 ,&
\qquad  x_5= \tilde{T}B_1= a_0b_0c_1 ,&
\qquad x_6=\tilde{T}B_2= a_0c_1d_0 ,\\
\qquad x_7=TB_1= a_{-1}b_0c_0 ,&
\qquad  x_8=TB_2=  a_{-1}c_0d_0.&
\qquad   \;
\end{array}
\ee
They of course correspond to the affine coordinates on $C(Q^{1,1,1})$, whose algebra is
\be\label{chiral ring equ q111}
\begin{array}{ccc}
\qquad x_1x_2-x_3x_4=0\;,&
\qquad x_1x_2-x_5x_8=0\;,&
\qquad x_1x_2-x_6x_7=0\;,\\
\qquad x_1x_3-x_5x_7=0\;,&
\qquad x_1x_6-x_4x_5=0\;,&
\qquad x_1x_8-x_4x_7=0\;,\\
\qquad x_2x_4-x_6x_8=0\;,&
\qquad x_2x_5-x_3x_6=0\;,&
\qquad x_2x_7-x_3x_8=0\;.
\end{array}
\ee
Remark that the affine coordinates have $U(1)_M$ charges
\be
\begin{tabular}{c|c c c c c c c c c}
            & $x_1$ & $x_2$ & $x_3$ & $x_4$ & $x_5$ & $x_6$ & $x_7$ & $x_8$  \\ \hline
$U(1)_M$ & $0$  & $0$  & $0$   &  $0$ &  $-1$  &  $-1$  & $1$   &  $1$
\end{tabular}
\ee
so the $U(1)_M$ fixed point locus is at $x_5=x_6=x_7=x_8=0$, $x_1x_2=x_3x_4=x_1x_3=x_2x_4=0$. This locus of fixed points has two branches:
\bea
1)\;  &x_1=x_4=0\,, \quad  x_5=x_6=x_7=x_8=0\,, \quad \forall x_2,\,  x_3  &\quad &\Longleftrightarrow \quad   a_0=a_{-1}=0 \\
2)\;  &x_2=x_3=0\,, \quad  x_5=x_6=x_7=x_8=0\,, \quad  \forall x_1,\,  x_4 &\quad &\Longleftrightarrow \quad c_0=c_1=0 \;.
\eea
It is easy to see that they descend to the toric divisors $\{a_0=0\}$ and $\{c_0=0\}$ in the conifold $\calC$. The D6-branes wrapping these divisors provide us with the chiral flavors in the quiver field theory.

Another quiver for the low energy field theory on M2-branes on $C(Q^{1,1,1})$ was proposed in  \cite{Franco:2008um}, and further studied in \cite{Franco:2009sp}. The quiver of \cite{Franco:2008um}, which has two double-bonds, is precisely the A-theory of our chirally flavored conifold theory.

\subsubsection{Flavoring the fields $A_1$ and $B_1$: the $D_3$ geometry} \label{subsubsec: D3 geom from ABJM}

Let us now couple a $U(1)$ flavor group to $A_1$ and a $U(1)$ flavor group to $B_1$, with $\delta W= p_1A_1q_1 +\tilde{p}_1B_1\tilde{q}_1$ and vanishing CS levels. In this case there is no induced gauge charge for the monopole operators, because there are as many incoming as outgoing arrows in each gauge group. We have the quantum relation $T\tilde{T}= A_1B_1$, which is solved by $A_1=a_0a_1$, $B_1=b_0b_1$, $T= a_0b_0$ and $\tilde{T}= a_1b_1$. The associated GLSM is
\be
\begin{tabular}{c|c c c c c c}
             & $a_0$ & $b_0$ & $c_0$ & $d_0$ & $a_1$ & $b_1$ \\ \hline
$U(1)_{B_1}$ & $1$   & $-1$   & $1$  & $-1$   & $0$   & $0$ \\
$U(1)_{B_2}$ & $1$  & $-1$   & $0$   & $0$  & $-1$   & $1$ \\ \hline
$U(1)_M$ & $1$  & $0$   & $0$   & $0$  & $-1$   & $0$
\end{tabular}
\ee
The toric diagram, shown in Fig. \ref{fig:ToricDiagD3}, is the one of the $D_3$ geometry. The generators of the chiral ring are
\be
x_1= \tilde{T}\;, \quad x_2=A_2B_2\;, \quad x_3=T\;, \quad x_4= A_1B_2\;, \quad x_5=A_2B_1\;.
\ee
As expected, they satisfy the defining equation of the $D_3$ singularity:
\be
x_1x_2x_3- x_4x_5=0 \;.
\ee
The locus of fixed points of $U(1)_M$ has two branches which descend to the two divisors $\{a_0=0\}$ and $\{b_0=0\}$ in the conifold.

\subsubsection{Flavoring $A_1$, $A_2$, $B_1$, $B_2$: the cubic conifold}

Consider coupling a $U(1)$ flavor group to each bifundamental field, with vanishing CS levels. The quantum relation is
\be
T\tilde{T} = A_1B_1A_2B_2\;.
\ee
One can check that the moduli space is described by the following GLSM:
\be
\begin{tabular}{c|c c c c c c c c}
             & $a_0$ & $b_0$ & $c_0$ & $d_0$ & $a_1$ & $b_1$ & $c_1$ & $d_1$ \\ \hline
$U(1)_{B_1}$ & $1$   & $-1$   & $1$  & $-1$   & $0$   & $0$ & $0$   & $0$ \\
$U(1)_{B_2}$ & $0$   & $0$   & $0$  & $0$   & $1$   & $-1$ & $1$   & $-1$\\
$U(1)_{B_3}$ & $1$   & $0$   & $0$  & $-1$   & $-1$   & $0$& $0$   & $1$ \\
$U(1)_{B_4}$ & $1$  & $-1$   & $0$   & $0$  & $-1$   & $1$ & $0$   & $0$\\ \hline
$U(1)_M$ & $1$  & $0$   & $0$   & $0$  & $-1$   & $0$ & $0$   & $0$
\end{tabular}
\ee
The toric diagram is shown in Fig. \ref{fig:ToricDiagCube}, and we will call this geometry the \emph{cubic conifold}.
The gauge invariant operators are
\be\nonumber
\begin{array}{ccc}
\qquad x_1=a_0b_0c_0d_0= T,&
\qquad x_2=a_1b_1c_1d_1=\tilde{T},&
\qquad x_3=a_0b_0a_1b_1=A_1B_1,\\
\qquad x_4=c_0d_0c_1d_1=A_2B_2,&
\qquad x_5=a_0d_0a_1d_1=A_1B_2,&
\qquad x_6=b_0c_0b_1c_1=A_2B_1,
\end{array}
\ee
satisfying the equations
\be
\label{defining equs cubic conifold}
x_1x_2-x_3x_4=0 \;, \qquad\qquad  x_1x_2-x_5x_6=0 \;.
\ee
This is a complete intersection. The $U(1)_M$ charges of $(x_1, \cdots, x_6)$ are $(1,-1,0,0,0,0)$.
The locus of fixed point is at $x_1=x_2=0$, $x_3x_4=x_5x_6=0$, which has four branches and descend to the four toric divisors of the conifold.


\subsection{Flavoring the modified $\bC\times\bC^2/\bZ_2$ theory}\label{subsec: expl of modified CC2/Z2}

\begin{figure}[t]
\begin{center}
\subfigure[\small The $A_1$ quiver.]{
\includegraphics[height=2.8cm]{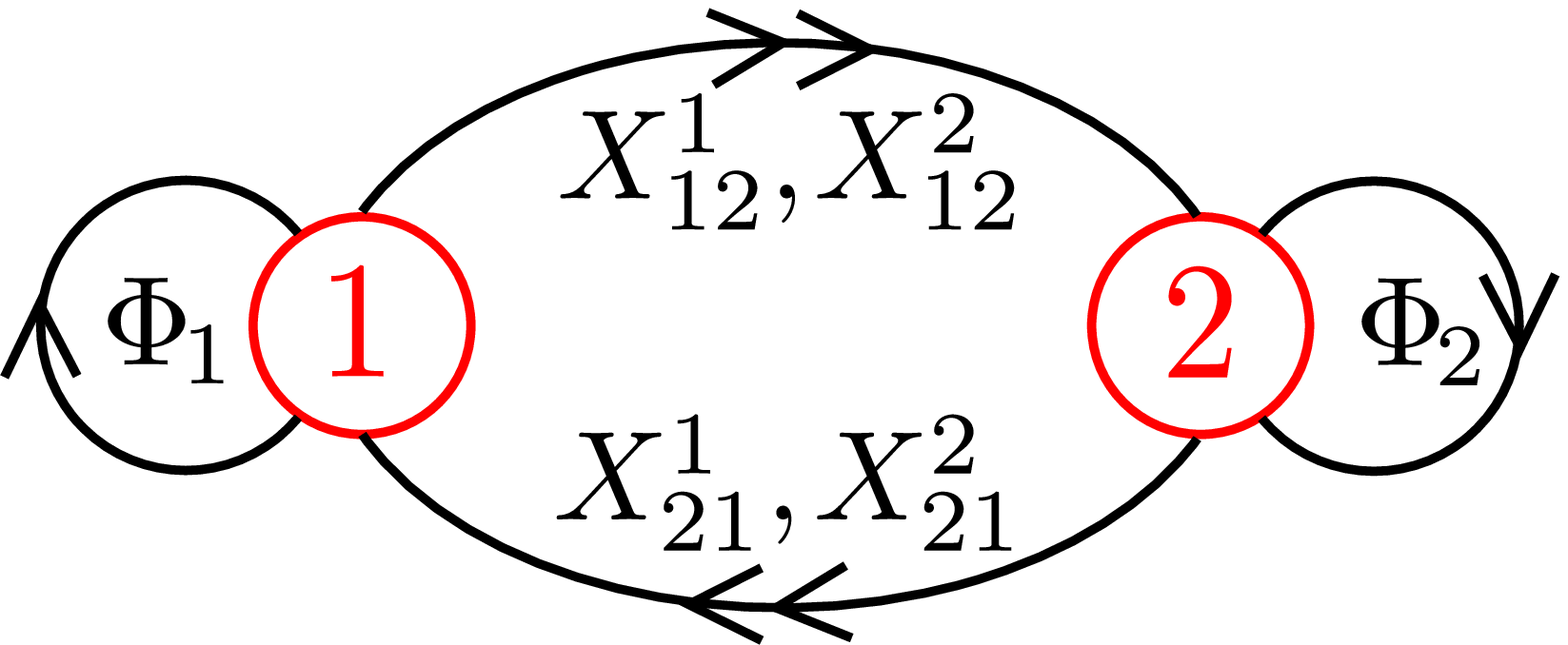}
\label{quiver_CC2Z2}
}\qquad\qquad
\subfigure[\small $\bbC\times \calC$.]{
\includegraphics[height=3.7cm]{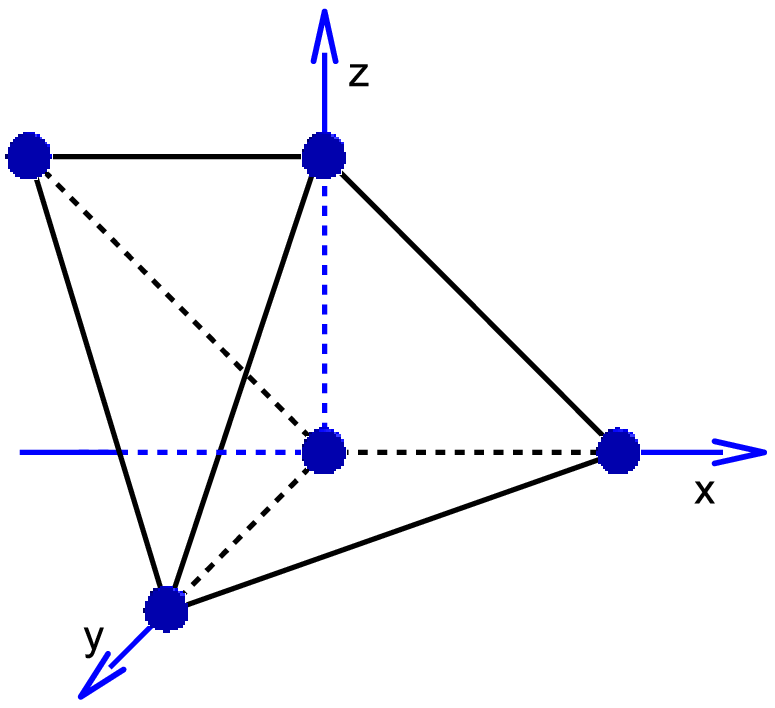}
\label{toric_modified_CC2Z2}
}
\caption{Quiver of the modified $\mathbb{C}\times\mathbb{C}^2/\mathbb{Z}_2$ model (CS levels $(1,-1)$), and moduli space.}
\end{center}
\end{figure}

In this section we add flavors to the so-called modified $\bC\times\bC^2/\bZ_2$ theory of \cite{Hanany:2008cd}. The quiver of the unflavored theory, Fig. \ref{quiver_CC2Z2}, is the one for D-branes at a $\bC\times\bC^2/\bZ_2$ singularity; we choose the height numbers $n_{ij}$ equal to $1$ for the bifundamental $X_{12}^1$ and $0$ otherwise, so that the two gauge groups have CS levels $(1,-1)$. The superpotential is
\be\label{superpot_A1}
W=\Phi_1(X_{12}^1 X_{21}^2 - X_{12}^2 X_{21}^1)-\Phi_2(X_{21}^2 X_{12}^1 - X_{21}^1 X_{12}^2)\;.
\ee
From the permanent of the Kasteleyn matrix,
\be
\mathrm{Perm}\, K = X_{21}^1 X_{21}^2 + X_{12}^2 X_{21}^2 \,x + X_{21}^1 X_{12}^1\, x^{-1}z + X_{12}^1 X_{12}^2 \,z + \Phi_1\Phi_2\, y\;,
\ee
we see that the perfect matchings are
\be
\begin{array}{cccc}
\quad a_0&=\{X_{21}^1, X_{21}^2\} = (0,0,0), & \qquad d_1&=\{X_{12}^1, X_{12}^2 \}=(0,0,1), \\
\quad b_0&=\{X_{12}^2, X_{21}^2\}= (1,0,0),  & \qquad e_0&=\{\Phi_1,\Phi_2 \}= (0,1,0), \\
\quad c_1&=\{X_{21}^1, X_{12}^1\}=(-1,0,1). &&
\end{array}
\ee
The 3d toric diagram, Fig. \ref{toric_modified_CC2Z2}, is the one of $\bC\times\cC$.
The F-term equations imply $X_{12}^1 X_{21}^2 = X_{12}^2 X_{21}^1$ and $\Phi_1=\Phi_2$ along the mesonic branch. They are solved by
\be\nn
X_{12}^1 = c_1 \, d_1 \;, \quad X_{12}^2 = b_0 \, d_1 \;, \quad X_{21}^1 = a_0 \, c_1 \;,\quad X_{21}^2 = a_0 \, b_0 \;,\quad \Phi_1=\Phi_2 = e_0 \;.
\ee
The face in the 3d toric diagram whose vertices are $\{a_0,c_1,d_1,b_0\}$ is vertical, therefore additional objects may appear in the type IIA background. Nevertheless, encouraged by the results of \cite{Hanany:2008cd} where the geometric moduli space was successfully matched with $\mathbb{C} \times \mathcal{C}$, we will trust the duality and add flavors to this model.

We will study three illustrative examples where two flavor pairs are added to this theory.


\subsubsection{$U(2)$ flavor group coupled to $X_{12}^1$: levels $(0,0)$}\label{doubleflavored_level0}

\begin{figure}[t]
\begin{center}
\subfigure[\small CS levels $(0,0)$. ]{
\includegraphics[width=3.6cm]{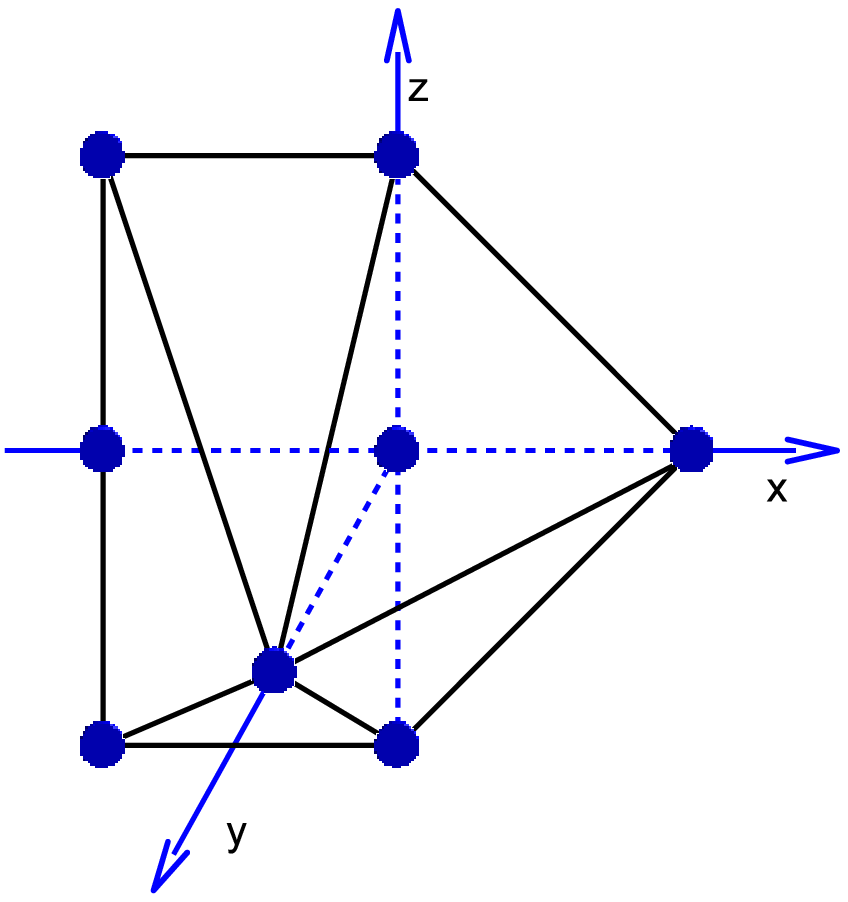}
\label{toric_modified_CC2Z2_2flav_bif_level0}
}
\subfigure[\small One flavored bifundamental.]{
\includegraphics[width=5.8cm]{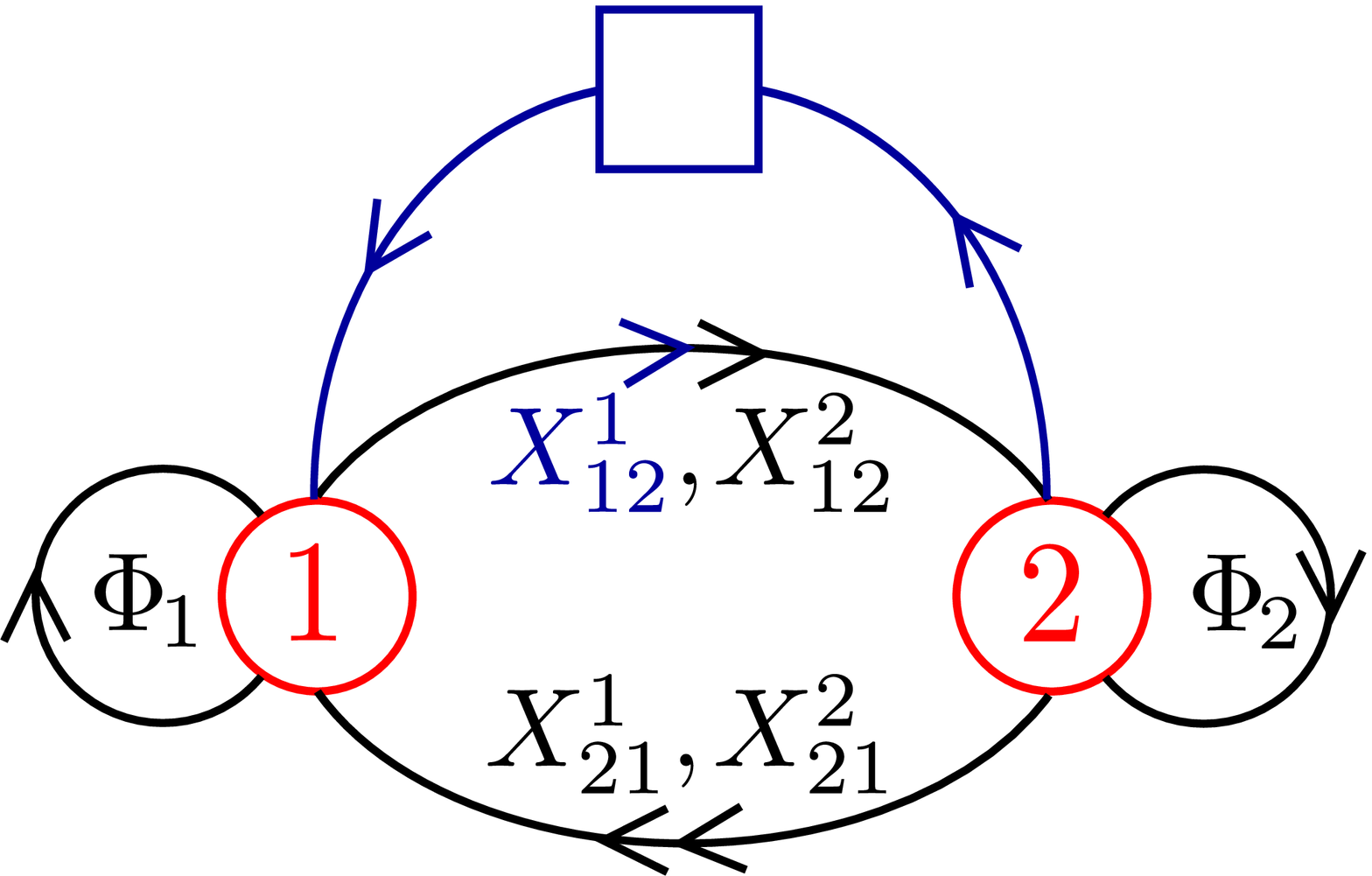}
\label{quiver_CC2Z2_flav_bif}
}
\subfigure[\small CS levels $(1,-1)$.]{
\includegraphics[width=3.6cm]{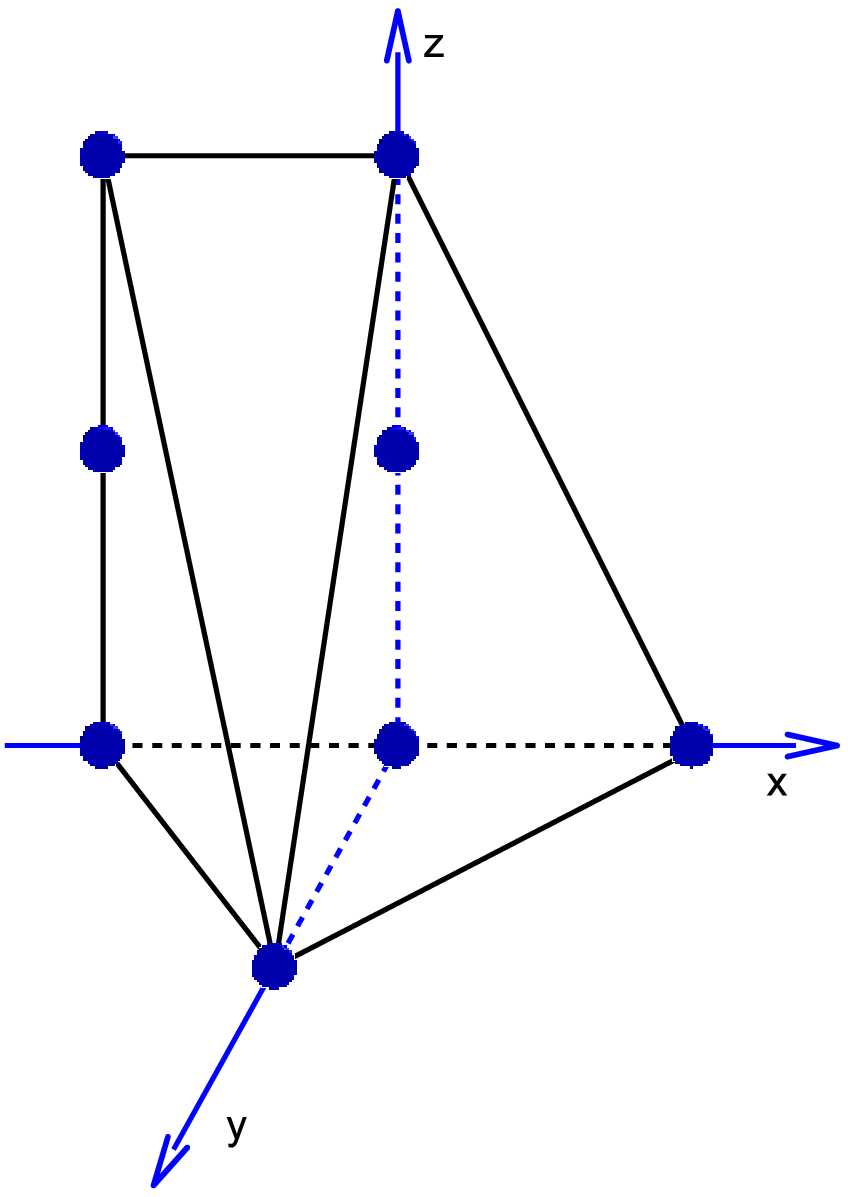}
\label{toric_modified_CC2Z2_2flav_bif_level1}
}
\caption{\small The $\bbC\times\bbC^2/\bbZ_2$ quiver with a $U(2)$ flavor group coupled to $X_{12}^{1}$. Those two toric quivers are obtained by fixing the CS levels as indicated.}
\end{center}
\end{figure}

We study two cases where we couple a $U(2)$ flavor group to $X_{12}^1$, as in Fig. \ref{quiver_CC2Z2_flav_bif}.
Consider first the case where the CS levels vanish. The bifundamental fields and monopole operators of the quiver theory have gauge charges
\be\label{charges_2flav_bif_level0}
\begin{tabular}{c|c c c c c}
              & $X_{12}$ & $X_{21}$ & $\Phi$ &  ${T}$ & $\tilde T$ \\ \hline
$U(1)_{0}$  &      $1$   &   $-1$   &   $0$  &  $1$ &   $1$ \\
$U(1)_{0}$  &     $-1$   &    $1$   &   $0$  & $-1$ &  $-1$
\end{tabular}
\ee
The gauge invariant operators in the geometric branch are $\Phi$, $X_{12}X_{21}$, $T X_{21}$, $\tilde{T} X_{21}$.

In the A-theory, this flavoring corresponds to replacing the edge $X_{12}^1$ with $n_{X_{12}^1}=1$ in the original brane tiling with a triple-bond with $n = -1,0,1$.
It amounts to considering a 3d toric diagram with the points $\{a_0,b_0,c_{-1},c_0,c_1,d_{-1},d_0,d_1,e_0 \}$ as in Fig. \ref{toric_modified_CC2Z2_2flav_bif_level0}.
We solve for the F-term relation and the quantum relation $T\tilde{T}= X^1_{12}$ by
\bea
X_{12}^1 &= c_{-1}\,c_0\,c_1\,d_{-1}\,d_0\,d_1 \;, \qquad & X_{12}^2 &= b_0\,d_{-1}\,d_0\,d_1 \;, \qquad & \\
X_{21}^1 &= a_0\,c_{-1}\,c_0\,c_1 \;, & X_{21}^2 &= a_0\,b_0 \;, & \Phi_1 &= \Phi_2=e_0 \;, \\
\tilde{T} &= c_0\,c_1^2\,d_0\,d_1^2\;, & T &= c_{-1}^2\,c_0\,d_{-1}^2\,d_0 \;.
\eea
The charges of the homogeneous coordinates of the four-fold and of the quiver theory fields under the associated $U(1)^5$ GLSM are
\be\nn
\label{charges_2flav_bif_level0_GLSM}
\begin{tabular}{c|c c c c c c c c c|c c c c c}
            &$a_0$&$b_0$&$c_1$&$d_1$&$e_0$&$c_0$&$d_0$&$c_{-1}$&$d_{-1}$& $X_{12}$ & $X_{21}$ & $\Phi$ &  ${T}$ & $\tilde T$ \\ \hline
$U(1)_{B1}$ & $1$ & $-1$& $-1$& $1$ & $0$ & $0$ & $0$ & $0$ & $0$ &    $0$   &    $0$   &   $0$  &  $0$ &   $0$ \\
$U(1)_{B2}$ & $1$ &  $0$& $1$ & $-1$& $0$ & $-1$& $0$ & $0$ & $0$ &   $-1$   &    $1$   &   $0$  & $-1$ &  $-1$ \\
$U(1)_{B3}$ & $1$ &  $0$&  $0$& $0$ & $0$ & $0$ &$-1$ & $0$ & $0$ &   $-1$   &    $1$   &   $0$  & $-1$ &  $-1$ \\
$U(1)_{B4}$ & $1$ &  $0$&  $0$& $0$ & $0$ &$-1$ & $0$ & $1$ &$-1$ &   $-1$   &    $1$   &   $0$  & $-1$ &  $-1$ \\
$U(1)_{B5}$ &$-2$ &  $0$&  $0$& $1$ & $0$ & $0$ & $0$ & $0$ & $1$ &    $2$   &   $-2$   &   $0$  &  $2$ &   $2$
\end{tabular}
\ee
matching the gauge charges \eqref{charges_2flav_bif_level0}.
The affine coordinates of the fourfold match the gauge invariant operators of the flavored quiver theory:
\be
\begin{split}
&x_1=e_0=\Phi_1=\Phi_2\;,\qquad\qquad\qquad\qquad\quad x_2=a_0\,b_0^2\,d_{-1}\,d_0\,d_1=X_{12}^2 X_{21}^2\;, \\
&x_3=a_0\,b_0\,c_0\,c_1^2\,d_0\,d_1^2= \tilde{T} X_{21}^2\;,\qquad\qquad\quad x_4=a_0\,c_{-1}\,c_0^2\,c_1^3\,d_0\,d_1^2= \tilde{T} X_{21}^1\;,\\
&x_5=a_0\,b_0\,c_{-1}\,c_0\,c_1\,d_{-1}\,d_0\,d_1= X_{12}^1 X_{21}^2=X_{12}^2 X_{21}^1\;, \\
&x_6=a_0\,c_{-1}^2\,c_0^2\,c_1^2\,d_{-1}\,d_0\,d_1=X_{12}^1 X_{21}^1\;,\\
&x_7=a_0\,b_0\,c_{-1}^2\,c_0\,d_{-1}^2\,d_0= T X_{21}^2\;, \qquad\quad\quad
x_8=a_0\,c_{-1}^3\,c_0^2\,c_1\,d_{-1}^2\,d_0= T X_{21}^1\;.
\end{split}
\ee


\subsubsection{$U(2)$ flavor group coupled to $X_{12}^1$: levels $(1,-1)$}
\label{doubleflavored_level1}

Consider now the case of CS levels $(1,-1)$. The gauge charges are:
\be\label{charges_2flav_bif_level1}
\begin{tabular}{c|c c c c c}
              & $X_{12}$ & $X_{21}$ & $\Phi$ &  $T$ & $\tilde{T}$ \\ \hline
$U(1)_{1}$  &      $1$   &   $-1$   &   $0$  &  2& $0$  \\
$U(1)_{-1}$ &     $-1$   &    $1$   &   $0$  &  2& $0$
\end{tabular}
\ee
The gauge invariant operators are $\Phi$, $X_{12}X_{21}$, $T (X_{21})^2$, $\tilde{T}$.

In the A-theory, this flavoring corresponds to replacing the edge $X_{12}^1$ with $n_{X_{12}^1}=1$ in the original brane tiling by a triple-bond with $n =0,1,2$. The GLSM field appearing in the 3d toric diagram, Fig. \ref{toric_modified_CC2Z2_2flav_bif_level1}, are $\{a_0,b_0,c_0,c_1,c_2,d_0,d_1,d_2,e_0 \}$.
We solve for the geometric moduli space by setting
\bea
X_{12}^1 &=c_0\,c_1\,c_2\,d_0\,d_1\,d_2 \;,\qquad & X_{12}^2 &=b_0\,d_0\,d_1\,d_2\;, \qquad &\\
X_{21}^1 &=a_0\,c_0\,c_1\,c_2\;, & X_{21}^2 &=a_0\,b_0 \;, & \Phi_1 &=\Phi_2=e_0 \;, \\
\tilde{T} &=c_1\,c_2^2\,d_1\,d_2^2 \;, & T &=c_0^2\,c_1\,d_0^2\,d_1 \;.
\eea
The charges of the homogeneous coordinates of the fourfold and of the quiver theory fields under the $U(1)^5$ GLSM are
\be\nn
\label{charges_2flav_bif_level1_GLSM}
\begin{tabular}{c|c c c c c c c c c|c c c c c}
            &$a_0$&$b_0$&$c_1$&$d_1$&$e_0$&$c_0$&$d_0$&$c_2$&$d_2$& $X_{12}$ & $X_{21}$ & $\Phi$ &  $T$ & $\tilde{T}$ \\ \hline
$U(1)_{B1}$ & $1$ & $-1$& $-1$& $1$ & $0$ & $0$ & $0$ & $0$ & $0$ &    $0$   &    $0$   &   $0$  &  $0$ &   $0$ \\
$U(1)_{B2}$ & $1$ &  $0$& $1$ & $-1$& $0$ & $-1$& $0$ & $0$ & $0$ &   $-1$   &    $1$   &   $0$  & $-2$ & $0$ \\
$U(1)_{B3}$ & $1$ &  $0$&  $0$& $0$ & $0$ & $0$ &$-1$ & $0$ & $0$ &   $-1$   &    $1$   &   $0$  & $-2$ &  $0$  \\
$U(1)_{B4}$ & $1$ &  $0$&  $0$&$-2$ & $0$ & $0$ & $0$ & $0$ & $1$ &   $-1$   &    $1$   &   $0$  & $-2$ &  $0$  \\
$U(1)_{B5}$ & $0$ &  $0$& $-2$& $0$ & $0$ & $1$ & $0$ & $1$ & $0$ &    $0$   &    $0$   &   $0$  &  $0$ &   $0$
\end{tabular}
\ee
matching the gauge charges \eqref{charges_2flav_bif_level1}.
The affine coordinates of the four-fold match the holomorphic gauge invariants  of the flavored quiver theory:
\be\nn
\begin{split}
&x_1=a_0\,b_0\,c_0\,c_1\,c_2\,d_0\,d_1\,d_2= X_{12}^1 X_{21}^2=X_{12}^2 X_{21}^1\;,\quad
x_2=e_0=\Phi_1=\Phi_2\;,\\
&x_3=a_0^2\,b_0^2\,c_0^2\,c_1\,d_0^2\,d_1= T (X_{21}^2)^2\;,\qquad\qquad\quad\quad\;\;\;\; \,x_4=a_0\,c_0^2\,c_1^2\,c_2^2\,d_0\,d_1\,d_2= X_{12}^1 X_{21}^1 \;,\\
&x_5=a_0^2\,b_0\,c_0^3\,c_1^2\,c_2\,d_0^2\,d_1= T X_{21}^1 X_{21}^2\;,\qquad\qquad\quad \;\;\; x_6=a_0\,b_0^2\,d_0\,d_1\,d_2= X_{12}^2 X_{21}^2 \;,\\
&x_7=a_0^2\,c_0^4\,c_1^3\,c_2^2\,d_0^2\,d_1= T (X_{21}^1)^2\;,\qquad\qquad\qquad\quad\; x_8=c_1\,c_2^2\,d_1\,d_2^2=\tilde{T}\;.
\end{split}
\ee


\subsubsection{$U(1)^2$ flavor groups coupled to $X_{12}^1$ and $X_{21}^1$: levels $(1,-1)$}

\begin{figure}[t]
\begin{center}
\subfigure[\small The quiver.]{
\includegraphics[width=4.6cm]{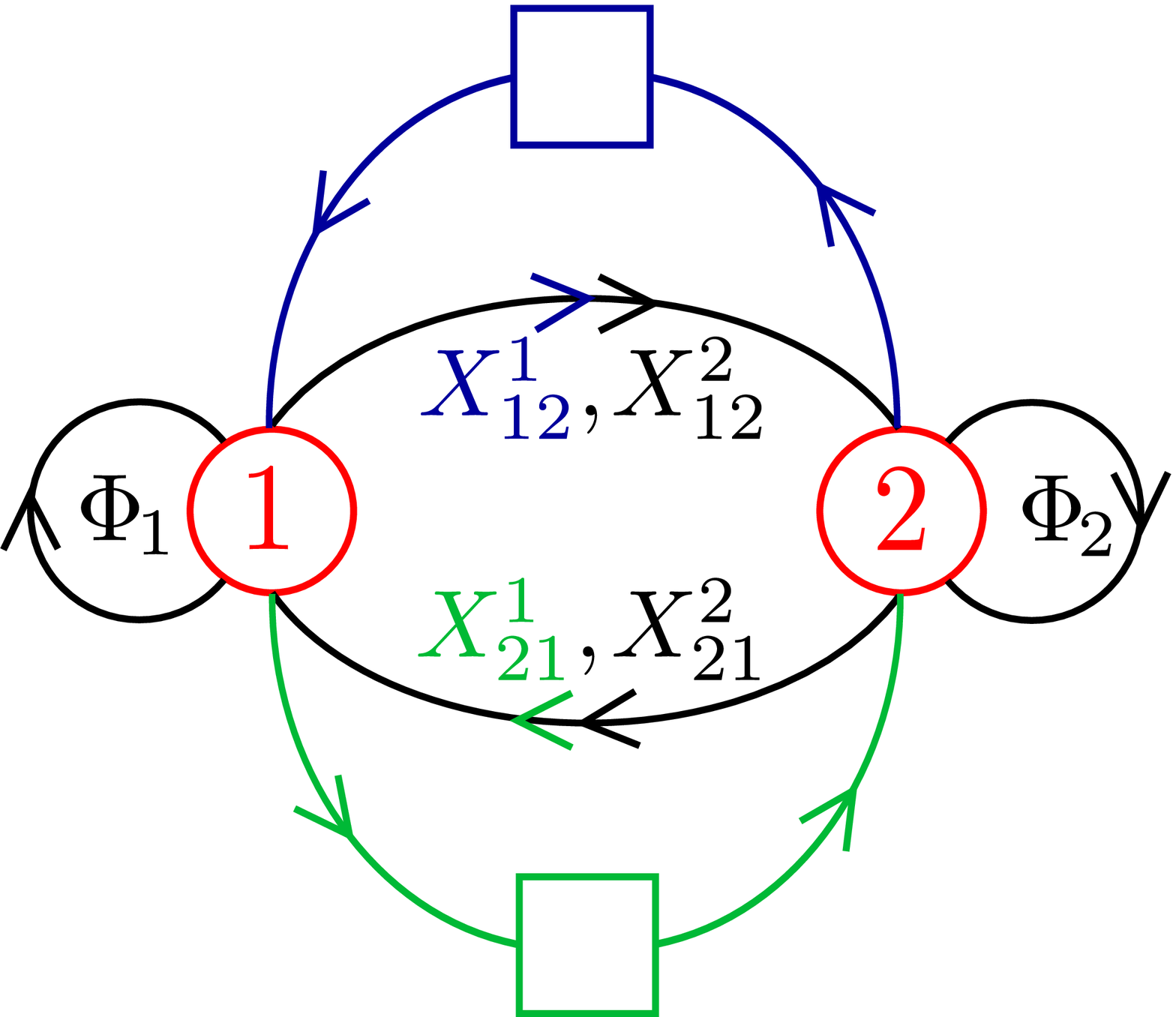}
\label{quiver_CC2Z2_flav_2bif}
}\qquad
\subfigure[\small Toric diagram when the CS levels are $(1,-1)$.]{
\qquad\qquad\qquad\includegraphics[width=3.6cm]{ToricDiagCConi_for_CCZ2_01}\qquad\qquad\qquad
\label{toric_modified_CC2Z2_flav_2bif_level1}
}
\caption{Modified $\mathbb{C}\times\mathbb{C}^2/\mathbb{Z}_2$ model with two flavored bifundamentals, and dual geometry.}
\end{center}
\end{figure}

Let us study a case where we couple a $U(1)$ flavor group to $X_{12}^1$ and a $U(1)$ flavor group to $X_{21}^1$, as in Fig. \ref{quiver_CC2Z2_flav_2bif}. The quantum relation reads
$ {T} \tilde T=X_{12}^1 X_{12}^2$.
We consider the case with CS levels
$(1,-1)$: bifundamentals and monopole operators charges are
\be\label{charges_flav_2bif_level1}
\begin{tabular}{c|c c c c c}
              & $X_{12}$ & $X_{21}$ & $\Phi$ &  ${T}$ & $\tilde T$ \\ \hline
$U(1)_{1}$  &      $1$   &   $-1$   &   $0$  & $1$ &   $-1$ \\
$U(1)_{-1}$ &     $-1$   &    $1$   &   $0$  &  $-1$ &  $1$
\end{tabular}
\ee
The gauge invariant operators are $\Phi$, $X_{12}X_{21}$, $T X_{21}$, $\tilde{T} X_{12}$.

In the A-theory, this flavoring corresponds to replacing the edge $X_{12}^1$ with $n_{X_{12}^1}=1$ in the brane tiling by a double-bond with $n =0,1$, and the edge $X_{21}^1$ with $n_{X_{21}^1}=0$ by another double-bond, with $n =-1,0$. All the other $n_{ij}$ vanish.
This gives a 3d toric diagram with points $\{a_{-1},a_0,b_0,c_{-1},c_0,c_0',c_1,d_0,d_1,e_0 \}$, Fig. \ref{toric_modified_CC2Z2_flav_2bif_level1}.
This is not a minimal presentation of the toric diagram. In particular, unlike for the other multiplicities, the distinction between $c_0$ and $c_0'$ is not needed to express the bifundamentals and monopole operators in terms of GLSM fields solving the F-term equations. It is possible to replace the two of them by a single field $\tilde{c}_0$ (setting $c_0\,c_0'= \tilde{c}_0$ in the formul\ae{} below), getting rid of a $U(1)$ in the GLSM. We will do that in the following. Keeping instead all the perfect matching fields of the A-theory may be useful in the study of partial resolutions dual to real mass terms.

We solve for the geometric moduli space by setting
\bea
X_{12}^1 &=c_{-1} \, \tilde{c}_0\, c_1\, d_0\, d_1 \;, \qquad & X_{12}^2 &=b_0\, d_0\, d_1\;, \qquad &\\
X_{21}^1 &=a_{-1}\, a_0\, c_{-1}\, \tilde{c}_0 \, c_1 \;, & X_{21}^2 &=a_{-1}\, a_0\, b_0\;, & \Phi_1 &=\Phi_2=e_0 \;,\\
\tilde{T} &=a_0\, \tilde{c}_0\, c_1^2\, d_1\;, & T &=a_{-1}\, c_{-1}^2\, \tilde{c}_0\, d_0 \;. \qquad &
\eea
The charges of the homogeneous coordinates of the four-fold and of the quiver theory fields under the resulting $U(1)^5$ GLSM are
\be\nn
\label{charges_flav_2bif_level1_GLSM}
\begin{tabular}{c|c c c c c c c c c|c c c c c}
            &$a_0$&$b_0$&$\tilde{c}_0$&$d_0$&$e_0$&$a_{-1}$&$c_1$&$c_{-1}$&$d_1$& $X_{12}$ & $X_{21}$ & $\Phi$ & $T$ & $\tilde{T}$ \\ \hline
$U(1)_{B1}$ & $1$ & $0$         & $0$ &$-1$ & $0$ &  $0$ &  $0$ &   $0$ &   $0$  & $-1$   &    $1$   &   $0$  &  $-1$ &  $1$ \\
$U(1)_{B2}$ &$-2$ & $1$         & $1$ & $0$ & $0$ &  $0$ &  $0$ &   $0$ &   $0$  &  $1$   &   $-1$   &   $0$  & $1$ &   $-1$ \\
$U(1)_{B3}$ & $1$ & $0$         &$-1$ & $0$ & $0$ &  $0$ &  $1$ &   $0$ &  $-1$  & $-1$   &    $1$   &   $0$  &  $-1$ &  $1$ \\
$U(1)_{B4}$ & $1$ & $0$         &$-1$ & $0$ & $0$ & $-1$ &  $0$ &   $1$ &   $0$  &  $0$   &    $0$   &   $0$  &  $0$ &   $0$ \\
$U(1)_{B5}$ & $0$ & $0$         & $2$ & $0$ & $0$ &  $0$ & $-1$ &  $-1$ &   $0$  &  $0$   &    $0$   &   $0$  &  $0$ &   $0$
\end{tabular}
\ee
matching the gauge charges \eqref{charges_flav_2bif_level1}.
The affine coordinates of the four-fold match the holomorphic gauge invariants of the flavored quiver theory:
\be
\begin{split}
&x_1=e_0=\Phi_1=\Phi_2\;,\qquad\qquad\qquad\qquad\qquad\; x_2=a_{-1}\, a_0\, b_0^2\, d_0\, d_1=  X_{12}^2 X_{21}^2\;,\\
&x_3=a_{-1}^2\, a_0\, c_{-1}^3\, c_0^2\, c_0'^2\, c_1\, d_0= T X_{21}^1\;,\qquad\qquad  x_4=a_0\, c_{-1}\, c_0^2\, c_0'^2\, c_1^3\, d_0\, d_1^2= \tilde{T} X_{12}^1\;,\\
&x_5=a_{-1}^2\, a_0\, b_0\,  c_{-1}^2\, c_0\, c_0'\, d_0 = T X_{21}^2\;,\qquad\qquad\, x_6=a_0\, b_0\, c_0\, c_0'\, c_1^2\, d_0\, d_1^2 = \tilde{T} X_{12}^2\;,\\
&x_7=a_{-1}\, a_0\, c_{-1}^2\, c_0^2\, c_0'^2\, c_1^2\, d_0\, d_1= X_{12}^1 X_{21}^1 = \tilde{T} T\;,\\
&x_8=a_{-1}\, a_0\, b_0\, c_{-1}\, c_0\, c_0'\, c_1\, d_0\, d_1= X_{12}^1 X_{21}^2=X_{12}^2 X_{21}^1\;.
\end{split}
\ee
The toric diagram of the CY$_4$ is the same as in the double-flavored $X_{12}^1$ model with CS levels $(0,0)$ studied in subsection \ref{doubleflavored_level0}: thus the geometric branches of the moduli spaces of these two theories are the same, although the manifest flavor symmetries of the quivers are different. Presumably, the M-theory backgrounds will differ in monodromies of the 3-form potential $C_3$.

The three double flavored models analyzed here for the modified $\bbC\times \bbC^2/\bbZ_2$ model lead to D6-branes along the same toric divisor inside the CY$_3$. However there are different gauge connections on the flavor branes, everywhere flat but at the tip, and gauge fluxes on the 2-cycles at the singularity. In spite of the D6-branes being identically embedded at the level of the complex structure, the type IIA/M-theory backgrounds differ, because the different gauge fluxes at the singularity generate RR fluxes that backreact onto the metric.

\subsection{Flavoring the $dP_0$ quiver}

\begin{figure}[t]
\begin{center}
\subfigure[\small The quiver.]{
\includegraphics[width=3.8cm]{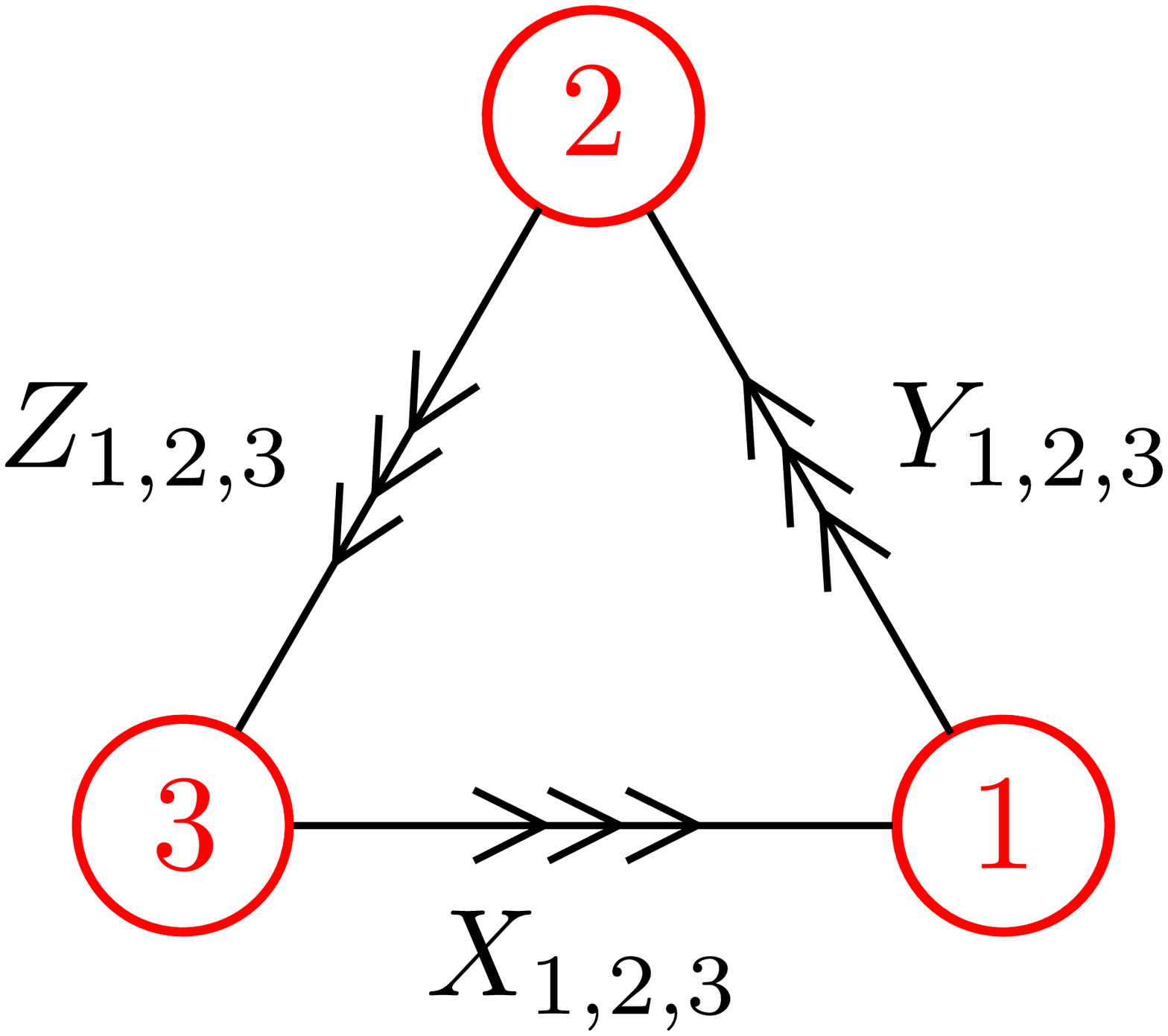}
\label{quiver: dP0}
}\qquad\qquad
\subfigure[\small The 2d toric diagram.]{
\qquad\includegraphics[width=3.4cm]{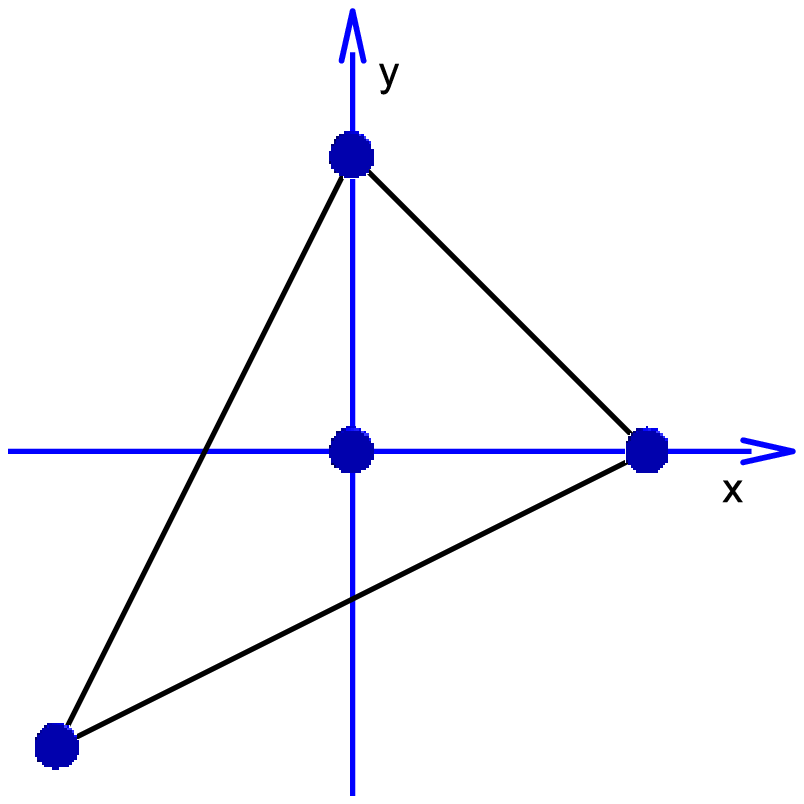}\qquad
\label{toric diag: dP0}
}
\caption{The $dP_0$ quiver and the 2d toric diagram.}
\end{center}
\end{figure}

The $dP_0$ quiver, Fig. \ref{quiver: dP0}, is the quiver for D-branes at a $\bbC^3/\bbZ_3$ singularity. It has three nodes and nine bifundamental fields, $X_i$, $Y_i$, $Z_i$, $i=1,2,3$. We choose to parametrize the CS levels by $(k_1,k_2,k_3)=(q-p,q,p-2q)$.
The charges under the $U(1)^3$ gauge group are
\be
\begin{tabular}{l|ccccc}
 & $X_i$ & $Y_i$ & $Z_i$ & $T$ & $\tilde{T}$ \\
\hline
 $U(1)_{q-p}$ & $-1$ & $1$ & $0$ & $q-p$ & $-q+p$ \\
 $U(1)_{q}$ & $0$ & $-1$ & $1$ & $q$ & $-q$ \\
 $U(1)_{p-2q}$ & $1$ &  $0$ & $-1$ & $p-2q$ & $-p+2q$
\end{tabular}
\ee
The superpotential is $W= X_iY_jZ_k \epsilon^{ijk}$, so the indices $ijk$ are fully symmetric in the chiral ring.
From the permanent of the Kasteleyn matrix,
\be\nonumber
\mathrm{Perm}\, K = X_1Y_1Z_1xz^{p}  + X_2Y_2Z_2x^{-1}y^{-1}  +X_3Y_3Z_3y   +X_1X_2X_3z^{p-q} +Y_1Y_2Y_3z^{q}   + Z_1Z_2Z_3,
\ee
we read off the perfect matchings and the coordinates of the points in the toric diagram:
\be
\begin{array}{cc}
\quad a_p=\{X_1,Y_1,Z_1\}  \,= \,  (1,0,p) \;,&
\quad e_{p-q}=\{X_1,X_2,X_3\} \,= \,  (0,0,p-q)\; ,\\
\quad b_0=\{X_2,Y_2,Z_2\}  \,= \,  (0,1,0) \;,&
\quad f_q=\{Y_1,Y_2,Y_3\} \,= \,  (0,0,q)\;, \\
\quad c_0=\{X_3,Y_3,Z_3\} \,= \,  (-1,-1,0) \;,&
\quad g_0=\{Z_1,Z_2,Z_3\} \,= \,  (0,0,0)\;.
\end{array}
\ee
The choice of $SL(4,\bbZ)$ frame is such that for $p,q >0$ we have the geometry $Y^{p,q}(\bC\bP^2)$ as presented in \cite{Martelli:2008rt}.
In particular, this family includes the geometry $M^{3,2}=Y^{3,2}(\bC\bP^2)$.
The perfect matching variables allow to solve the F-term relations as
\be\label{Solution Fterm for dP0 dWeq0}
\begin{array}{ccc}
\qquad X_1=a_pe_{p-q}\;,&
\qquad Y_1=a_pf_q\;,&
\qquad Z_1=a_pg_0\;,\\
\qquad X_2=b_0e_{p-q}\;,&
\qquad Y_2=b_0f_q\;,&
\qquad Z_2=b_0g_0\;,\\
\qquad X_3=c_0e_{p-q}\;,&
\qquad Y_3=c_0f_q\;,&
\qquad Z_3=c_0g_0\;,
\end{array}
\ee
and the redundancies in this parametrization correspond to a non-minimal GLSM for the toric geometry. We couple chiral flavors to bifundamental fields in the $dP_0$ quiver, and consider a few simple but interesting examples, flavoring the theory with vanishing CS levels $p=q=0$. The 2d diagram is shown in Fig. \ref{toric diag: dP0}.

\subsubsection{$U(1)$ flavor group coupled to $X_1$}

Let us couple one flavor to the field $X_1$ in the quiver with vanishing CS levels, which induces CS levels $(-\frac{1}{2},0,\frac{1}{2})$. The quantum relation is $T\tilde{T}= X_1$, and the gauge charges of the fields and monopole operators are:
\be
\label{charges of fields in dP0 with one flavor}
\begin{tabular}{l|ccccc}
 & $X_i$ & $Y_i$ & $Z_i$ & ${T}$& $\tilde T$  \\
\hline
 $U(1)_{-\frac{1}{2}}$ & $-1$ & $1$ & $0$  & $-1$ & $0$ \\
 $U(1)_{0}$ & $0$ & $-1$ & $1$  & $0$ & $0$\\
 $U(1)_{\frac{1}{2}}$ & $1$ &  $0$ & $-1$ & 1& $0$
\end{tabular}
\ee
To find the geometric branch of the moduli space, we solve both the F-terms and the quantum relation by adding
two new variables $a_1$ and $e_1$ to the solution (\ref{Solution Fterm for dP0 dWeq0}):
\be
\begin{array}{cccc}
\qquad X_1=a_0a_1e_0e_1\;,&
\qquad Y_1=a_0a_1f_0\;,&
\qquad Z_1=a_0a_1g_0\;,&
\qquad T = a_0e_0 \;,\\
\qquad X_2=b_0e_0e_1\;,&
\qquad Y_2=b_0f_0\;,&
\qquad Z_2=b_0g_0\;,&
\qquad \tilde T = a_1e_1 \;,\\
\qquad X_3=c_0e_0e_1\;,&
\qquad Y_3=c_0f_0\;,&
\qquad Z_3=c_0g_0\;.
\end{array}
\ee
The associated GLSM is
\be
\begin{tabular}{l|c c c c c c cc}
       & $a_0$ & $b_0$ & $c_0$ & $e_0$ & $f_0$ & $g_0$ & $a_1$ & $e_1$ \\ \hline
$U(1)_{B_1}$ &   $1$  & $1$  & $1$   &  $-2$ &  $0$ &  $-1$  &  $0$  &  $0$  \\
$U(1)_{B_2}$ &   $1$  & $1$  & $1$   &  $-1$ &  $-2$ &  $0$  &  $0$  &  $0$  \\
$U(1)_{B_3}$ &   $1$  & $1$  & $1$   &  $0$ &  $-1$ &  $-2$  &  $0$  &  $0$  \\
$U(1)_{B_4}$&   $1$  & $0$  & $0$   &  $-1$ &  $0$ &  $0$  &  $-1$  &  $1$  \\\hline
$U(1)_{M}$ & $1$  & $0$  & $0$   &  $0$ &  $0$ &  $0$  &  $-1$ &  $0$
\end{tabular}
\ee
The three first rows correspond to the gauge group $U(1)^3$ of the quiver. This GLSM is a non-minimal presentation of the toric geometry of Fig. \ref{quiver: dP0 1 flav}, corresponding to adding two points $a_1$ and $e_1$ as suggested by the A-theory. We have also specified the $Q_M$ charges. Gauging $U(1)_M$ leads to the CY$_3$ $\bbC^3/\bbZ_3$, and the locus of fixed points projects to the non-compact divisor $\{a_0=0\}$.
Let us check that the gauge invariant operators match the affine coordinates of the toric variety. There are 10 operators of the form $XYZ$, 6 of the form $TYZ$, and $\tilde{T}$, but the quantum relation makes $X_1YZ= \tilde T T YZ$ redundant, so that we are left with 11 generators of the chiral ring. We can check that they match all the gauge invariant functions of the GLSM:
\be\nn
\label{gauge inv ops dP0 case 1}
\begin{array}{ccc}
        x_1=TY_1Z_1=a_0^3e_0f_0g_0a_1^2\;,&
\qquad x_6=TY_2Z_2=a_0b_0^2e_0f_0g_0\;, & \quad x_{11}=\tilde{T} =a_1e_1\;.
\\
        x_2=X_2Y_2Z_2 =b_0^3e_0f_0g_0e_1\;,&
\qquad x_7=X_2Y_2Z_3  =b_0^2c_0e_0f_0g_0e_1\;,\\
        x_3=X_3Y_3Z_3= c_0^3e_0f_0g_0e_1\;,&
\qquad x_8=TY_3Z_3 = a_0c_0^2e_0f_0g_0\;,\\
        x_4=TY_1Z_2 =a_0^2b_0e_0f_0g_0a_1\;,&
\qquad x_9=X_2Y_3Z_3 =b_0c_0^2e_0f_0g_0e_1\;,\\
        x_5=TY_1Z_3 =a_0^2c_0e_0f_0g_0a_1\;,&
\qquad x_{10}=TY_2Z_3 = a_0b_0c_0e_0f_0g_0\;,
\end{array}
\ee

\subsubsection{$U(1)^2$ flavor groups coupled to $X_1$ and $Y_1$}

\begin{figure}[t]
\begin{center}
\subfigure[\small]{
\includegraphics[width=4cm]{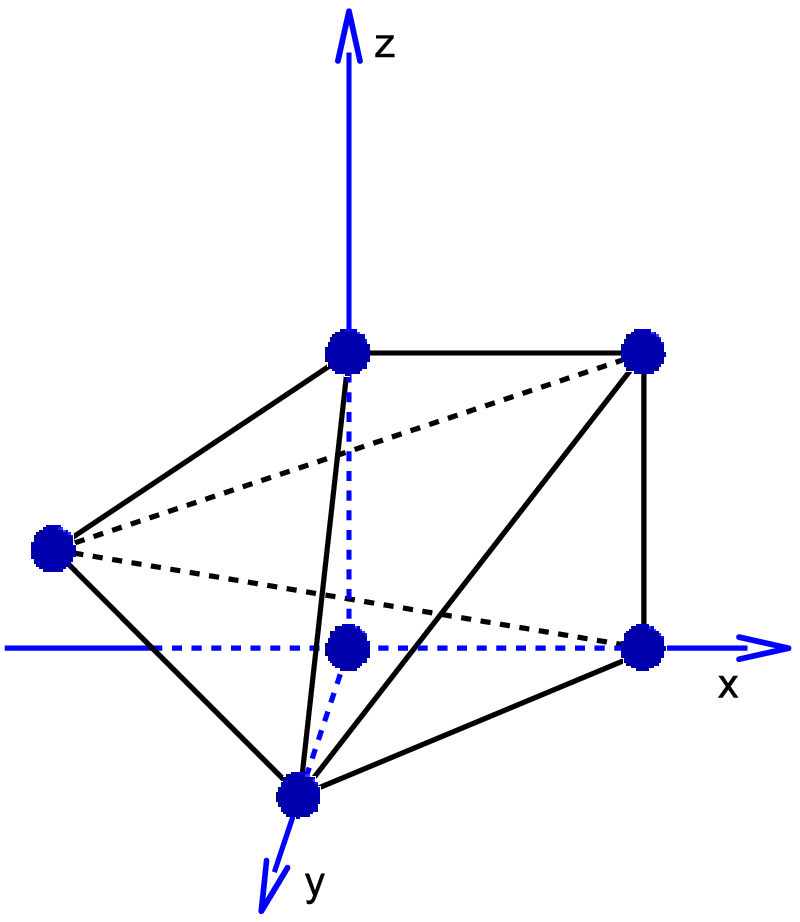}
\label{quiver: dP0 1 flav}
}\qquad\quad
\subfigure[\small]{
\includegraphics[width=4cm]{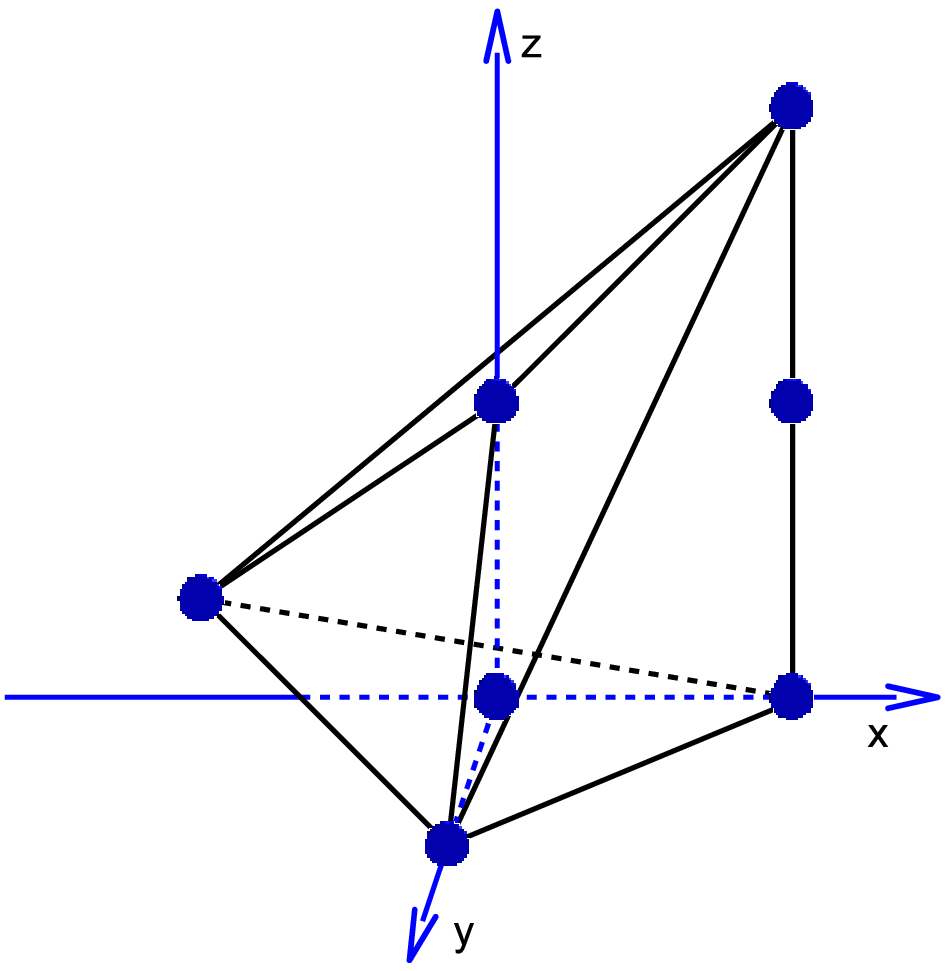}
\label{toric diag: dP0 2 flav}
}
\caption{Toric diagram obtained by flavoring one or two fields in the $dP_0$ quiver.}
\end{center}
\end{figure}

Consider flavoring $X_1=ae$ and $Y_1=ef$. There are two possible CS levels, but let us consider the case $(0,-\frac{1}{2},\frac{1}{2})$ corresponding to adding four perfect matching variables $a_1$, $a_2$, $e_1$, $f_1$. The toric diagram is in Fig. \ref{toric diag: dP0 2 flav}.
The field theory gauge charges are
\be
\begin{tabular}{l|ccccc}
 & $X_i$ & $Y_i$ & $Z_i$ &${T}$& $\tilde T$  \\
\hline
 $U(1)_{0}$ & $-1$ & $1$ & $0$  & $0$ & $0$ \\
 $U(1)_{-\frac{1}{2}}$ & $0$ & $-1$ & $1$  & $-1$ & 0\\
 $U(1)_{\frac{1}{2}}$ & $1$ &  $0$ & $-1$ &  1 & 0
\end{tabular}
\ee
The quantum relation is $T\tilde{T}=X_1Y_1$. There are again 11 gauge invariant operators: $X_iY_jZ_k$, $TZ_i$ and $\tilde{T}$, but the three operators $X_1Y_1Z_i$ are redundant due to the quantum relation. We can solve the moduli space equations by
\be\nn
\label{field in term of pms dP0 02}
\begin{array}{cccc}
X_1=a_0a_1a_2e_0e_1\;,&
\qquad Y_1=a_0a_1a_2f_0f_1\;,&
\qquad Z_1=a_0a_1a_2g_0\;,&
\qquad T=a_0^2a_1 e_0f_0\;, \\
X_2=b_0e_0e_1\;,&
\qquad Y_2=b_0f_0f_1 \;,&
\qquad Z_2=b_0g_0\;,&
\qquad \tilde{T}= a_1a_2^2e_1f_1 \;. \\
X_3=c_0e_0e_1\;,&
\qquad Y_3=c_0f_0f_1\;,&
\qquad Z_3=c_0g_0\;,
\end{array}
\ee
and the associated GLSM is
\be
\begin{tabular}{l|c c c c c c cc c c}
            & $a_0$ & $b_0$ & $c_0$ & $e_0$ & $f_0$ & $g_0$ & $a_1$ & $e_1$ & $a_2$& $f_1$ \\ \hline
$U(1)_{B_1}$ &         $1$  & $1$  & $1$   &  $-2$ &  $0$ &  $-1$  &  $0$  &  $0$ &  $0$  &  $0$ \\
$U(1)_{B_2}$ &         $1$  & $1$  & $1$   &  $-1$ &  $-2$ &  $0$  &  $0$  &  $0$ &  $0$  &  $0$ \\
$U(1)_{B_3}$ &         $1$  & $1$  & $1$   &  $0$ &  $-1$ &  $-2$  &  $0$  &  $0$ &  $0$  &  $0$ \\
$U(1)_{B_4}$ &         $1$  & $0$  & $0$   &  $0$ &  $0$ &  $0$  &  $-2$  &  $0$  &  $1$  &  $0$\\
$U(1)_{B_5}$ &         $1$  & $0$  & $0$   &  $-1$ &  $0$ &  $0$  &  $-1$  &  $1$  &  $1$  &  $0$\\
$U(1)_{B_6}$ &         $1$  & $0$  & $0$   &  $0$ &  $-1$ &  $0$  &  $-1$  &  $0$  &  $0$  &  $1$
\end{tabular}
\ee
The map between affine coordinates and gauge invariant operators is
\be\nonumber
\begin{array}{cc}
\qquad        x_1=TZ_1  =a_0^3a_1^2a_2e_0f_0g_0  \;,&
\qquad \qquad x_6=X_1Y_2Z_2  =a_0a_1a_2b_0^2e_0e_1f_0f_1g_0      \;,\\
\qquad        x_2=X_2Y_2Z_2	=b_0^3e_0e_1f_0f_1g_0			\;,&
\qquad \qquad x_7=X_2Y_2Z_3	=b_0^2c_0e_0e_1f_0f_1g_0  			\;,\\
\qquad        x_3=X_3Y_3Z_3	=c_0^3e_0e_1f_0f_1g_0 			\;,&
\qquad \qquad x_8=X_1Y_3Z_3	=a_0a_1a_2c_0^2e_0e_1f_0f_1g_0			\;,\\
\qquad        x_4=TZ_2	=a_0^2a_1b_0e_0f_0g_0			\;,&
\qquad \qquad x_9=X_2Y_3Z_3	=b_0c_0^2e_0e_1f_0f_1g_0  			\;,\\
\qquad        x_5=TZ_3	=a_0^2a_1ce_0f_0g_0 		\;,&
\qquad \qquad x_{10}=X_1Y_2Z_3 =a_0a_1a_2b_0c_0 e_0e_1f_0f_1g_0 			\;,\\
\qquad        &
\qquad \qquad x_{11}=\tilde{T} = a_1a_2^2e_1f_1 \;.
\end{array}
\ee

\subsection{Flavoring the $dP_1$ quiver}

\begin{figure}[t]
\begin{center}
\subfigure[\small The quiver.]{
\includegraphics[height=4.5cm]{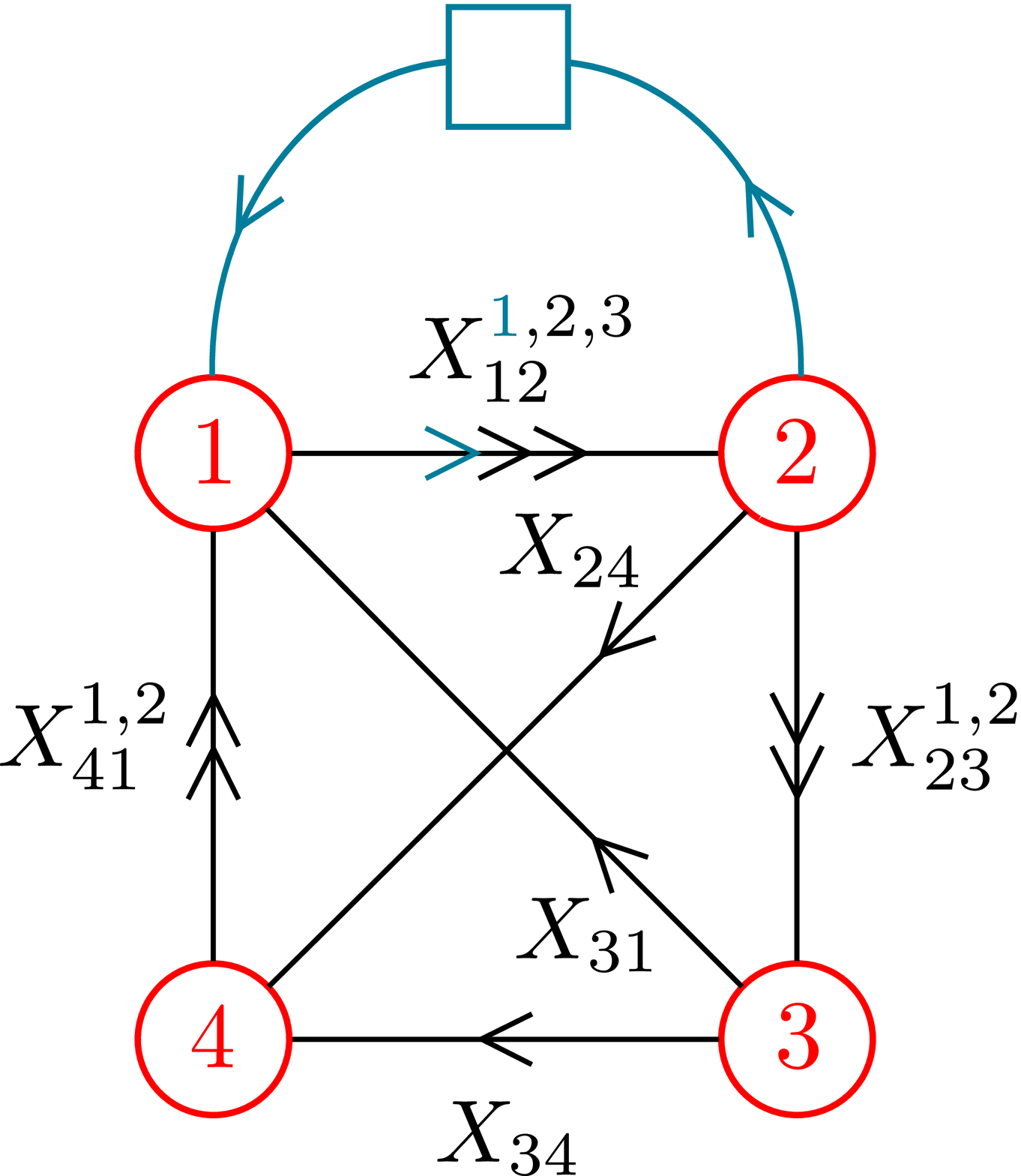}
\label{fig: dP1 quiver with 1 flav}
}\qquad
\subfigure[\small Toric diagram.]{
\includegraphics[height=4.2cm]{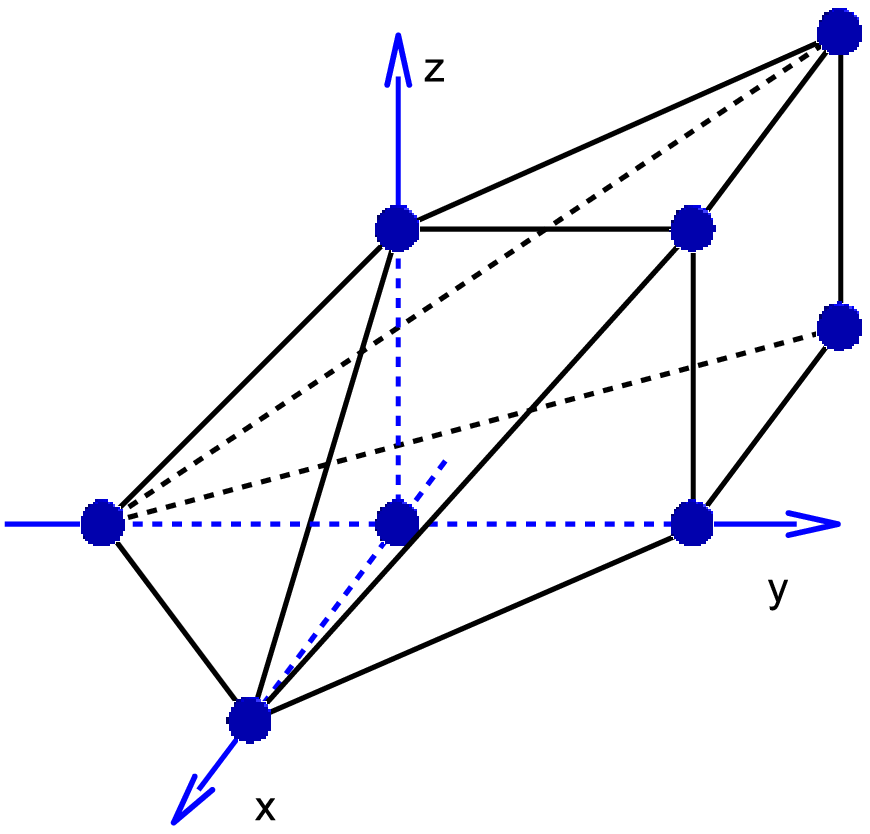}
\label{toric diag: dP1 one flav}
}
\caption{The $dP_1$ quiver with one flavor. Flavoring $X_{12}^1$ has the effect of adding a new point above two external points and one internal point in the toric diagram.}
\end{center}
\end{figure}

The $dP_1$ quiver describes D-branes at the $C(Y^{2,1})$ CY$_3$ singularity. The quiver has 4 nodes and 10 bifundamental fields, as reviewed in Appendix \ref{sec: pm}. The brane tiling is shown in Fig. \ref{fig: dP1} and its perfect matchings are given in (\ref{pm of dP1 in Appendix}).
Consider coupling a single flavor to the field $X_{12}^1$, as in Figure \ref{fig: dP1 quiver with 1 flav}. This time the field we flavor corresponds to two external points $b_0$ and $c_0$, , as well as an internal point $e_0$, in the toric diagram of $dP_1$. The Chern-Simons levels are $(\frac{1}{2}, -\frac{1}{2}, 0,0)$, which corresponds to adding three points $b_1$, $c_1$ and $e_1$ in the toric diagram, as shown in Figure \ref{toric diag: dP1 one flav}.
\be
\begin{tabular}{l|cccccccc}
                          & $X_{12}^j$ & $X_{23}^i$ & $X_{41}^i$& $X_{31}$ & $X_{24}$ & $X_{34}$  &${T}$& $\tilde T$   \\
\hline
 $U(1)_{\frac{1}{2}}$ &    $1$   &    $0$ &        $-1$  &     $-1$ &     $0$ &        $0$  &   1&  $0$     \\
 $U(1)_{\frac{1}{2}}$ &    $-1$   &   $1$ &        $0$  &     $0$ &     $1$ &        $0$  &   $-1$ &  $0$ \\
 $U(1)_{0}$ &              $0$   &    $-1$ &        $0$  &     $1$ &     $0$ &        $1$  &     $0$ &     $0$ \\
 $U(1)_{0}$ &              $0$   &    $1$ &        $1$  &     $0$ &     $-1$ &        $-1$  &     $0$ &     $0$
\end{tabular}
\ee
The quantum relation is  $T\tilde{T}= X_{12}^1$. The F-term equations are solved by
\be\nn
\begin{array}{cccc}
 X_{12}^1= b_0b_1c_0c_1e_0e_1 \;,& \qquad X_{41}^1= c_0c_1h_0 \;,& \qquad X_{23}^1= c_0c_1f_0\;,& \qquad X_{34}= b_0b_1g_0 \;, \\
 X_{12}^2 = a_0b_0b_1e_0e_1\;,& \qquad X_{41}^2= a_0h_0\;,& \qquad X_{23}^2=  a_0f_0\;,& \qquad  T= b_0c_0e_0  \;,\\
 X_{12}^3= d_0e_0e_1 \;,& \qquad X_{31}= d_0g_0h_0 \;,& \qquad X_{24}= d_0f_0g_0 \;,& \qquad \tilde{T}= b_1c_1e_1 \;.
\end{array}
\ee
The GLSM is
\be
\begin{tabular}{l|c c c c c c cc c cc}
                    & $a_0$ & $b_0$ &$c_0$ &$d_0$ & $e_0$ & $f_0$ & $g_0$ & $h_0$ & $b_1$& $c_1$ & $e_1$ \\ \hline
$U(1)_{B_1}$ &         $0$  & $0$  & $0$   &  $0$ &  $1$ &  $0$  &  $0$  &  $-1$ &  $0$  &  $0$ &  $0$ \\
$U(1)_{B_2}$ &         $0$  & $0$  & $0$   &  $0$ &  $-1$ & $1$  &  $0$  &  $0$ &  $0$  &  $0$ &  $0$\\
$U(1)_{B_3}$ &         $0$  & $0$  & $0$   &  $0$ &  $0$ &  $-1$  &  $1$  &  $0$ & $0$  &  $0$ &  $0$\\
$U(1)_{B_4}$ &         $1$  & $0$  & $1$   &  $1$ &  $-1$ & $0$  &  $-1$  &  $-1$&  $0$  &  $0$&  $0$\\
$U(1)_{B_5}$ &         $0$  & $1$  & $0$   &  $1$ &  $-1$ & $-1$  & $0$  &  $0$  &  $0$  &  $0$&  $0$\\
$U(1)_{B_6}$ &         $0$  & $1$  & $-1$  &  $0$ &  $0$ &  $0$  &  $0$  &  $0$  &  $-1$  &  $1$&  $0$\\
$U(1)_{B_7}$ &         $0$  & $1$  & $0$   &  $0$ &  $-1$ & $0$  &  $0$  &  $0$  &  $-1$  &  $0$&  $1$
\end{tabular}
\ee
The three first lines correspond to the gauge charges under the first three gauge groups. Using the F-term relations together with $T\tilde{T}= X_{12}^1$, one can show that there are only 10 independent generators of the chiral ring,
\be\nn
\begin{array}{cccc}
x_1= X_{12}^3X_{24}X_{41}^1 \;,& \qquad x_4= TX_{24}X_{41}^2 \;,& \qquad x_7= TX_{23}^1X_{34}X_{41}^2 \;,& \quad x_{10}= \tilde{T} \;. \\
x_2= X_{12}^3X_{24}X_{41}^2 \;,& \qquad x_5= X_{12}^3X_{23}^2X_{34}X_{41}^2 \;,&\qquad x_8=	 TX_{23}^2X_{34}X_{41}^2 \;, \\
x_3= TX_{24}X_{41}^1 \;,&\qquad x_6= TX_{23}^1X_{34}X_{41}^1 \;,&\qquad x_9=   X_{12}^2X_{23}^2X_{34}X_{41}^2 \;.
\end{array}
\ee
and that they match the 10 affine coordinates of the toric geometry of Figure \ref{toric diag: dP1 one flav}.


\section{Conclusions}
\label{sec: conclusions}

In this paper we studied the chiral ring of CFTs describing the IR fixed point of general 3d $\cN=2$ supersymmetric quiver gauge theories with chiral flavors, with or without CS terms, focusing on the toric case. These CFTs are conjectured to be holographically dual to M-theory on $AdS_4\times H_7$ backgrounds.

We have generalized the stringy derivation of the quiver theories \cite{Aganagic:2009zk} to cases where the M-theory circle degenerates at complex codimension-two loci in the toric $CY_4$ cone, leading to flavor D6-branes wrapping toric divisors of the fibered $CY_3$ in type IIA string theory. The holomorphic embedding of flavor branes determines the superpotential couplings between the (anti)fundamental flavor superfields and bifundamental matter in the dual theory, whereas the RR $F_2$ fluxes contributed by D6-branes shift the CS levels.

Conversely, we have studied the addition of flavors coupled to bifundamental fields in toric 3d Abelian quiver theories. Flavoring is accompanied by shifts of some CS levels in order to balance the parity anomaly. We proved that the geometric branch of the moduli space (the one where flavor fields do not acquire a VEV) of the chirally flavored quiver theories is a toric conical $CY_4$, and provided a recipe for deriving the toric diagram, exploiting an auxiliary quiver theory whose brane tiling has multi-bonds instead of flavors. The derivation of the moduli space relies on the existence of a non-trivial holomorphic OPE between BPS diagonal monopole operators, that we conjecture to appear at the quantum level since it is consistent with all global and gauge symmetries of the theory. Applying the reduction of \cite{Aganagic:2009zk} to the $CY_4$ branch, we can provide a stringy derivation of the proposed flavored gauge theories, closing the circle.

Firstly, it would be interesting to explore the Higgs branches of the flavored theories. In the presence of intersecting D6-branes, it will be crucial to understand whether new superpotential interactions arising from flavor branes intersections can appear and be marginal at the IR fixed point.
The issue may be addressed using orbifold techniques and following the result of partial resolutions, as suggested in \cite{Franco:2006es}.

Secondly, it would be nice to understand whether the auxiliary multi-bond brane tilings are dual to the flavored quiver theories we studied. This issue requires the study of the full flavored theory and A-theory moduli spaces. Partial resolutions, interpreted as Higgsings (removal of one edge in a multi-bond) in the A-theory, correspond to explicit breaking of the flavor groups due to real mass terms in the flavored theory. Even though this is reminiscent of mirror symmetry, the P- and A-theory are not geometric dual in the sense of \cite{Jensen:2009xh}: they correspond to the same M-theory reduction. The stringy derivation naturally leads to the flavored theory. Moreover, adding multi-bonds or flavorings are local operations in the brane tiling/quiver, therefore any duality between the two theories must be a local operation as well. Finally, notice that giving a VEV to a bifundamental field in the flavored theory not only Higgses the gauge groups but also gives mass to all flavors coupled to it. In the brane tiling of the A-theory, all the edges between two vertices are removed.

It would also be interesting to extend our analysis to the full class of $ADE$ singularities in M-theory, which goes beyond the toric case: $D$-type singularities descend to orientifolds in type IIA. One could also consider the addition of a Romans mass to the type IIA gravity duals with D6-branes, contributing a CS term to the diagonal gauge group \cite{Gaiotto:2009mv,Gaiotto:2009yz}: this would be particularly interesting for models with no CS terms, since it would provide a manifestly conformal action in the sense of ABJM \cite{Aharony:2008ug}. To study a large number of D6-branes, a smeared setup \cite{Acharya:2006ne} could be useful.
Finally, one could apply the projection of \cite{Forcella:2009jj} to identify $\cN=1$ dual pairs with flavors.


\section*{Acknowledgments}

We would like to thank Daniel Jafferis, Igor Klebanov, Alberto Mariotti and Yuji Tachikawa for interesting conversations, and Ofer Aharony, Riccardo Argurio, Chethan Krishnan, Peter Ouyang and Chris Herzog for various discussions on related topics. F.B. would like to thank the KITP for the kind hospitality. F.B. is supported in part by the US NSF Grant No.~PHY-0756966 and by the US NSF Grant No.~PHY-0844827.
C.C. is a Boursier FRIA-FNRS. The research of C.C. is also supported by IISN - Belgium (convention 4.4505.86) and by the
Belgian Federal Science Policy Office through the Interuniversity Attraction Pole P5/27. The work of S.C. is supported in part by the Israeli Science Foundation center of excellence, by the Deutsch-Israelische Projektkooperation (DIP), by the US-Israel Binational Science Foundation (BSF), and by the German-Israeli Foundation (GIF).


\appendix


\section{Brane tilings and the Kasteleyn matrix algorithm}
\label{sec: pm}

\begin{figure}[t]
\begin{center}
\hspace{\stretch{1}}
\begin{minipage}[b]{.28\textwidth}
\includegraphics[width=\textwidth]{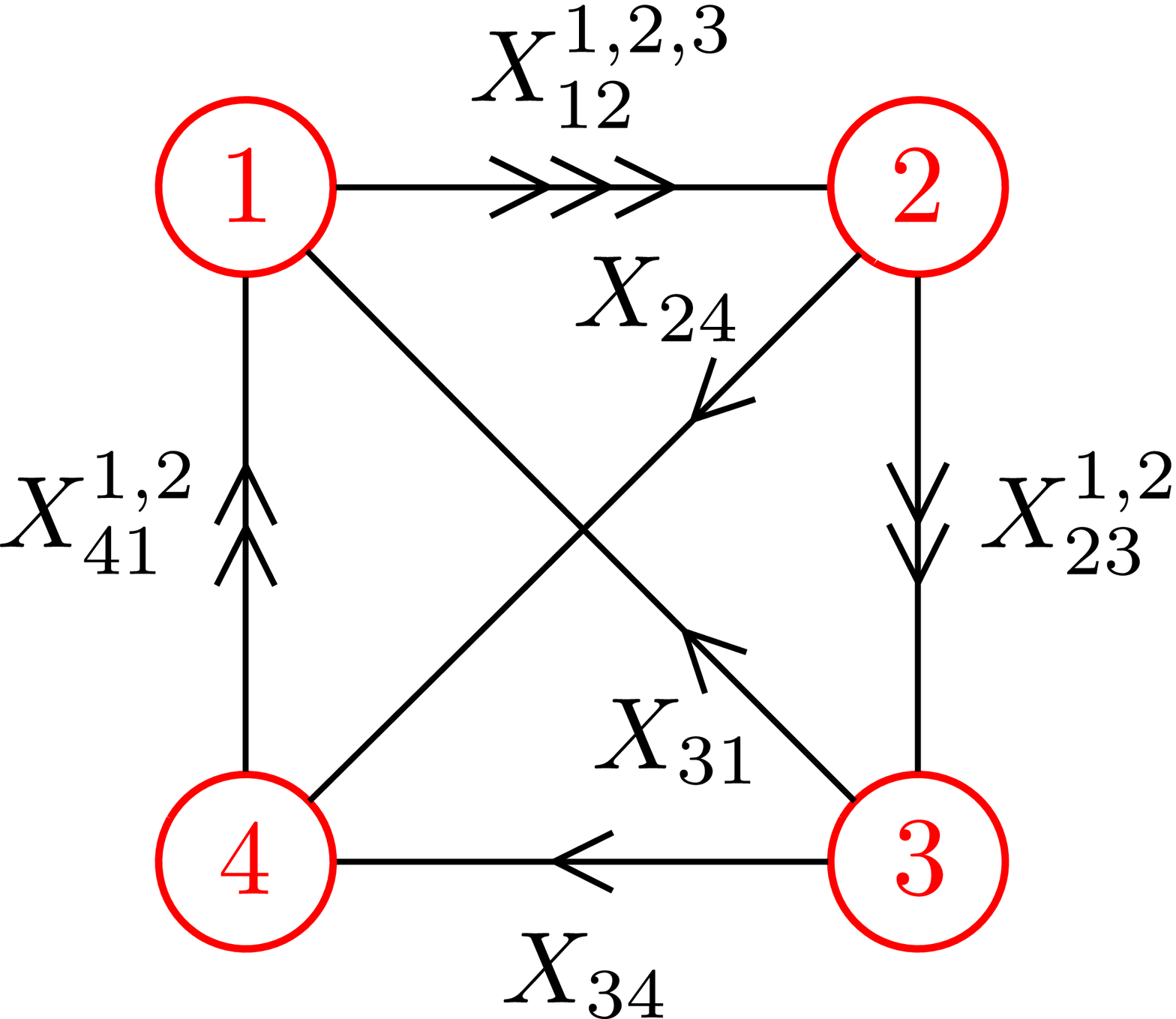}

\vspace{1ex}

\includegraphics[width=.9\textwidth]{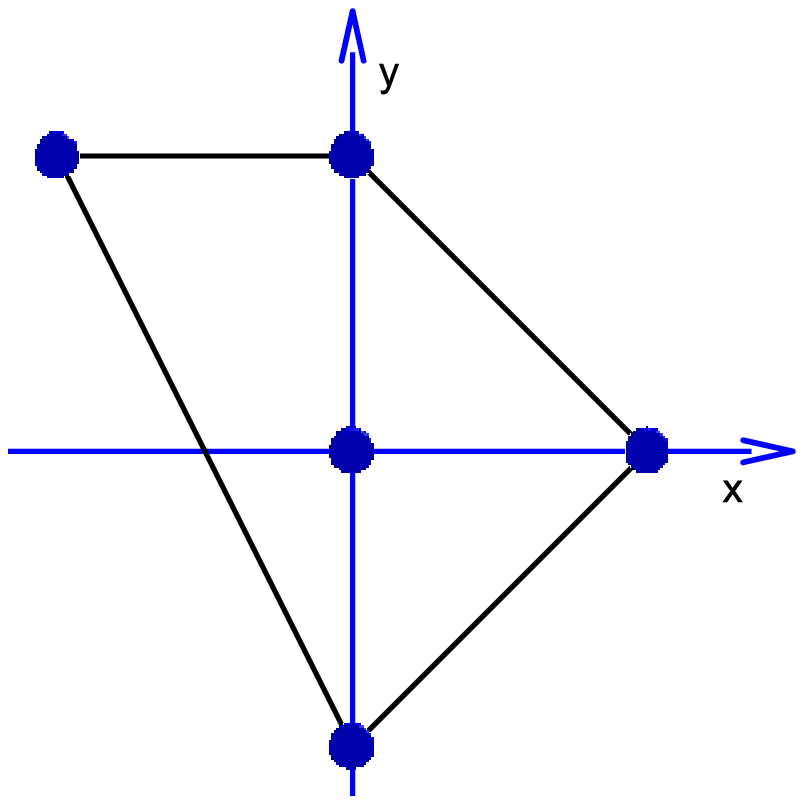}
\end{minipage}
\hspace{\stretch{1}}
\raisebox{5ex}{\includegraphics[width=.6\textwidth]{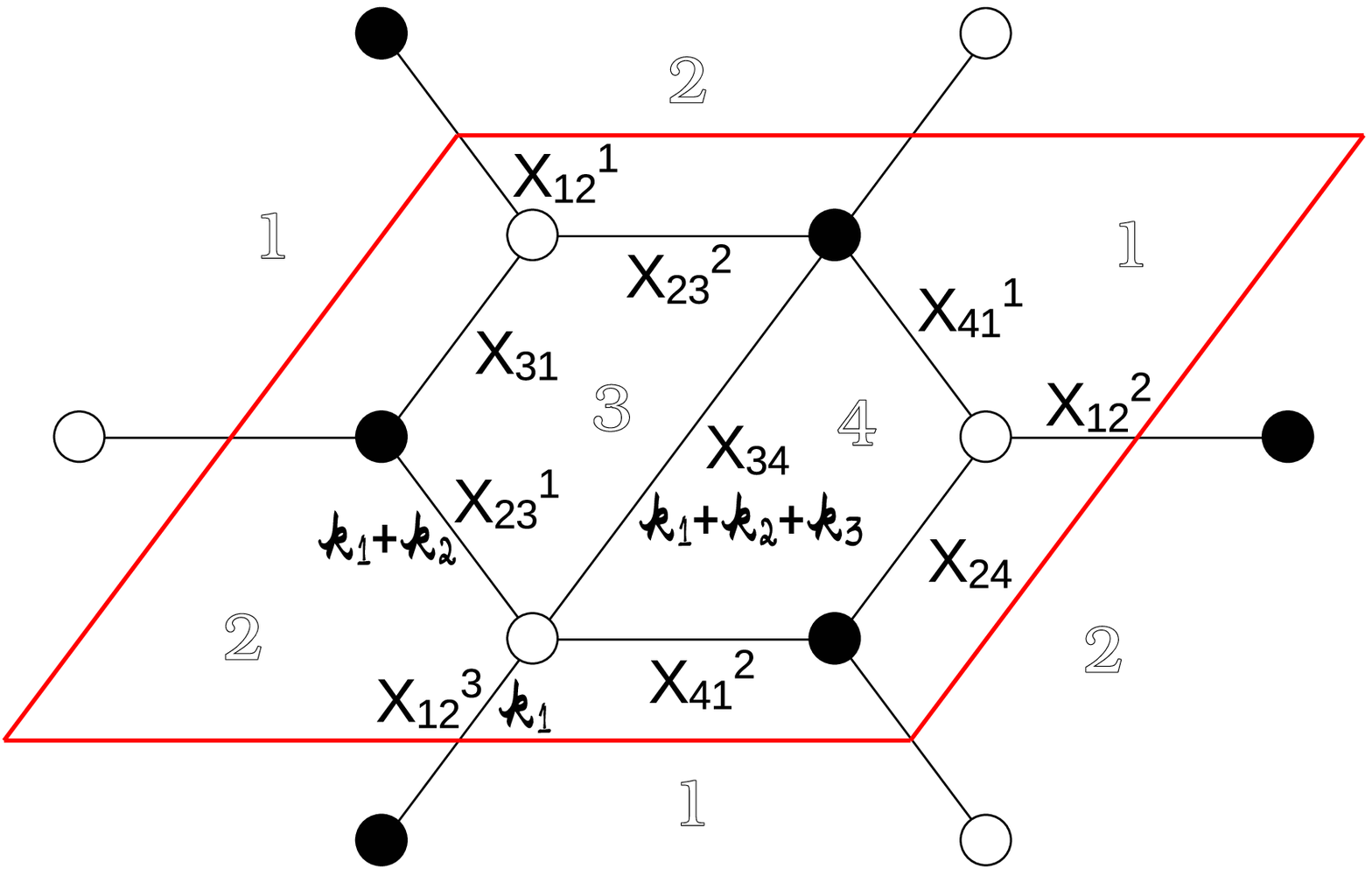}}
\hspace{\stretch{1}}

$ W = \epsilon^{ab} X_{12}^a X_{23}^b X_{31} + \epsilon^{ab} X_{41}^a X_{12}^b X_{24} + \epsilon^{ab} X_{12}^3 X_{23}^a X_{34} X_{41}^b $
\caption{Quiver diagram, toric diagram, brane tiling (refined by the $n_{ij}$ integers encoding Chern-Simons level) and superpotential of the $dP_1$ theory (also called $Y^{2,1}$). \label{fig: dP1}}
\end{center}
\end{figure}

An $\cN=1$ quiver gauge theory in 3+1 dimensions is specified by a collection of gauge groups, that we will consider all of the same type $SU(N)$, a collection of bifundamental chiral fields $X_{ij}$ in the fundamental of $SU(N)_i$ and anti-fundamental of $SU(N)_j$, and a superpotential. For a subclass of quiver theories, this information is encoded into a \emph{brane tiling}: a bipartite graph on the torus $T^2$. The graph has white and black nodes in equal number, and non-intersecting edges connecting a white and a black node. Each face represents a gauge group. Each edge represents a chiral superfield, in the fundamental of the face (gauge group) on its right looking towards the white node, and in the anti-fundamental of the face on its left. Each white (black) node represents a single-trace superpotential term, with the fields appearing in clockwise (counter-clockwise) order, and a plus (minus) sign in front. In Figure \ref{fig: dP1}, as an example, we report the superpotential, brane tiling (with fundamental domain) and quiver diagram of the $dP_1$ theory, which is dual to D3-branes probing the CY complex cone over the first del Pezzo surface, or real cone over $Y^{2,1}$. As a consequence, the superpotential of a brane tiling theory has specific properties: each field appears linearly in exactly two terms, with opposite signs. Moreover gauge anomalies vanish, and the number of gauge groups plus the number of superpotential terms equals the number of chiral fields.

A quiver theory which is a brane tiling theory (under some conditions such that all groups reach an IR conformal fixed point, see \eg{} \cite{Gulotta:2008ef}) has a mesonic moduli space which is the symmetric product of $N$ copies of a toric CY$_3$. An easy way to compute its 2d toric diagram is through the Kasteleyn matrix $K$. Each row of this matrix represents a white node, each column a black node. Each entry is a sum of monomials
\be
K_{ab} = \sum_{\gamma \,\in\, \{a \,\rightarrow\, b\}} x^{m_x^{(\gamma)}} y^{m_y^{(\gamma)}} X_\gamma
\ee
where we sum over the fields $X_\gamma$ from the white node $a$ to the black node $b$; then $x,y$ are formal parameters and $m_x^{(\gamma)}$, $m_y^{(\gamma)}$ are the number of times, with sign, the field $X_\gamma$ crosses the $x$ and $y$ boundaries of the fundamental domain. The permanent%
\footnote{The permanent of an $n \times n$ matrix $M$ is defined, similarly to the determinant, as
\be
\perm M = \sum_{i_1,\dots, i_n} (\epsilon^{i_1 \dots i_n})^2 M_{i_11} \dots M_{i_nn} \;.
\ee
}
of $K$ is a sum of terms in $x^{m_x} y^{m_y}$: each of them represents a point in the 2d toric diagram, with coordinates $(m_x, m_y)$. The ambiguity in the choice of fundamental domain translates to $SL(3,\bZ)$ transformations of the toric diagram. For the $dP_1$ example, the Kasteleyn matrix is
\be
K = \begin{pmatrix} X_{23}^2 & X_{31} & X_{12}^1 \, x^{-1} y \\ X_{34} + X_{12}^3 \, y^{-1} & X_{23}^1 & X_{41}^2 \\ X_{41}^1 & X_{12}^2 \, x & X_{24} \end{pmatrix}
\ee
and its permanent is
\begin{multline}
\perm K = \big( X_{12}^1 X_{12}^2 X_{12}^3 + X_{23}^1 X_{23}^2 X_{24} + X_{24} X_{31} X_{34} + X_{31} X_{41}^1 X_{41}^2 \big) \,+ \\
+  X_{12}^2 X_{23}^2 X_{41}^2 \, x+ X_{12}^3 X_{24} X_{31} \, y^{-1} + X_{12}^1 X_{12}^2 X_{34} \, y + X_{12}^1 X_{23}^1 X_{41}^1 \, x^{-1} y \;.
\end{multline}
The resulting 2d toric diagram (made of 5 points) is in Figure \ref{fig: dP1}.

Each coefficient of a term $x^{m_x} y^{m_y}$ is in turn a sum of monomials in the fields. Each term is a \emph{perfect matching} $t_\rho$, \ie{} a choice of edges in the tiling such that each vertex is touched once and only once. Notice that multiple perfect matchings can lie on the same point in the toric diagram. We write the relations between fields and perfect matchings as:
\be
t_\rho \,\subset\, \{X_a \;|\; a \in R^{-1}(\rho) \} \qquad\quad \leftrightarrow \qquad\quad X_a = \prod_{\rho \,\in R(a)} t_\rho \;.
\ee
With a little abuse of notation, $R^{-1}(\rho)$ gives the subset of fields associated to $t_\rho$ by $\perm K$, whilst its ``inverse'' $R(a)$ gives the subset of perfect matchings to which $X_a$ is associated.
Perfect matchings are useful because they provide a parametrization $X_a = \prod_{\rho \,\in R(a)} t_\rho$ that automatically solves the F-term equations; they can then be used as fields of a GLSM that reproduces the CY$_3$ as its moduli space.

\

An $\cN=2$ Chern-Simons quiver gauge theory in 2+1 dimensions is specified by the same data as before, this time considering gauge groups $U(N)_i$, plus a collection of CS levels $k_i$. Under the condition $\sum_i k_i = 0$, the CS levels can be included in the brane tiling by assigning a number $n_{ij}$ to each chiral field $X_{ij}$:
\be
k_i = \sum\nolimits_j (n_{ij} - n_{ji}) \;.
\ee
This means that each edge contributes $+n_{ij}$ ($-n_{ij}$) to the group on its right (left), looking towards the white node. In Figure \ref{fig: dP1} we already refined the $dP_1$ brane tiling with the integers $n_{ij}$.

A Chern-Simons quiver theory which is a brane tiling has a toric CY$_4$ geometric moduli space. Its 3d toric diagram can be computed with a refined version of the Kasteleyn matrix algorithm explained above. Each entry $K_{ab}$ is a sum of monomials
\be
K_{ab} = \sum_{\gamma \,\in\, \{a \,\rightarrow\, b\}} x^{m_x^{(\gamma)}} y^{m_y^{(\gamma)}} z^{n^{(\gamma)}} X_\gamma
\ee
where the new formal parameter $z$ is weighted by the number $n_{ij}$ associated with $X_{ij}$. As before, $\perm K$ is a sum of terms $x^{m_x} y^{m_y} z^{m_z}$: each of them represents a point of the 3d toric diagram, of coordinates $(m_x, m_y, m_z)$. The ambiguity in the choice of fundamental domain and of the numbers $n_{ij}$ translates to $SL(4,\bZ)$ transformations of the toric diagram. For the $dP_1$ example, the permanent is
\be
\begin{split}
\perm K &=  X_{31} X_{41}^1 X_{41}^2 + X_{12}^2 X_{23}^2 X_{41}^2 \, x + X_{12}^1 X_{12}^2 X_{41}^1 \, x^{-1} y \,+ \\
&\quad + X_{12}^1 X_{12}^2 X_{12}^3 \, z^{k_1} + X_{12}^3 X_{24} X_{31} \, y^{-1} z^{k_1} + X_{23}^1 X_{23}^2 X_{24} \, z^{k_1 + k_2} \,+\\
&\quad + X_{24} X_{31} X_{34} \, z^{k_1 + k_2 + k_3} + X_{12}^1 X_{12}^2 X_{34} \, y z^{k_1 + k_2 + k_3}\;.
\end{split}
\ee
Notice that setting $z=1$ we reproduce the same algorithm as before. Thus the projection of the 3d toric diagram of the 2+1 dimensional theory on the plane $z=0$ is the 2d toric diagram of the 3+1 dimensional theory.

\subsection{Perfect matchings and toric divisors}
\label{sec: pm divisors}

Each field $X_{ij}$ appears in at least one strictly external perfect matching; when it appears in more than one, the perfect matchings are consecutive along the perimeter of the 2d toric diagram. For fields appearing in a single strictly external perfect matching, the dibaryon is dual to a D3-brane wrapping the radial section of the toric divisor.
For fields appearing in more than one perfect matching, the dibaryon is dual to a union of pairwise intersecting D3-branes.

In the $dP_1$ example, the perfect matchings and the corresponding points in the toric diagram are
\bea
\label{pm of dP1 in Appendix}
a_0 &= \{X_{12}^2,X_{23}^2,X_{41}^2 \}  \,=\,  (1,0,0) \;, &
e_0 &= \{X_{12}^1,X_{12}^2,X_{12}^3\} \,=\,  (0,0,0) \;, \\
b_0 &= \{X_{12}^1,X_{12}^2,X_{34}\}  \,=\,  (0,1,0) \;, &
f_0 &= \{X_{23}^1,X_{23}^2,X_{24}\} \,=\,  (0,0,0) \;, \\
c_0 &= \{X_{12}^1,X_{23}^1,X_{41}^1\} \,=\, (-1,1,0) \;, &
g_0 &= \{X_{24},X_{31},X_{34}\} \,=\,  (0,0,0) \;, \\
d_0 &= \{X_{12}^3,X_{24},X_{31}\} \,=\, (0,-1,0) \;, \qquad &
h_0 &= \{X_{31},X_{41}^1,X_{41}^2\} \,=\,  (0,0,0) \;.
\eea
The F-term relations of the theory are solved by
\bea
X_{12}^1 &= b_0c_0e_0 \;, & X_{41}^1 &= c_0h_0 \;, & X_{23}^1 &= c_0f_0 \;, & X_{34} &= b_0g_0 \;, \\
X_{12}^2 &= a_0b_0e_0 \;, & X_{41}^2 &= a_0h_0 \;, & X_{23}^2 &= a_0f_0 \;, \\
X_{12}^3 &= d_0e_0 \;, & X_{31} &= d_0g_0h_0 \;, & X_{24} &= d_0f_0g_0 \;.
\eea
Therefore, some fields represent an irreducible toric divisor: $\{a_0 = 0\} \leftrightarrow \{X_{23}^2, X_{41}^2\}$, $\{b_0 = 0\} \leftrightarrow \{X_{34}\}$, $\{c_0 = 0\} \leftrightarrow \{X_{23}^1, X_{41}^1\}$, $\{d_0 = 0\} \leftrightarrow \{X_{12}^3, X_{24}, X_{31}\}$; other fields represent a collection of pairwise intersecting toric divisors: $\{a_0 b_0 = 0\} \leftrightarrow \{X_{12}^2\}$, $\{b_0 c_0 = 0\} \leftrightarrow \{X_{12}^1\}$.


\section{Moduli space of flavored quivers and the A-theory}
\label{sec: proof}


In this appendix we prove that the geometric moduli space of the A-theory is the same as $\cM_\mr{flav}$ in (\ref{flavored moduli space}).

Consider a single bifundamental $X_\alpha \equiv X_{ij}$ flavored by $h_\alpha$ quarks $(p_\alpha, q_\alpha)$ in the flavored theory. In the A-theory $X_{ij}$ has been substituted by $h_\alpha + 1$ bifundamentals $C_{i1}$, $C_{12}$, \dots, $C_{h_\alpha j}$, $h_\alpha$ new gauge groups $U(1)^{(l)}_1$ with $l = 1,\dots, h_\alpha$ have been added, and the other two gauge groups involved have CS levels $k_i - \gamma_\alpha$ and $k_j + \gamma_\alpha - h_\alpha$, with $0 \leq \gamma_\alpha \leq h_\alpha$, in terms of the levels $k_i$ and $k_j$ before flavoring.
As we showed in Section \ref{sec: no flav moduli space}, the geometric moduli space of the A-theory is the K\"ahler quotient
\be
\cM_\mr{A-theory} = \{ X_a, R, \tilde R \;|\; dW_A = 0, \; R \tilde R = 1 \} // U(1)^{\tilde G} \;,
\ee
where $R, \tilde R$ are the monopoles in the A-theory, $W_A$ its superpotential and $\tilde G = G + h_\alpha$ is the total number of gauge groups.

The only fields charged under the $h_\alpha$ new groups $U(1)^{(l)}_1$ are $C_{i1}, C_{12}, \dots, C_{h_\alpha j}$, $R$ and $\tilde R$. Their charges, including $U(1)_{k_i -\gamma_\alpha}$ and $U(1)_{k_j + \gamma_\alpha - h_\alpha}$, are:
\be
\nn
\begin{array}{c|ccccccc}
& C_{i1} & C_{12} & \dots & C_{h_\alpha -1, h_\alpha} & C_{h_\alpha j} & R & \tilde R \\
\hline
\tabs U(1)_{k_i - \gamma_\alpha} & 1 & 0 & \dots & 0 & 0 & k_i - \gamma_\alpha & - k_i + \gamma_\alpha \\
\tabs U(1)^{(1)}_1 & -1 & 1 & & 0 & 0 & 1 & -1 \\
\tabs U(1)^{(2)}_1 & 0  &-1 & & 0 & 0 & 1 & -1 \\
\vdots & \vdots & \vdots & & \vdots & \vdots & \vdots & \vdots \\
U(1)^{(h_\alpha-1)}_1 & 0 & 0 & & 1 & 0 & 1 & -1 \\
U(1)^{(h_\alpha)}_1   & 0 & 0 & & -1 & 1 & 1 & -1 \\
U(1)_{k_j + \gamma_\alpha - h_\alpha} & 0 & 0 & \dots & 0 & -1 & \;\; k_j + \gamma_\alpha - h_\alpha \;\; & -k_j - \gamma_\alpha + h_\alpha
\end{array}
\ee
Let us perform the K\"ahler quotient by the complexified gauge group $\prod_{l=1}^{h_\alpha} U(1)^{(l)}_1$ only: it is done by introducing gauge invariants and relations between them. The independent gauge invariants are:
\bea\label{equation T in term of R and C}
T &\equiv R \, (C_{i1})^{h_\alpha} (C_{12})^{h_\alpha - 1} \dots (C_{h_\alpha -1, h_\alpha})^1 \qquad\qquad & R \tilde R &= 1 \\
\tilde T &\equiv \tilde R \, (C_{12})^1 (C_{23})^2 \dots (C_{h_\alpha j})^{h_\alpha} & X_{ij} &\equiv C_{i1} C_{12} \dots C_{h_\alpha j} \;,
\eea
where we dubbed one of them as the old field $X_{ij}$. The only relation is
\be
T \tilde T = (X_{ij})^{h_\alpha} \;.
\ee
We see that, after quotienting, the new monopole operators are $T$, $\tilde T$ and obey a ``quantum'' F-term relation. The charges of $X_{ij}$, $T$, $\tilde T$ under the remaining two groups are:
\be
\nn
\begin{array}{c|ccc}
& X_{ij} & T & \tilde T \\
\hline
\tabs U(1)_{k_i - \gamma_\alpha}            & 1  & \;\; k_i -\gamma_\alpha + h_\alpha \;\; & - k_i + \gamma_\alpha \\
\tabs U(1)_{k_j + \gamma_\alpha - h_\alpha} & -1 & k_j + \gamma_\alpha - h_\alpha & - k_j -\gamma_\alpha
\end{array}
\ee
We want to compare these charges with those in the flavored theory. When the flavored theory is being flavored by $h_\alpha$ quarks, its CS levels have to be shifted by $\delta k_l = \big( \frac{h_\alpha}2 - \gamma_\alpha \big) \, g_l[X_{ij}]$ (\ref{CS shift h}), where $\gamma_\alpha$ is a choice of theory. Plugging into (\ref{gauge charge of diag monopole}) we get:
\be
g_l[T] = k_l + (h_\alpha - \gamma_\alpha) \, g_l[X_{ij}] \qquad\qquad g_l[\tilde T] = - k_l + \gamma_\alpha \, g_l[X_{ij}] \;.
\ee
This precisely agrees with the table above if we identify the choice of $0 \leq \gamma_\alpha \leq h_\alpha$ between the flavored and A-theory.
So the quotient by $\prod_{l=1}^{h_\alpha} U(1)^{(l)}_1$ gives the A-theory monopoles the same quantum charges as in the flavored theory.%
\footnote{The CS levels are different in the flavored and A-theory, but this does not matter. What matters for the moduli space are the charges of chiral fields.}

Let us now consider the classical F-term relations. In the A-theory, the F-terms are of two sorts: differentiating $W_A$ by a field which is not $C_{i1}, \dots, C_{h_\alpha j}$ we get the same equation as in the flavored theory, but with $X_{ij} \to C_{i1} \dots C_{h_\alpha j}$; differentiating $W_A$ by one of $C_{i1}, \dots, C_{h_\alpha j}$ we get the same equation as in the flavored theory, but multiplied by the other $C$ fields:
\be
\big( \prod C \big) \, (\text{flavored theory relation}) = 0 \;.
\ee
As long as no more than one of the $C$ fields vanishes, we exactly reproduce the same F-terms as in the flavored theory.
When more than one $C$ field vanishes, all equations become trivial and the A-theory could develop a branch which is not contained in the geometric moduli space of the flavored theory. However, the geometric moduli space of the A-theory (which is the CY$_4$) is the one where the F-terms are solved by the parametrization $X_a = \prod_{\rho \,\in R(a)} t_\rho$ (\ref{pm redefinitions}), thus if the flavored theory relations are satisfied at $C \neq 0$, they are satisfied also at $C = 0$.

We have thus shown that:
\be
\begin{split}
\cM_\mr{A-theory} &= \{ X_a^A \;|\; dW_A = 0 \} // U(1)^{\tilde G-2} = \\
&= \{ X_a^A, R, \tilde R \;|\; dW_A = 0 ,\; R \tilde R = 1\}// U(1)^{\tilde G} = \\
&= \{ X_a^F, T, \tilde T \;|\; dW_F = 0,\; T \tilde T = \prod X_\alpha^{h_\alpha} \}//U(1)^G = \cM_\mr{flav} \;.
\end{split}
\ee
The argument is straightforwardly generalized to the case that we flavor multiple fields $X_\alpha$, each with its own $h_\alpha$ quarks. This concludes the proof.

Let us conclude with a remark.
Suppose that the theory before flavoring has some global Abelian symmetry, under which $X_{ij}$ has charge $Q$. Then also the A-theory has such a symmetry, if we assign charges $Q/(h_\alpha +1)$ to $C_{i1}, \dots, C_{h_\alpha j}$. It is easy to compute that, after modding out by $\prod_l U(1)^{(l)}_1$, both $T$ and $\tilde T$ have charge $h_\alpha Q/2$. This reproduces the quantum formul\ae{} (\ref{flav charges}) and (\ref{R charge}).

\bibliographystyle{utphys}
\bibliography{bib}{}

\end{document}